\documentclass[11pt, abstract=on]{article}
\usepackage[utf8]{inputenc}
\usepackage[T1]{fontenc}
\usepackage{latexsym}
\usepackage{amssymb}
\usepackage{amsmath}
\usepackage{graphicx}
\usepackage{colortbl}
\usepackage{enumerate}
\usepackage{tikz}
\usepackage{empheq}
\usetikzlibrary{decorations.pathreplacing,shapes.misc}
\usepackage[pdfpagelabels=true]{hyperref}
\usepackage{geometry}

\usepackage[square,comma]{natbib} % this gives author [year] or [author, year]

\begin{document}

\def\Nset{\mathbb{N}}
\def\Ascr{\mathcal{A}}
\def\Bscr{\mathcal{B}}
\def\Cscr{\mathcal{C}}
\def\Dscr{\mathcal{D}}
\def\Escr{\mathcal{E}}
\def\Fscr{\mathcal{F}}
\def\Hscr{\mathcal{H}}
\def\Iscr{\mathcal{I}}
\def\Lscr{\mathcal{L}}
\def\Mscr{\mathcal{M}}
\def\Nscr{\mathcal{N}}
\def\Pscr{\mathcal{P}}
\def\Qscr{\mathcal{Q}}
\def\Rscr{\mathcal{R}}
\def\Sscr{\mathcal{S}}
\def\Wscr{\mathcal{W}}
\def\Xscr{\mathcal{X}}
\def\cupp{\stackrel{.}{\cup}}

\newcommand{\boldheader}[1]{\smallskip\noindent{\bold #1:}\quad}
\newcommand{\PP}{\mbox{\slshape P}}
\newcommand{\NP}{\mbox{\slshape NP}}
\newcommand{\opt}{\mbox{\scriptsize\rm OPT}}
\newcommand{\lp}{\mbox{\scriptsize\rm LP}}
\newcommand{\inn}{\mbox{\rm in}}
\newcommand{\MAXSNP}{\mbox{\slshape MAXSNP}}
\newtheorem{theorem}{Theorem}
\newtheorem{lemma}[theorem]{Lemma}
\newtheorem{corollary}[theorem]{Corollary}
\newtheorem{proposition}[theorem]{Proposition}
\newtheorem{definition}[theorem]{Definition}
\def\prove{\par \noindent \hbox{\bf Proof:}\quad}
\def\endproof{\eol \rightline{$\Box$} \par}
\renewcommand{\endproof}{\hspace*{\fill} {\boldmath $\Box$} \par \vskip0.5em}
\newcommand{\mathendproof}{\vskip-1.8em\hspace*{\fill} {\boldmath $\Box$} \par \vskip1.8em}
\def\cupp{\stackrel{.}{\cup}}

\definecolor{orange}{rgb}{1.0,0.5,0}
\definecolor{violet}{rgb}{0.6,0,0.8}
\definecolor{darkgreen}{rgb}{0,0.5,0}
\definecolor{grey}{rgb}{0.35,0.35,0.35}

\newcommand{\red}{\textcolor{red}}
\renewcommand{\epsilon}{\varepsilon}
\newcommand{\citegenitiv}[1]{\citeauthor{#1}' [\citeyear{#1}]}
\newcommand{\citegenitivs}[1]{\citeauthor{#1}'s [\citeyear{#1}]}
\newcommand{\sfrac}[2]{{\textstyle\frac{#1}{#2}}} 

\defaulthyphenchar=127

\title{Beating the integrality ratio for $s$-$t$-tours in graphs}
\author{Vera Traub \and Jens Vygen}
\date{\small Research Institute for Discrete Mathematics, University of Bonn \\
\texttt{\{traub,vygen\}@or.uni-bonn.de}}

\maketitle

\begin{abstract}
Among various variants of the traveling salesman problem,  
the $s$-$t$-path graph TSP has the special feature that we know
 the exact integrality ratio, $\frac{3}{2}$, and an approximation algorithm matching this ratio. In this paper, we go below this threshold:
 we devise a polynomial-time algorithm for the $s$-$t$-path graph TSP with approximation ratio $1.497$. 
 Our algorithm can be viewed as a refinement of the $\frac{3}{2}$-approximation algorithm by \cite{SebV12}, but we introduce several completely new techniques.
 These include a new type of ear-decomposition, an enhanced ear induction that reveals a novel connection to matroid union,
 a stronger lower bound, and a reduction of general instances to instances in which $s$ and $t$ have small distance (which works for general metrics).
\end{abstract}

\section{Introduction}

Since 2010, there have been many interesting results on approximation algorithms for variants of the traveling salesman problem 
(see e.g.\ the survey by \cite{Vyg12}).
This includes in particular better approximation algorithms for
the graph TSP (\cite{SebV12}), the asymmetric TSP (\cite{SveTV18}), and the $s$-$t$-path TSP (\cite{Zen18}), 

Almost all the algorithms work with the classical linear programming relaxations, 
with an LP solution as starting point of the algorithm or at least for the analysis.
Although these LPs have been studied intensively for decades,
both for the symmetric and the asymmetric TSP, we still do not know their integrality ratios.
For the (symmetric) $s$-$t$-path TSP we are quite close: the integrality ratio is between $1.5$ and $1.5284$ (\cite{SebvZ16}, \cite{TraV18b}), and in the 
$s$-$t$-path graph TSP (a well-studied special case), it is indeed $\frac{3}{2}$,  
and an approximation guarantee matching the integrality ratio is known (\cite{SebV12}).

In this paper, we go below this threshold: we show that the $s$-$t$-path graph TSP has a $1.497$-approximation algorithm.

The $s$-$t$-path graph TSP is defined as follows.
Given a connected undirected graph $G$ and two vertices $s$ and $t$, find a shortest tour from $s$ to $t$
that visits all vertices. Such a tour can be described as a sequence
$v_0,v_1,\ldots,v_k$ of vertices such that $v_0=s$, $v_k=t$, every vertex appears at least once, and $\{v_{i-1},v_i\}\in E(G)$ for all $i$ (and we minimize $k$).
Equivalently (via Euler's theorem), it can be described as a subset $F$ of $2E(G)$ (the multi-set that contains two copies of every edge) such that
$(V(G),F)$ is a connected multi-graph in which all vertices have even degree except $s$ and $t$ (which have odd degree unless $s=t$). 
Such a set $F$ is called an \emph{$s$-$t$-tour}; the goal is to minimize $|F|$. 

The $s$-$t$-path graph TSP is a special case of the general $s$-$t$-path TSP (in which the edges have arbitrary nonnegative costs).
It has been studied mainly because the well-known examples that yield the lower bounds for the integrality ratio 
($\frac{4}{3}$ for $s=t$ and $\frac{3}{2}$ for $s\not=t$) are in fact graph instances. 
Moreover, some ideas developed for the graph case triggered progress on general weights later on; an example is 
\citegenitivs{Gao13} new proof of the integrality ratio $\frac{3}{2}$ for the $s$-$t$-path graph TSP 
and its use by \cite{GotV16} for the general $s$-$t$-path TSP. 

For $s\not=t$, Christofides' algorithm yields only a $\frac{5}{3}$-approximation, even for the $s$-$t$-path graph TSP, as \cite{Hoo91} showed.
Within nine months in 2011--2012, the approximation ratio for the $s$-$t$-path graph TSP was improved four times: 
first to 1.586 by \cite{MomS16}, then  to 1.584 by \cite{Muc12}, then to 1.578 by \cite{AnKS15}, and finally to 1.5 by \cite{SebV12}.
This approximation ratio $\frac{3}{2}$ matches the integrality ratio lower bound.
However, in this paper, we improve on this. 

We present a polynomial-time algorithm that guarantees to find, for every instance of the $s$-$t$-path graph TSP, 
an $s$-$t$-tour with at most $1.497\,\opt$ edges, where $\opt$ is the minimum number of edges of such a tour.

\subsection{Preliminaries on \boldmath{$T$}-tours}\label{section:preliminary_T_tours}

Given a vertex set $V$ and a set $T\subseteq V$ of even cardinality, 
a \emph{$T$-join} is a (multi)set $J$ of edges such that $T$ is exactly the set of odd-degree vertices in $(V,J)$.
A \emph{$T$-tour} in a graph $G$ is a $T$-join $J\subseteq 2 E(G)$ such that $(V(G), J)$ is connected.
We allow taking two copies of an edge, but more than two are never useful.
Henceforth we speak of edge sets and graphs even if they contain parallel edges.
It is obvious that a $T$-tour exists if and only if $G$ is connected and $|T|$ is even.
The $s$-$t$-path graph TSP asks for an $(\{s\}\triangle\{t\})$-tour of minimum cardinality in a given connected graph $G$.
Instead of $(\{s\}\triangle\{t\})$-tour we will use the shorter name \emph{$s$-$t$-tour}.
The $s$-$t$-path graph TSP is NP-hard, and unless P=NP there is no polynomial-time
algorithm with better approximation ratio than $\frac{685}{684}$ (\cite{KarS15}).

The more general $T$-tour problem (in graphs or in general) has been introduced by \cite{SebV12}; 
they gave a $\frac{3}{2}$-approximation algorithm for finding a smallest $T$-tour in a graph, 
and a $\frac{7}{5}$-approximation algorithm if $T=\emptyset$.
\cite{Seb13} showed an approximation ratio of $\frac{8}{5}$ for finding a minimum-weight $T$-tour in a weighted graph.
These approximation ratios have not been improved since then.
Section \ref{section:ear_induction} of our paper works for general $T$-tours, 
but later parts do not seem to extend beyond constant $|T|$.

For a graph $G$ and a set $T\subseteq V(G)$ with $|T|$ even and a set $W\subseteq V(G)$, 
let $(G, T)/W$ be the instance of the $T$-tour problem arising by \emph{contraction} of $W$. 
More precisely, we define $(G, T)/W$ to be the instance of the $T$-tour problem where we are 
looking for a $T'$-tour in the graph $G/W$ and $T'$ contains all elements of $T \setminus W$
and contains in addition the vertex arising from the contraction if $|T\cap W|$ is odd.

Without loss of generality one may assume that the input graph is 2-vertex-connected because
if $G=G[W_1]\cup G[W_2]$, where $W_1$ and $W_2$ share a single vertex, then
it suffices to solve the instances $(G,T)/W_1$ and $(G,T)/W_2$ (cf.\ \cite{SebV12}).

We denote $n:=|V(G)|$ throughout this paper. 
As an obvious lower bound, note that every $T$-tour has at least $n-1$ edges (for any $T$).
Another (often better) lower bound is given by the classical linear program, which, for $T=\{s,t\}$, is
\begin{align*}
 \min \bigl\{x(E(G)) : 
& \ x(\delta(U)) \geq 2 \ (\emptyset \subset U \subseteq V(G)\setminus\{s,t\}),\ \\
& \ x(\delta(U)) \geq 1 \ (\{s\} \subseteq U \subseteq V(G)\setminus\{t\}), \ \\
& \ x_e \geq 0 \ (e\in E(G))
\bigr\},
\end{align*}
where $\delta(U)$ denotes the set of edges with exactly one endpoint in $U$,
and $x(F):=\sum_{e\in F}x_e$ for $x\in\mathbb{R}^{E(G)}$ and $F\subseteq E(G)$.
Let $\lp$ denote the value of this linear program. The \emph{integrality ratio} is the supremum 
of $\frac{\opt}{\lp}$ over all instances.
If $G$ is a circuit and $s$ and $t$ have distance $\frac{n}{2}$ in $G$, then 
the value of this LP is $n$, but every $s$-$t$-tour has at least $\frac{3}{2}n-2$ edges.
This classical example shows that the integrality ratio is at least $\frac{3}{2}$.

\subsection{Preliminaries on ear-decompositions}

For a finite sequence $P_0,P_1,\ldots,P_l$ of graphs, 
let $V_i = V(P_0) \cup V(P_1) \cup \dots \cup V(P_i)$ and $G_i=(V_i,E(P_1)\cup\dots\cup E(P_i))$.
If $P_0$ has a single vertex, and each $P_i$ is either a circuit with $|V(P_i)\cap V_{i-1}|=1$ 
or a path such that exactly its endpoints belong to $V_{i-1}$ ($i=1,\ldots,l$),
then $P_0,P_1,\ldots,P_l$ is called an \emph{ear-decomposition} of $G_l$.

The graphs $P_1,\dots,P_l$ are called \emph{ears}. 
The vertices of $V(P_i)\cap V_{i-1}$ are called \emph{endpoints} of $P_i$, the other vertices of $P_i$ are its \emph{internal vertices}.
We denote the set of internal vertices of $P_i$ by $\inn(P_i)$.
We always have $ |\inn(P_i)|=|E(P_i)| -1$.
An ear is called \emph{open} if it is $P_1$ or it is a path. Other ears are called \emph{closed}.
%Every ear is a circuit (with one endpioint, called a \emph{closed} ear) or a path (with two endpoints, called an \emph{open} ear).
If all ears are open, the ear-decomposition is called \emph{open}.
An ear is called an \emph{$r$-ear} if it has exactly $r$ edges. 
We denote the number of $r$-ears of a fixed ear-decomposition by $k_r$.
The 1-ears are also called \emph{trivial} ears; they have no internal vertices.
A \emph{short} ear is a 2-ear or 3-ear. Ears with more than three edges are called \emph{long}.
An ear is \emph{odd} if the number of its edges is odd, otherwise \emph{even}.

Figure \ref{fig:ear_induction_example_connectivity}(a) shows an ear-decomposition with $P_0$ colored black, 
a 6-ear $P_1$ colored brown, a closed 6-ear $P_2$ colored blue,
an open 5-ear $P_3$ colored green, an open 6-ear colored cyan, and five 2-ears (gray, dotted, ignore the orientation).
Every vertex is an internal vertex of exactly one ear (except for the vertex of $P_0$) and is colored accordingly in this figure.

We say that an ear $P$ is \emph{attached to} an ear $Q$ (at $v$) if $v$ is an internal vertex of $Q$ and an endpoint of $P$.
A vertex is \emph{pendant} if it is not an endpoint of any nontrivial ear, and an ear is \emph{pendant} if it is nontrivial and all its internal vertices are pendant.
Having a fixed vertex set $T$, we call an ear $P$ \emph{clean} if it is short and $|T\cap \inn(P)|=\emptyset$, 
i.e.\ none of its internal vertices is contained in $T$.

\subsection{Simple ear induction}\label{section:simple_ear_induction}

The following lemma tells how to construct a $T$-tour by considering the nontrivial ears in reverse order.
For any nontrivial ear $P_i$, note that $G_i/V_{i-1}$ is a circuit with $|E(G_i/V_{i-1})| = |E(P_i)|$.

\begin{lemma}[\cite{SebV12}]\label{lemma:standard_ear_induction}
 Let $P$ be a circuit and $T_P\subseteq V(P)$ with $|T_P|$ even. 
 Then there exists a $T_P$-join $F\subseteq 2 E(P)$ such that the graph $(V(P), F)$ is connected and
 \begin{equation}
 \label{eq:standard_ear_induction}
|F| \ \le \ \sfrac{3}{2} (|E(P)|-1) - \sfrac{1}{2} + \gamma,
\end{equation} 
where $\gamma=1$ if $|E(P)|\le 3$ and $T_P=\emptyset$, and $\gamma=0$ otherwise.
\end{lemma}
\prove
If $T_P=\emptyset$, then $F=E(P)$ does the job because $|E(P)| -1 \ge 3$ or $\gamma=1$.

Let now $T_P\ne\emptyset$. 
The vertices of $T_P$ subdivide $P$ into subpaths. Color these paths alternatingly red and blue.
Let $E_R$ and $E_B$ denote the set of edges of red and blue subpaths, respectively.
Without loss of generality $|E_R| \le |E_B|$. Then we take two copies of each edge in $E_R$ and 
one copy of each edge in $E_B$. Note that $E_R \ne \emptyset$, and remove one pair of parallel edges.
This yields $F\subseteq 2 E(P)$ with 
\[|F| \ = \ |E_B|+2|E_R|-2 \ \le \ \sfrac{3}{2}|E(P)| - 2 \ = \ \sfrac{3}{2} (|E(P)|-1) - \sfrac{1}{2}.\]
\endproof

This can be used to construct a $T$-tour as follows.
Let $P_1,\ldots,P_l$ be the nontrivial ears of an ear-decomposition of $G$ (trivial ears can be deleted beforehand).
Starting with $T_l:=T$ and $F:=\emptyset$, we do the following for $i=l,\ldots,1$.
Apply Lemma  \ref{lemma:standard_ear_induction} to $(G_i, T_i)/V_{i-1}$ and obtain a set $F_i\subseteq 2E(P_i)$. 
Set $F:=F\cup F_i$ and $T_{i-1}:=T_i\triangle \{v\in V(P_i): |\delta_{F_i}(v)| \text{ odd}\}$.
Then the union of $F_i$ and any $T_{i-1}$-tour in $G_{i-1}$ is a $T_i$-tour in $G_i$.
By induction, $F_1 \cup \dots \cup F_l$ is a $T$-tour in $G$. 
Since $|F_i| \le  \sfrac{3}{2} (|E(P_i)|-1) - \sfrac{1}{2} + \gamma_i = \sfrac{3}{2} |\inn(P_i)| - \sfrac{1}{2} + \gamma_i$,
this $T$-tour has at most $\sfrac{3}{2} (n-1)$ edges if $\gamma_i=0$ for at least half of the nontrivial ears,
in particular if most ears are long.

\subsection{Outline of the Seb\H{o}--Vygen algorithm} 

The previously best approximation algorithm for the graph $s$-$t$-path TSP, due to \cite{SebV12},
is the basis of our work. Let us briefly review this algorithm before we explain how to improve on it.
The previous section shows already why short ears (of length 2 and 3) need special attention. 

The first step is to compute a \emph{nice} ear-decomposition: one with minimum number of even ears,
in which all short ears are pendant, and internal vertices of distinct short ears are non-adjacent.
We will present a strengthening of this in Section \ref{section:ear_decomposition}.

The second step is to re-design the short ears so that as many of them as possible are part of a forest
(i.e., help connecting vertices that are not internal vertices of short ears). 
Re-designing a short ear means changing its endpoints by replacing its first and/or last edge by an edge of a trivial ear.
This can be reduced to a 
matroid intersection problem (with a graphic matroid and a partition matroid). For every short ear
that is not part of this forest, we can raise the lower bound.
We will present a refinement of this step in Section \ref{section:improving_lower_bound}.

Finally, two simple algorithms are applied to the resulting ear-decomposition.
If at least half of the nontrivial ears are long, simple ear induction (Lemma \ref{lemma:standard_ear_induction}) yields a short tour.
Otherwise one can complete the forest of short ears to a spanning tree, so that internal vertices
of short ears in the forest (which are pendant) keep their even degree, and then do parity correction like Christofides.
The catch is that this parity correction can be done in the ear-decomposition without the short ears
in the forest, and hence significantly cheaper if there are many of these.

Here we will combine the two algorithms in the last step to a single new step, which will be described in Section \ref{section:ear_induction}. 

The critical case, when this algorithm has no better approximation ratio than $\frac{3}{2}$,
is when (essentially) all ears are even (note that we save $\frac{1}{2}$ more for odd ears in \eqref{eq:standard_ear_induction}
by rounding down the right-hand side), half of the ears are 2-ears, and the 2-ears form a forest.

\subsection{Well-oriented ear-decompositions}

Given an ear-decomposition, let $\Fscr$ be a subset of the pendant ears that form a forest.
Let $ear(v)$ denote the index of the ear that contains $v$ as an internal vertex.
A \emph{rooted orientation} of $\Fscr$ is an orientation of the edges of the ears in $\Fscr$ such that each connected component is an arborescence
whose root is a vertex $v$ with $ear(v)$ minimum. Then every ear of $\Fscr$ is a directed path.
A \emph{well-oriented ear-decomposition} consists of an ear-decomposition and a rooted orientation of a subset of pendant ears that form a forest.
See Figure \ref{fig:ear_induction_example_connectivity} (a) for an example; the dotted, gray 2-ears are pendant and have a rooted orientation.

We denote by $r(w)$ the root of the connected component of the branching of oriented edges that contains $w$.
We say that an ear $Q$ \emph{enters} another ear $P$ if $Q\in\Fscr$ and there is an oriented edge $(v,w)$ of $Q$ such that $w\in\inn(P)$.
If any ear enters $P$ we call $P$ \emph{entered}; other nontrivial ears are called \emph{non-entered}.
In particular, all oriented ears are pendant and hence non-entered.

\subsection{Summary of new techniques and structure of the paper}

Although our proof can be viewed as a refined version of \cite{SebV12}, we need many new ideas,
some of which may be of independent interest or have further applications.

In Section \ref{section:ear_induction}, we describe a more sophisticated ear induction, and we 
assume that we have a well-oriented ear-decomposition in which the oriented ears are precisely the short ears.
In particular, we assume the short ears to form a forest.
First we will directly use the connectivity service of the short ears, exploiting their orientation and
revealing a novel connection to matroid union (Section \ref{sect:enhanced_ear_induction}).
This saves in many cases but does not always help. 
Therefore, we also propose a second new way to benefit from the 2-ears:
instead of taking a 2-ear as it is, one can also double one edge and discard the other, changing the parity at the endpoints.
Combining those two different possibilities of exploiting the short ears, either for connectivity or for parity,
we obtain the main result of Section \ref{section:ear_induction}.
Our ear induction algorithm saves at least $\frac{1}{26}$ for every non-entered ear, compared to $\sfrac{3}{2}(n-1)$, 
unless most of the long ears are 4-ears (Theorem \ref{thm:result_ear_induction}).

Therefore, in addition to the properties of short ears, we need an ear-decomposition with extra properties of 4-ears.
In Section \ref{section:ear_decomposition} we show that one can always obtain such an ear-decomposition in polynomial time.
In particular there will be only four types of 4-ears: pendant, blocked (with a closed ear attached to it), horizontal, or vertical,
and at most one third of the long ears can be blocked 4-ears.

Then, in Section \ref{section:improving_lower_bound} we re-design short ears but also vertical 4-ears. 
Again we can use matroid intersection, one matroid is again graphic,  
but the other one is now a laminar matroid (instead of a partition matroid as in \cite{SebV12}).
We can raise the lower bound not only for short ears that are not part of the forest, 
but also for horizontal and vertical 4-ears.

By this we remove the assumptions that the short ears form a forest and there are not too many 4-ears. See Section \ref{section:many_pendant}.
We are done unless there are only few non-entered ears.
Then there are few nontrivial ears at all because every entered ear is entered by a non-entered (short) ear.
But then it is quite easy to obtain a better approximation ratio than $\frac{3}{2}$
if in addition there is a short $s$-$t$-path $P$ in $G$.
To see this, let $G'$ result from $G$ by deleting the trivial ears (note that $|E(G')|$ is $n-1$ plus the number of nontrivial ears), and let 
the vector $x\in\mathbb{R}^{E(G)}$ be the sum of the incidence vectors of $P$ and $G'$, both multiplied by $\frac{1}{3}$.
Then one can easily show that $x$ is in the convex hull of $T$-joins for $T=\{s\}\triangle\{t\}\triangle\{v: \text{$v$ has odd degree in $G'$}\}$, and hence 
adding a minimum $T$-join results in an $s$-$t$-tour with $\frac{4}{3}|E(G')|+\frac{1}{3}|E(P)|$ edges.
One can do a bit better by applying the removable-pairing technique of \cite{MomS16} in a slightly novel way; see Section \ref{section:few_pendant}.
We prove a variant of their lemma for well-oriented ear-decompositions that works without the 2-vertex-connectivity assumption.

The only remaining case is when the distance from $s$ to $t$ in $G$ is large (close to $\frac{n}{2}$), 
but then one can apply recursive dynamic programming similar to 
\cite{Orienteering} and \cite{TraV18} (Section \ref{section:large_distance}).
This recursive dynamic programming algorithm yields a general statement about approximation
algorithms for the $s$-$t$-path TSP, not only applying to the graph case:
we show that for finding a polynomial-time $\alpha$-approximation 
for some constant $\alpha > 1$, it is sufficient to consider the special case 
where the distance of the vertices $s$ and $t$ is at most $\frac{1}{3}+ \delta$ times the cost of an 
optimum solution for some arbitrary constant $\delta >0$. 

The case in which our ear-decomposition has only very few non-entered ears is the only case in which we do not 
compare our solution to the optimum LP value but to the optimum $s$-$t$-tour.
Here, the dynamic programming algorithm allows us to bound the number of edges of our $s$-$t$-tour with respect to $\opt$ (rather than the LP value), 
which we need to obtain an approximation ratio below the integrality ratio of the LP.

The different sections of this paper can be read mostly independently of each other. 
Later sections make use only of the main result of previous sections; summarized in one theorem each:
Theorem \ref{thm:result_ear_induction} states the result of our ear induction algorithm described in Section \ref{section:ear_induction},
Theorem \ref{earmain} states the properties of the initial ear-decomposition that we construct in Section \ref{section:ear_decomposition}, and 
Theorem \ref{thm:raise_lb} gives the optimized and well-oriented ear-decomposition and the raised lower bound, as shown in Section \ref{section:improving_lower_bound}.
The rest of the paper will be relatively short.
In Section \ref{section:many_pendant} we combine the previous sections to obtain a good bound if we have many non-entered ears in our well-oriented 
ear-decomposition (Theorem \ref{thm:many_pendant}). 
In Section \ref{section:few_pendant} we use the removable-pairing technique for the case where there are few non-entered ears and the distance 
of $s$ and $t$ is small.
Combining these results we then 
obtain a $1.497$-approximation on instances in which the distance of $s$ and $t$ is small (Theorem \ref{thm:approx_guarantee}). 
Finally, in Section \ref{section:large_distance}, we show via dynamic programming that this is sufficient for obtaining a 
$1.497$-approximation for the general case of the $s$-$t$-path graph TSP (Theorem \ref{thm:long_s_t_dist_easy}).

Let us review the overall algorithm: first use the reduction to the case where $s$ and $t$ have small distance (Theorem \ref{thm:long_s_t_dist_easy}).
To solve this case, first compute an initial ear-decomposition as in Theorem \ref{earmain}. 
Based on this, compute an optimized and well-oriented ear-decomposition as in Theorem \ref{thm:raise_lb}.
If there are many non-entered ears, we get a short tour by enhanced ear induction (Theorem \ref{thm:result_ear_induction}).
If there are few non-entered ears, we obtain a short tour by the removable-pairing technique; see Lemma \ref{lemma:MS_few_ears}.
Our presentation follows a different order because each section is motivated by the previous ones.

\section{Enhanced ear induction}\label{section:ear_induction}

\subsection{Outline of our ear induction algorithm}

In this section we describe an ear induction algorithm that computes a $T$-tour, 
where $T\subseteq V(G)$ is a given even-cardinality set.
Our goal is to obtain an upper bound on the number of edges where we gain some constant amount per non-entered ear,
compared to $\frac{3}{2}(n-1)$.

For the entire Section \ref{section:ear_induction} we will assume 
that we are given a well-oriented ear-decomposition in which all short ears are clean and the oriented ears are precisely the clean ears.
In particular, the clean ears are all pendant and form a forest.
Later (in Section \ref{section:improving_lower_bound}) we consider the general case.

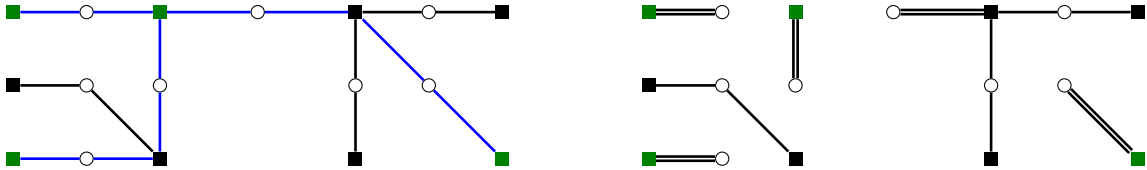
\begin{figure}
  \begin{center}
   \begin{tikzpicture}[scale=0.65]
  \tikzstyle{vertex}=[rectangle,fill,minimum size=5,inner sep=0pt]
  \tikzstyle{pendant}=[circle,draw,minimum size=5,inner sep=0pt] 
  \tikzstyle{change}=[rectangle, fill, darkgreen,minimum size=5,inner sep=0pt] 
  \tikzstyle{main}=[line width=1]
  \tikzstyle{two}=[double,line width=1.0]
  \tikzstyle{short}=[line width=1.3]
  
  \node[vertex] (a) at (2,0) {};
  \node[change] (b) at (5,0) {};
  \node[vertex] (c) at (5,3) {};
  \node[vertex] (d) at (2,3) {};
  \node[change] (e) at (-2,3) {};
  \node[change] (f) at (-5,3) {};
  \node[vertex] (g) at (-2,0) {};
  \node[vertex] (h) at (-5,1.5) {};
  \node[change] (i) at (-5,0) {};
  
   \node[pendant] (pa) at (2, 1.5) {}; 
  \node[pendant] (pb) at (3.5, 1.5) {}; 
  \node[pendant] (pc) at (3.5, 3) {}; 
  \node[pendant] (pd) at (0, 3) {}; 
  \node[pendant] (pf) at (-3.5, 3) {}; 
  \node[pendant] (pg) at (-2, 1.5) {}; 
  \node[pendant] (ph) at (-3.5, 1.5) {}; 
  \node[pendant] (pi) at (-3.5, 0) {}; 
  
  \draw[main] (a) -- (pa) -- (d);  
  \draw[main, blue] (b) -- (pb) -- (d);
  \draw[main] (c) -- (pc) -- (d);
  \draw[main, blue] (d) -- (pd) -- (e);
  \draw[main, blue] (f) -- (pf) -- (e);  
  \draw[main, blue] (g) -- (pg) -- (e);
  \draw[main] (h) -- (ph) -- (g);
  \draw[main, blue] (i) -- (pi) -- (g);
  
  \begin{scope}[shift={(13,0)}]
    
  \node[vertex] (a) at (2,0) {};
  \node[change] (b) at (5,0) {};
  \node[vertex] (c) at (5,3) {};
  \node[vertex] (d) at (2,3) {};
  \node[change] (e) at (-2,3) {};
  \node[change] (f) at (-5,3) {};
  \node[vertex] (g) at (-2,0) {};
  \node[vertex] (h) at (-5,1.5) {};
  \node[change] (i) at (-5,0) {};
  
  \node[pendant] (pa) at (2, 1.5) {}; 
  \node[pendant] (pb) at (3.5, 1.5) {}; 
  \node[pendant] (pc) at (3.5, 3) {}; 
  \node[pendant] (pd) at (0, 3) {}; 
  \node[pendant] (pf) at (-3.5, 3) {}; 
  \node[pendant] (pg) at (-2, 1.5) {}; 
  \node[pendant] (ph) at (-3.5, 1.5) {}; 
  \node[pendant] (pi) at (-3.5, 0) {}; 
  
  \draw[main] (a) -- (pa) -- (d);  
  \draw[two] (b) -- (pb);
  \draw[main] (c) -- (pc) -- (d);
  \draw[two] (d) -- (pd) ;
  \draw[two] (f) -- (pf);  
  \draw[two] (pg) -- (e);
  \draw[main] (h) -- (ph) -- (g);
  \draw[two] (i) -- (pi);
  \end{scope}
   \end{tikzpicture}
  \end{center}
  \caption{Using 2-ears for parity correction: The filled squares denote internal vertices of long ears, 
  the circles internal vertices of 2-ears. Edges of long ears and the orientation of short ears are not shown here.
  The vertex set $T'$ is shown in green. The blue edges (in the left picture)
  show a $T'$-join in this component of $G_{\gamma}$. 
  The right picture shows the same component after ``flipping'' the $T'$-join. 
  Compared to the left picture, precisely the parity of the degree of the vertices in $T'$ are changed. \label{fig:flipping_T_join}}
 \end{figure}

Let $G_{\gamma}=(V(G),E_{\gamma})$ be the spanning subgraph of $G$ that contains only the edges of clean ears. 
Due to the rooted orientation this is a branching.
Every connected component of $G_{\gamma}$ will be used either for connectivity or for parity correction.
If we use a connected component of $G_{\gamma}$ for connectivity, we add all edges of the component to our $T$-tour.
However, we can instead use a component $C$ of $G_{\gamma}$ consisting of only 2-ears for parity correction as follows:
Let $T'$ be a set of vertices that are contained in the component $C$, but are not internal vertices of short ears
(see Figure \ref{fig:flipping_T_join}). If $|T'|$ is even, we can change the parity of exactly the vertices in $T'$ by ``flipping''
the 2-ears that are part of a $T'$-join in $C$, i.e. we take two copies of one edge instead of one copy of each of the 
edges of the ``flipped'' 2-ears (see Figure \ref{fig:flipping_T_join}). 
As a consequence, we can choose the parity of all vertices that have an entering clean ear in $C$ and then 
fix the parity at the root of the component $C$ such that the set $T'$ of vertices where we need to change the 
parity of the degree has even cardinality.
We can also flip 3-ears, but then we need four instead of three edges.

If we could bound the number of edges that we need during ear induction (as in Section \ref{section:simple_ear_induction})
for every ear $P$ by $\sfrac{3}{2}|\inn(P)|$, we would obtain a $T$-tour with at most $\sfrac{3}{2} (n-1)$ edges.
Lemma \ref{lemma:standard_ear_induction} yields an even better bound for long ears (of length at least four); so we can 
gain some constant amount per long (non-entered) ear.
However, for (clean) 2-ears we need two edges, which is $\frac{1}{2}$ more than $\frac{3}{2}|\inn(P)|$, and also for 
(clean) 3-ears we can not improve over $\frac{3}{2}|\inn(P)|=3$.
To make up for this, we would like to improve over $\sfrac{3}{2}|\inn(P)|$ for long ears by some constant amount 
for each short (oriented) ear entering the ear $P$. 
In order to gain from a short ear entering $P$ at some vertex $w$, we then either exploit that the clean ears connect 
$w$ to the root $r(w)$ (if the component of $G_{\gamma}$ containing $w$ is used for connectivity) or make use of the fact
that we can choose the parity of the vertex $w$ (by possibly changing the parity at $r(w)$).

In Section \ref{sect:enhanced_ear_induction} we use the matroid union theorem to prove our main lemma for enhanced ear induction, 
which allows us to benefit from the connectivity service of the clean ears in many cases. 
We call the ears to which this lemma applies good ears.
In Section \ref{sect:first_ear_ind} we describe our ear induction algorithm that makes use of the short ears for connectivity.
However, some connected components of $G_{\gamma}$ will not be used for connectivity and we instead use them for 
parity correction in a post-processing step improving the $T$-tour found by ear induction 
(see proof of Lemma \ref{lemma:prelim_result_ear_ind_1}).
This allows us to decrease the number of edges used from so-called special ears.
The resulting $T$-tour will be short if there are many clean ears entering good or special ears 
(or many non-entered long ears; these are good ears).

The bad ears are the long ears that are neither good nor special.
If there are many bad ears, the $T$-tour resulting from Section \ref{sect:first_ear_ind} has too many edges.
To deal with this case we compute a second $T$-tour, again by ear induction. 
In contrast to the ear induction in Section \ref{sect:first_ear_ind}, we now use all components of $G_{\gamma}$ for parity correction. 
Here we gain some constant amount for every clean (oriented) ear entering a bad ear of length at least five. 
Taking the better of the two constructed $T$-tours, we can improve upon $\frac{3}{2} (n-1)$ by $\sfrac{1}{26}$ per non-entered ear,
unless a large fraction of the long ears are 4-ears.
This is our main result of Section \ref{section:ear_induction}, summarized in Theorem \ref{thm:result_ear_induction}.

\subsection{Using clean ears for connectivity via matroid union}\label{sect:enhanced_ear_induction}

In this section we prove our main lemma for enhanced ear induction.
It yields a better bound than Lemma \ref{lemma:standard_ear_induction} in many cases by 
making use of the contribution of the clean ears to connectivity.
In the proof we will need a statement about matroids that follows from the matroid union theorem.
Recall the contraction operation in matroids: if $\Mscr=(E,\Fscr)$ is a matroid and $F\in\Fscr$ is an independent set,
then $\Mscr/F:=(E\setminus F, \{Z\subseteq E\setminus F: Z\cup F\in\Fscr\})$ is well-known to be a matroid.

\begin{lemma}
\label{matroidcoloring}
Let $\Mscr=(E,\Fscr)$ be a matroid with rank function $r$, and a partition of $E$ into sets $R$, $B$, and $U$ (red, blue, and uncolored).
Then there is a partition of $U$ into sets $X$ and $Y$ such that
$r(R\cup X)+r(B\cup Y) \ge r(R\cup B)+r(U)$.
\end{lemma}

\prove
Let $R'\subseteq R$, $B'\subseteq B$, and $U'\subseteq U$ such that
$r(R\cup B)=r(R'\cup B')=|R'\cup B'|$ and $r(U)=r(U')=|U'|$.

For every $S\subseteq U'$ we have 
\begin{eqnarray*}
|U'\setminus S| + r_{\Mscr/R'}(S) +  r_{\Mscr/B'}(S)
&=& |U'\setminus S| + r(S\cup R') - |R'| + r(S\cup B') - |B'| \\
&\ge& |U'| - |S| + r(S) + r(S\cup R'\cup B') - |R'| - |B'| \\
&=& |U'| + r(S\cup R'\cup B') - |R'\cup B'| \\
&\ge& |U'|. 
\end{eqnarray*}
We used submodularity of $r$ in the first inequality.
By the matroid union theorem (\cite{Edm68}),
the minimum of the left-hand side over all $S\subseteq U'$ is the rank of $U'$ in the union of $\Mscr/R'$ and $\Mscr/B'$ (which is also a matroid).
Hence there is a partition $U'=X'\cupp Y'$ such that $X'$ is independent in $\Mscr/R'$ and $Y'$ is independent in $\Mscr/B'$.
Then any partition $U=X\cupp Y$ with $X'\subseteq X$ and $Y'\subseteq Y$ satisifies
\begin{eqnarray*}
r(R\cup X)+r(B\cup Y) 
&\ge& r(R'\cup X') + r(B'\cup Y') \\
&=& |X'| + |R'| + |Y'| + |B'| \\
&=& |R'\cup B'| + |U'| \\
&=& r(R\cup B)+r(U).
\end{eqnarray*}
\mathendproof

The special case when $R\cup B$ and $U$ are bases of $\Mscr$ 
is a well-known theorem by \cite{Bry73}, \cite{Gre73}, and \cite{Woo74};
see (42.13) in \citegenitivs{Sch03} book.

We apply the lemma in the following context:

\begin{lemma}
\label{forestcoloring}
Let $(V,E)$ be a graph (possibly with parallel edges) and a partition of $E$ into nonempty sets $R,B,U$ such that
$(V,U)$ is a forest and $(V,R\cup B)$ is a circuit.
Then there is a partition of $U$ into sets $U_R$, $U_B$, and $Z$ such that  
$(V,R\cup U_R)$ is a forest, $(V,B\cup U_B)$ is a forest, and $Z$ contains at most one element.
\end{lemma}

\prove
Apply Lemma \ref{matroidcoloring} to the cycle matroid of $(V,E)$; for its rank function $r$ we have 
$r(R\cup B)+r(U) = |R|+|B|+|U|-1$. 
We get a partition $U=X\cupp Y$ with $r(R\cup X)+r(B\cup Y)\ge |R\cup X| + |B\cup Y| - 1$.

If $r(R\cup X)=|R\cup X|$, set $U_R:=X$.
If $r(R\cup X)=|R\cup X|-1$, there is an element $z\in X$ such that $R\cup (X\setminus\{z\})$ is independent
because $(V,R)$ is a forest; then set $U_R:=X\setminus\{z\}$.

Set $U_B$ analogously, and $Z:=U\setminus(U_R\cup U_B)$.
\endproof

 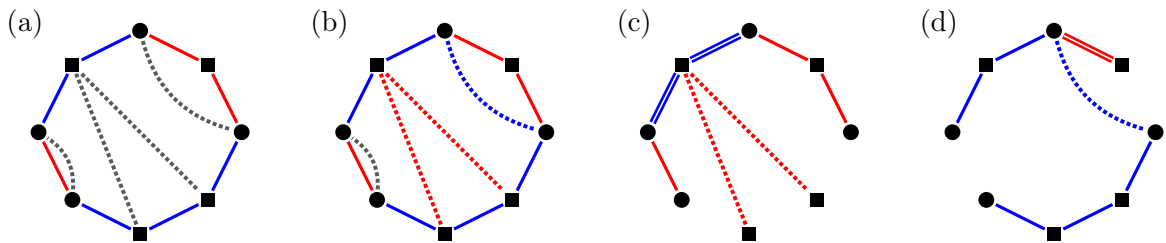
\begin{figure}
\begin{center}
 \begin{tikzpicture}[scale=0.9]

  \tikzstyle{tvertex}=[outer sep = 1pt, circle, fill,minimum size=6,inner sep=0pt]
  \tikzstyle{vertex}=[outer sep =1pt, minimum size=5,fill,inner sep=0pt]
  \tikzstyle{forest}=[grey, line width=1.5, densely dotted]
  \tikzstyle{main}=[line width=1.2]
  \tikzstyle{two}=[double,line width=1.0]
  \tikzstyle{no}=[white]
  
  \begin{scope}[shift={(0,4)}]
  \node at (-1.7,1.6) {(a)};
   \node[vertex] (a) at (0, -1.5) {};
   \node[tvertex] (b) at (-1, -1) {};
   \node[tvertex] (c) at (-1.5, 0) {};
   \node[vertex] (d) at (-1, 1) {};
   \node[tvertex] (e) at (0, 1.5) {};
   \node[vertex] (f) at (1, 1) {};
   \node[tvertex] (g) at (1.5, 0) {};
   \node[vertex] (h) at (1, -1) {};
   \draw[main, blue] (g) --(h) --(a) -- (b);
   \draw[main, red] (b) -- (c);
   \draw[main, blue] (c) -- (d) -- (e);
   \draw[main, red] (e) -- (f) -- (g);
   \draw[forest, bend right] (b) to (c);
   \draw[forest] (a) to (d);
   \draw[forest] (d) to (h);
   \draw[forest, bend right] (e) to (g);
  \end{scope}
  
    \begin{scope}[shift={(4.5,4)}]
    \node at (-1.7,1.6) {(b)};
   \node[vertex] (a) at (0, -1.5) {};
   \node[tvertex] (b) at (-1, -1) {};
   \node[tvertex] (c) at (-1.5, 0) {};
   \node[vertex] (d) at (-1, 1) {};
   \node[tvertex] (e) at (0, 1.5) {};
   \node[vertex] (f) at (1, 1) {};
   \node[tvertex] (g) at (1.5, 0) {};
   \node[vertex] (h) at (1, -1) {};
   \draw[main, blue] (g) --(h) --(a) -- (b);
   \draw[main, red] (b) -- (c);
   \draw[main, blue] (c) -- (d) -- (e);
   \draw[main, red] (e) -- (f) -- (g);
     \draw[forest, bend right] (b) to (c);
   \draw[forest, red] (a) to (d);
   \draw[forest, red] (d) to (h);
   \draw[forest, blue, bend right] (e) to (g);
  \end{scope}
  
   \begin{scope}[shift={(9,4)}]
    \node at (-1.7,1.6) {(c)};
    \node[vertex] (a) at (0, -1.5) {};
   \node[tvertex] (b) at (-1, -1) {};
   \node[tvertex] (c) at (-1.5, 0) {};
   \node[vertex] (d) at (-1, 1) {};
   \node[tvertex] (e) at (0, 1.5) {};
   \node[vertex] (f) at (1, 1) {};
   \node[tvertex] (g) at (1.5, 0) {};
   \node[vertex] (h) at (1, -1) {};
   \draw[main, red] (b) -- (c);
   \draw[main, red] (e) -- (f) -- (g);
   \draw[two, blue] (c) -- (d) -- (e);
   \draw[forest, red] (a) to (d);
   \draw[forest, red] (d) to (h);
  \end{scope}
  
    \begin{scope}[shift={(13.5,4)}]
    \node at (-1.7,1.6) {(d)};
   \node[vertex] (a) at (0, -1.5) {};
   \node[tvertex] (b) at (-1, -1) {};
   \node[tvertex] (c) at (-1.5, 0) {};
   \node[vertex] (d) at (-1, 1) {};
   \node[tvertex] (e) at (0, 1.5) {};
   \node[vertex] (f) at (1, 1) {};
   \node[tvertex] (g) at (1.5, 0) {};
   \node[vertex] (h) at (1, -1) {};
   \draw[main, blue] (g) --(h) --(a) -- (b);
   \draw[main, blue] (c) -- (d) -- (e);
   \draw[two, red] (e) -- (f);
   \draw[forest, blue, bend right] (e) to (g);
  \end{scope}
  
 \end{tikzpicture}
 \end{center}
 \caption{Enhanced ear induction in Lemma \ref{lemma:enhanced_ear_induction}: 
 (a) ear $P$ is drawn with solid lines, dotted lines indicate the edges in $U$,
black squares are elements of $V(P)\setminus T_P$; 
black circles are elements of $V(P)\cap T_P$;
(b) coloring the edges of $U$ red and blue;
(c) the red solution $F_R$;
(d) the blue solution $F_B$. \label{fig:enhanced_ear_induction}}
 \end{figure}

An \emph{antipodal pair} in a circuit $P$ is a set of two vertices of $P$ that have distance $\frac{|E(P)|}{2}$ in $P$.
Obviously, only even circuits have antipodal pairs. 
We use the following lemma to exploit the connectivity service of clean ears during ear induction.
The clean ears entering $P$ are represented by the edge set $U$ in this lemma.
A subset $C$ of $U$ will be used for connectivity.

\begin{lemma}\label{lemma:enhanced_ear_induction}
 Let $P$ be a circuit with at least four edges and $T_P\subseteq V(P)$ with $|T_P|$ even. 
 Let $(V(P),U)$ be a forest. Then one of the following is true:
 \begin{itemize}
\item[\rm (i)] $|E(P)|=4$ and $|U|=1$ and $T_P=\emptyset$;
\item[\rm (ii)] $T_P$ is an antipodal pair, and $U$ consists of a single edge with endpoints $T_P$;
\item[\rm (iii)] There exists a $T_P$-join $F\subseteq 2 E(P)$ and a subset $C \subseteq U$ such that
the graph $(V(P), F\cup C)$ is connected, 
\begin{equation}\label{eq:lower_bound_gain}
|F| \ \le \ \sfrac{3}{2}(|E(P)|-1) - \sfrac{1}{2} |U| -\sfrac{1}{2} \max\{1,\, |U| -1\},
\end{equation}
and $|C| \le 2 \left(\sfrac{3}{2}(|E(P)|-1) - \sfrac{1}{2} |U| - |F|\right)$.
\end{itemize}
\end{lemma}

\prove 
Similarly to the proof of Lemma \ref{lemma:standard_ear_induction}, we distinguish two cases.
\\[2mm]
\textbf{Case 1:} $T_P\ne \emptyset$. \\[2mm]
The vertices of $T_P$ subdivide $P$ into subpaths, alternatingly colored red and blue.
Let $E_R$ and $E_B$ denote the set of edges of red and blue subpaths, respectively.
Let $T_R$ and $T_B$ be the set of vertices having odd degree in $(V(P_i),E_R)$ and $(V(P_i),E_B)$,
respectively. Note that $\{E_R,E_B\}$ is a partition of $E(P)$, both sets are nonempty, and
$T_R=T_B=T_P$.
Color the edges in $U$ red and blue according to Lemma \ref{forestcoloring};
one edge may remain uncolored. Let $C$ be the set of colored edges in $U$.
See Figure \ref{fig:enhanced_ear_induction}.

We consider two solutions:
To construct $F_R$, we take $E_R$ plus two copies of some edges of $E_B$.
Since $E_R$ plus the red elements of $U$ form a forest, the number of blue edges needed for connectivity (with two copies each) 
is at most $|E_B|-1$ minus the number of red elements of $U$.
To construct $F_B$, we exchange the roles of red and blue.

The smaller of the two has at most
 $\frac{1}{2}(3|E(P)|-4 - 2|C|) \le
 \frac{1}{2}(3|E(P)|-4 - (|U|-1) - |C|)
 = \frac{3}{2}(|E(P)|-1) - \frac{1}{2}|U| -\frac{1}{2}|C|$
edges.
Since $|C| \ge |U|-1$, we are done if $|U| > 1$ or $|U|= |C| = 1$. 

Now let $|U|\le 1$. 
If $U=\emptyset$, then the smaller of the sets $F_B$ and $F_R$ has size at most 
 $\frac{1}{2}(3|E(P)|-4) = \frac{3}{2}(|E(P)|-1) - \frac{1}{2}$.
  
Now consider the remaining case that $|U|=1$ and $C=\emptyset$, i.e.,
 the only edge in $U$ cannot be colored.
This means that the endpoints of this edge are connected by a path in $E_R$ and a path in $E_B$, and hence 
$T_P$ consists of exactly these two elements. If (ii) does not hold, $T_P$ is not an antipodal pair, and hence $|E_R| \ne |E_B|$.
Then the smaller of the sets $F_B$ and $F_R$ has size at most
$ |E(P)| + \min \{|E_R|, |E_B|\} - 2 \le \frac{3}{2}|E(P)| - \frac{1}{2} - 2 = \frac{3}{2}(|E(P)|-1) - \frac{1}{2}|U| - \frac{1}{2}$.
\\[2mm]
\textbf{Case 2:} $T_P = \emptyset$. \\[2mm]
 We can set $F=E(P)$ and $C= \emptyset$, but instead we can also set $C = U$ and
 take all but $|U|+1$ edges, each with two copies, making $2(|E(P)|-|U|-1)$ edges.
 The smaller of the two choices for $F$ has at most
  $\frac{1}{2}(3|E(P)|-2|U|-2) = \frac{3}{2} (|E(P)|-1) - \frac{1}{2}|U| - \frac{1}{2}(|U|-1)$ edges.
  If $|U| > 1$, this implies \eqref{eq:lower_bound_gain}.
  If $|U| > 1$ and both solutions have the same number of edges, $F=E(P)$ and $C= \emptyset$ fulfills (iii).
  If $|U| > 1$ and one of the two solutions for $F$ has fewer edges, the smaller of the two choices for $F$ has at most
  $\frac{1}{2}(3|E(P)|-2|U|-3) = \frac{3}{2}(|E(P)|-1) - \frac{1}{2}|U| - \frac{1}{2}|U|$ edges.
  Then we also have $|C| \le |U| = 2 \left(\sfrac{3}{2}(|E(P)|-1) - \sfrac{1}{2} |U| - |F|\right)$.

 If $|U|\le 1$, we have (i) or $|E(P)| -1 \ge 3+|U|$.
 In the latter case,
 $|E(P)| \le  \frac{3}{2}(|E(P)|-1) - \frac{1}{2}|U| - \frac{1}{2}$.
 Hence, we can set $F=E(P)$ and $C= \emptyset$.
\endproof

\subsection{Bad, special, and good ears}\label{sect:first_ear_ind}

Let $P_1$, \dots, $P_l$ be the long ears of our ear-decomposition, i.e., the ears of length at least four.
Roughly speaking, a long ear is good if we can apply Lemma \ref{lemma:enhanced_ear_induction} (iii) to it.
Let us consider the exceptions. We again refer to our orientation of the clean ears.

\begin{definition}\label{def:bad_pair}
 Call a pair $(P, T_P)$ for an ear $P$ and a set $T_P \subseteq \inn(P)$ a \emph{bad pair} 
 if exactly one clean ear $Q$ enters $P$, $|E(P)|$ is even and $P$ fulfills one of the following properties:
 \begin{enumerate}[(a)]
  \item $|E(P)| = 4$ and $T_P=\emptyset$,
  \item $|E(P)| > 4$, $Q$ enters $P$ at its middle internal vertex $w$, $T_P = \{w\}$,
        and $r(w)$ is not an internal vertex of $P$.
 \end{enumerate}
\end{definition}
\begin{definition}
 Let $P$ be an even ear with at least six edges such that  exactly one clean ear $Q$ enters $P$.
 Denote by $w$ the internal vertex of $P$ where $Q$ enters $P$.
 If $r(w)\in \inn(P)$ and if $r(w)$ and $w$ have distance $\frac{|E(P)|}{2}$ in $P$, we call the pair 
 $(P, \{r(w), w\})$ \emph{special}.

Call a pair $(P, T_P)$ for a long ear $P$ and a set $T_P \subseteq \inn(P)$ a \emph{good pair} if it is neither bad nor special.
\end{definition}

%NEU:
 \begin{figure}
\begin{center}
 \begin{tikzpicture}[scale=0.5]

  \tikzstyle{tvertex}=[outer sep=1pt, circle,fill,minimum size=6,inner sep=0pt]
  \tikzstyle{vertex}=[outer sep =1pt, minimum size=5,fill,inner sep=0pt]
  \tikzstyle{forest}=[grey, line width=1.5, densely dotted]
  \tikzstyle{main}=[line width=1.2]
  \tikzstyle{two}=[double,line width=1.0]
  \tikzstyle{no}=[white]
  
\begin{scope}[shift={(0,0)}]
\node at (3,10) {(a)};
   \node[vertex] (a0) at (7, -2) {};
   \node[vertex, brown] (a1) at (9, -2) {};
   \node[vertex, brown] (a2) at (9, 0) {};
   \node[vertex, brown] (a3) at (8, 2) {};
   \node[vertex, brown] (a4) at (6, 2) {};
   \node[vertex, brown] (a5) at (6, 0) {};
   \draw[main, brown] (a0) -- (a1);
   \draw[main, brown] (a1) -- (a2);
   \draw[main, brown] (a2) -- (a3);
   \draw[main, brown] (a3) -- (a4);
   \draw[main, brown] (a4) -- (a5);
   \draw[main, brown] (a5) -- (a0);
   \node[vertex, blue] (b1) at (7, 3.2) {};
   \node[vertex, blue] (b2) at (7, 4.8) {};
   \node[vertex, blue] (b3) at (8, 6) {};
   \node[vertex, blue] (b4) at (9, 4.8) {};
   \node[vertex, blue] (b5) at (9, 3.2) {};
   \draw[main, blue] (a3) -- (b1);
   \draw[main, blue] (b1) -- (b2);
   \draw[main, blue] (b2) -- (b3);
   \draw[main, blue] (b3) -- (b4);
   \draw[main, blue] (b4) -- (b5);
   \draw[main, blue] (b5) -- (a3);
   \node[vertex, darkgreen] (c1) at (11.5, 3) {};
   \node[vertex, darkgreen] (c2) at (13, 6) {};
   \node[vertex, darkgreen] (c3) at (11, 6.5) {};
   \node[vertex, darkgreen] (c4) at (9.5, 6.5) {};
   \draw[main, darkgreen] (a3) -- (c1);
   \draw[main, darkgreen] (c1) -- (c2);
   \draw[main, darkgreen] (c2) -- (c3);
   \draw[main, darkgreen] (c3) -- (c4);
   \draw[main, darkgreen] (c4) -- (b3);
   \node[vertex, cyan] (d1) at (4, 3) {};
   \node[vertex, cyan] (d2) at (4, 6) {};
   \node[vertex, cyan] (d3) at (5.5, 9) {};
   \node[vertex, cyan] (d4) at (8.5, 10) {};
   \node[vertex, cyan] (d5) at (12.5, 9) {};
   \draw[main, cyan] (a5) -- (d1);
   \draw[main, cyan] (d1) -- (d2);
   \draw[main, cyan] (d2) -- (d3);
   \draw[main, cyan] (d3) -- (d4);
   \draw[main, cyan] (d4) -- (d5);
   \draw[main, cyan] (d5) -- (c2);
   \node[vertex, gray] (e) at (5.75, 6) {};
   \draw[forest,->] (a4) -- (e);
   \draw[forest,->] (e) -- (d3);
   \node[vertex, gray] (f) at (10.5, 8.5) {};
   \draw[forest,->] (d3) -- (f);
   \draw[forest,->] (f) -- (c2);
   \node[vertex, gray] (g) at (13, 1) {};
  \draw[forest,->] (c2) -- (g);
  \draw[forest,->] (g) -- (a1);
   \node[vertex, gray] (i) at (10.75, 4.5) {};
   \draw[forest,->] (c4) -- (i);
   \draw[forest,->] (i) -- (c1);
   \node[vertex, gray] (j) at (6.75, 7.5) {};
   \draw[forest,->] (d3) to (j);
   \draw[forest,->] (j) to (b3);
\end{scope}

\begin{scope}[shift={(16.5,0)}]
\node at (3,10) {(b)};
   \node[vertex] (a0) at (7, -2) {};
   \node[vertex, brown] (a1) at (9, -2) {};
   \node[vertex, brown] (a2) at (9, 0) {};
   \node[vertex, brown] (a3) at (8, 2) {};
   \node[vertex, brown] (a4) at (6, 2) {};
   \node[vertex, brown] (a5) at (6, 0) {};
   \draw[main, brown] (a0) -- (a1);
   \draw[two, brown] (a1) -- (a2);
   \draw[no] (a2) -- (a3);
   \draw[two, brown] (a3) -- (a4);
   \draw[main, brown] (a4) -- (a5);
   \draw[main, brown] (a5) -- (a0);
   \node[vertex, blue] (b1) at (7, 3.2) {};
   \node[vertex, blue] (b2) at (7, 4.8) {};
   \node[vertex, blue] (b3) at (8, 6) {};
   \node[vertex, blue] (b4) at (9, 4.8) {};
   \node[vertex, blue] (b5) at (9, 3.2) {};
   \draw[two, blue] (a3) -- (b1);
   \draw[no] (b1) -- (b2);
   \draw[two, blue] (b2) -- (b3);
   \draw[main, blue] (b3) -- (b4);
   \draw[main, blue] (b4) -- (b5);
   \draw[main, blue] (b5) -- (a3);
   \node[vertex, darkgreen] (c1) at (11.5, 3) {};
   \node[vertex, darkgreen] (c2) at (13, 6) {};
    \node[vertex, darkgreen] (c3) at (11, 6.5) {};
   \node[vertex, darkgreen] (c4) at (9.5, 6.5) {};
   \draw[main, darkgreen] (a3) -- (c1);
   \draw[no] (c1) -- (c2);
   \draw[main, darkgreen] (c2) -- (c3);
   \draw[main, darkgreen] (c3) -- (c4);
   \draw[no] (c4) -- (b3);
   \node[vertex, cyan] (d1) at (4, 3) {};
   \node[vertex, cyan] (d2) at (4, 6) {};
   \node[vertex, cyan] (d3) at (5.5, 9) {};
   \node[vertex, cyan] (d4) at (8.5, 10) {};
   \node[vertex, cyan] (d5) at (12.5, 9) {};
   \draw[two, cyan] (a5) -- (d1);
   \draw[two, cyan] (d1) -- (d2);
   \draw[no] (d2) -- (d3);
   \draw[main, cyan] (d3) -- (d4);
   \draw[main, cyan] (d4) -- (d5);
   \draw[main, cyan] (d5) -- (c2);
   \node[vertex, gray] (e) at (5.75, 6) {};
   \draw[forest] (a4) -- (e);
   \draw[forest] (e) -- (d3);
   \node[vertex, gray] (f) at (10.5, 8.5) {};
   \draw[forest] (d3) -- (f);
   \draw[forest] (f) -- (c2);
   \node[vertex, gray] (g) at (13, 1) {};
   \draw[forest] (c2) -- (g);
   \draw[forest] (g) -- (a1);
   \node[vertex, gray] (i) at (10.75, 4.5) {};
   \draw[forest] (c4) -- (i);
   \draw[forest] (i) -- (c1);
   \node[vertex, gray] (j) at (6.75, 7.5) {};
   \draw[forest] (d3) to (j);
   \draw[forest] (j) to (b3);
\end{scope}

\begin{scope}[shift={(-4.5,-12.5)}]
\node at (7.5,7.5) {(c)};
   \node[outer sep = 1pt, minimum size=15, gray, circle, fill] (contracted) at (8, 4) {};
   \node[tvertex, darkgreen] (c1) at (11.5, 3) {};
   \node[tvertex, darkgreen] (c2) at (13, 6) {};
    \node[vertex, darkgreen] (c3) at (11, 6.5) {};
   \node[tvertex, darkgreen] (c4) at (9.5, 6.5) {};
   \draw[main, darkgreen] (contracted) -- (c1);
   \draw[main, darkgreen] (c1) -- (c2);
   \draw[main, darkgreen] (c2) -- (c3);
   \draw[main, darkgreen] (c3) -- (c4);
   \draw[main, darkgreen] (c4) -- (contracted);
   \draw[forest] (c1) to (c4);
   \draw[forest] (contracted) -- (c2);
\end{scope}
\begin{scope}[shift={(-4.5,-19)}]
\node at (7.5,7) {(c1)};
   \node[outer sep = 1pt, minimum size=15, gray, circle, fill] (contracted) at (8, 4) {};
    \node[tvertex, darkgreen] (c1) at (11.5, 3) {};
   \node[tvertex, darkgreen] (c2) at (13, 6) {};
    \node[vertex, darkgreen] (c3) at (11, 6.5) {};
   \node[tvertex, darkgreen] (c4) at (9.5, 6.5) {};
   \draw[main, red] (contracted) -- (c1);
   \draw[main, blue] (c1) -- (c2);
   \draw[main, red] (c2) -- (c3);
   \draw[main, red] (c3) -- (c4);
   \draw[main, blue] (c4) -- (contracted);
   \draw[forest, blue] (c1) to (c4);
   \draw[forest, red] (contracted) -- (c2);
\end{scope}
\begin{scope}[shift={(4,-19)}]
\node at (7.5,7) {(c2)};
   \node[outer sep = 1pt, minimum size=15, gray, circle, fill] (contracted) at (8, 4) {};
    \node[tvertex, darkgreen] (c1) at (11.5, 3) {};
   \node[tvertex, darkgreen] (c2) at (13, 6) {};
    \node[vertex, darkgreen] (c3) at (11, 6.5) {};
   \node[tvertex, darkgreen] (c4) at (9.5, 6.5) {};
   \draw[main, red] (contracted) -- (c1);
   \draw[main, blue] (c1) -- (c2);
   \draw[main, red] (c2) -- (c3);
   \draw[main, red] (c3) -- (c4);
   \draw[main, blue] (c4) -- (contracted);
   \draw[forest, red] (c1) to (c4);
   \draw[forest] (contracted) -- (c2);
\end{scope}

\begin{scope}[shift={(16.5,-15)}]
\node at (3,10) {(d)};
   \node[vertex] (a0) at (7, -2) {};
   \node[vertex, brown] (a1) at (9, -2) {};
   \node[vertex, brown] (a2) at (9, 0) {};
   \node[vertex, brown] (a3) at (8, 2) {};
   \node[vertex, brown] (a4) at (6, 2) {};
   \node[vertex, brown] (a5) at (6, 0) {};
   \draw[main, brown] (a0) -- (a1);
   \draw[main, brown] (a1) -- (a2);
   \draw[main, brown] (a2) -- (a3);
   \draw[main, brown] (a3) -- (a4);
   \draw[main, brown] (a4) -- (a5);
   \draw[main, brown] (a5) -- (a0);
   \node[vertex, blue] (b1) at (7, 3.2) {};
   \node[vertex, blue] (b2) at (7, 4.8) {};
   \node[vertex, blue] (b3) at (8, 6) {};
   \node[vertex, blue] (b4) at (9, 4.8) {};
   \node[vertex, blue] (b5) at (9, 3.2) {};
   \draw[two, blue] (a3) -- (b1);
   \draw[no] (b1) -- (b2);
   \draw[two, blue] (b2) -- (b3);
   \draw[main, blue] (b3) -- (b4);
   \draw[main, blue] (b4) -- (b5);
   \draw[main, blue] (b5) -- (a3);
   \node[vertex, darkgreen] (c1) at (11.5, 3) {};
   \node[vertex, darkgreen] (c2) at (13, 6) {};
   \node[vertex, darkgreen] (c3) at (11, 6.5) {};
   \node[vertex, darkgreen] (c4) at (9.5, 6.5) {};

   \draw[main, darkgreen] (a3) -- (c1);
   \draw[no] (c1) -- (c2);
   \draw[main, darkgreen] (c2) -- (c3);
   \draw[main, darkgreen] (c3) -- (c4);
   \draw[no] (c4) -- (b3);
   \node[vertex, cyan] (d1) at (4, 3) {};
   \node[vertex, cyan] (d2) at (4, 6) {};
   \node[vertex, cyan] (d3) at (5.5, 9) {};
   \node[vertex, cyan] (d4) at (8.5, 10) {};
   \node[vertex, cyan] (d5) at (12.5, 9) {};
   \draw[two, cyan] (a5) -- (d1);
   \draw[two, cyan] (d1) -- (d2);
   \draw[no] (d2) -- (d3);
   \draw[main, cyan] (d3) -- (d4);
   \draw[main, cyan] (d4) -- (d5);
   \draw[main, cyan] (d5) -- (c2);
   \node[vertex, gray] (e) at (5.75, 6) {};
   \draw[no] (a4) -- (e);
   \draw[forest, two] (e) -- (d3);
   \node[vertex, gray] (f) at (10.5, 8.5) {};
   \draw[no] (d3) -- (f);
   \draw[forest, two] (f) -- (c2);
   \node[vertex, gray] (g) at (13, 1) {};
   \draw[no] (c2) -- (g);
   \draw[forest, two] (g) -- (a1);
   \node[vertex, gray] (i) at (10.75, 4.5) {};
   \draw[forest] (c4) -- (i);
   \draw[forest] (i) -- (c1);
   \node[vertex, gray] (j) at (6.75, 7.5) {};
   \draw[forest] (d3) to (j);
   \draw[forest] (j) to (b3);
\end{scope}
 \end{tikzpicture}
 \end{center}
 \caption{An example with $T=\emptyset$. (a): ears with oriented short ears (gray, dotted). (b): result of enhanced ear induction, using short ears for connectivity.
 The cyan ear is bad, the green ear is good, the blue ear is bad, and with the choices made as in the figure, the brown ear is special.
 (c): Applying Lemma \ref{lemma:enhanced_ear_induction} to the good (green) ear. (c1) and (c2) show two different ways 
 to color the edges in $U$. In (c2) one edge remains uncolored (although it would be possible to color it).
 (d): The $T$-tour after applying the modification described in the proof of Lemma \ref{lemma:prelim_result_ear_ind_1}
 if the coloring as in (c2) is chosen. If the coloring of the set $U$ is the one shown in (c1),
 the algorithm described in the proof of Lemma \ref{lemma:prelim_result_ear_ind_1} won't modify the $T$-tour.
 \label{fig:ear_induction_example_connectivity}}
 \end{figure}
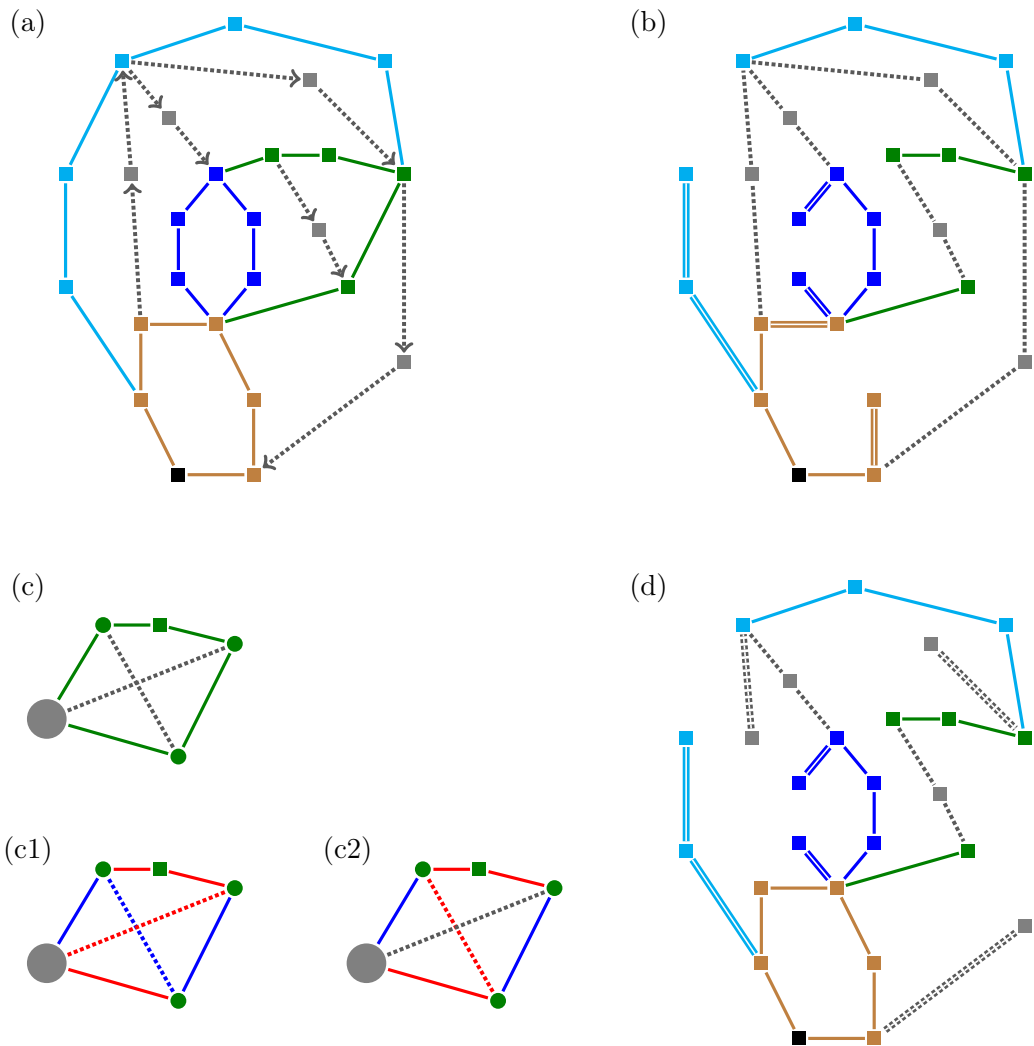

We now describe an algorithm that computes a short $T$-tour if we have many ears that are good or special.
To this end we first use ear induction to obtain a short $T$-tour if many ears are good and we can 
thus apply Lemma \ref{lemma:enhanced_ear_induction} (iii) often.
We then show that we can afterwards improve the resulting $T$-tour if there are many special ears
(cf.\ the proof of Lemma \ref{lemma:prelim_result_ear_ind_1}).

For a long ear $P_i$ (with $i\in \{1, \dots, l\}$) let $h_i$ denote the number of clean ears entering $P_i$.
For a multi-set $F \subseteq 2E(P_i)$ let 
\begin{equation*}
 \text{gain}_i(F) \ := \ \sfrac{3}{2}|\inn(P_i)| - \sfrac{1}{2} h_i - |F|.
\end{equation*}

\begin{lemma}\label{lemma:counting_ear_ind_1}
Let $F_i \subseteq 2E(P_i)$ be a multi-set for every $i=1, \dots l$. Then
 \[ |E_{\gamma}| + \sum_{i=1}^l |F_i|  \ = \ \sfrac{3}{2} (n-1) - \sfrac{1}{2} k_3 - \sum_{i=1}^l \textnormal{gain}_i(F_i). \]
\end{lemma}
\prove
 \begin{align*}
 |E_{\gamma}| + \sum_{i=1}^l |F_i|   
\ & = \ 2 \cdot k_2 + 3\cdot k_3  + \sum_{i=1}^l |F_i| \\
  & = \ 2 \cdot k_2 + 3\cdot k_3  + \sum_{i=1}^l \left(\sfrac{3}{2}|\inn(P_i)| - \sfrac{1}{2} h_i -  \text{gain}_i(F_i) \right) \\
  & = \ \sum_{P\text{ short ear}}\sfrac{3}{2} |\inn(P)| + \sfrac{1}{2} k_2  
  + \sum_{i =1 }^l \left(\sfrac{3}{2}|\inn(P_i)| - \sfrac{1}{2} h_i \right) 
  - \sum_{i =1 }^l \text{gain}_i(F_i)\\
  & = \ \sfrac{3}{2} (n-1) - \sfrac{1}{2} k_3  - \sum_{i =1 }^l \text{gain}_i(F_i).
 \end{align*}
\endproof

For a subset $C$ of the clean ears, let $E_{\gamma}(C)$ be the union of the edge sets of all connected components of $G_{\gamma}$
that contain a clean ear in $C$. 
Let $T_l:= T \triangle \{v \in V : |\delta_{E_{\gamma}}(v)|\text{ odd}\} \subseteq V_l$.
We consider the long ears in reverse order, starting from $P_l$ and 
apply the following lemma to obtain a multi-set $F_i \subseteq 2E(P_i)$ and a set $C_i$ of clean ears.

\begin{lemma}\label{lemma:construction_F_i}
 Given a set $T_i \subseteq V_i$ with $|T_i|$ even, we can construct a multi-set $F_i \subseteq 2E(P_i)$ 
 and a set $C_i$ of clean ears such that 
\begin{enumerate}
 \item[\rm (i)] $\{ v\in \inn(P_i) : |\delta_{F_i}(v)|\text{ odd} \} = T_i \cap \inn(P_i)$, and
 \item[\rm (ii)] $(V_i \cup \{v\in V(Q): E(Q) \subseteq E_{\gamma}(C_i) \}, E_{\gamma}(C_i) \cup F_i)/V_{i-1}$ is connected, and
 \item[\rm (iii)] $\textnormal{gain}_i(F_i) \ge 0$, and 
 \item[\rm (iv)] $|C_i| \le 2\cdot \textnormal{gain}_i(F_i)$.
 \item[\rm (v)] Moreover, we have:
 \begin{itemize}
  \item 
  if $(P_i,T_i \cap \inn(P_i))$ is good, then
  $\textnormal{gain}_i(F_i) \ge \frac{1}{2} \max\{1, h_i-1\}$; 
    \item 
        if $(P_i,T_i \cap \inn(P_i))$ is special, then
        $\textnormal{gain}_i(F_i) =0$,
        $C_i =\emptyset$, and the two edges of $P_i$ incident
        to an endpoint of $P_i$ are contained exactly once in $F_i$. 
  
 \end{itemize}
\end{enumerate}
\end{lemma}
We keep track of the clean ears $C_i$ used for connectivity, 
since we will later (in the proof of Lemma~\ref{lemma:prelim_result_ear_ind_1}) 
make use of the remaining clean ears (that we did not use for connectivity)
to further improve the $T$-tour we computed. We define 
\[ T_{i-1} \ := \ T_i \triangle \{ v\in V : |\delta_{F_i}(v)|\text{ odd} \}.\]
Note that $ T_{i-1} \subseteq  V_{i-1}$ since $\{ v\in \inn(P_i) : |\delta_{F_i}(v)|\text{ odd} \} = T_i \cap \inn(P_i)$.
Moreover, since the symmetric difference of two even-cardinality sets is even, $|T_{i-1}|$ is even.

\medskip
\par \noindent \hbox{\bf Proof of Lemma \ref{lemma:construction_F_i}:}\quad
If $(P_i, T_i \cap \inn(P_i))$ is a bad pair, we set  $C_i :=\emptyset$. To obtain $F_i$ we then
apply Lemma \ref{lemma:standard_ear_induction} to $(G_i, T_i)/V_{i-1}$. 
(Then the circuit $P$ is $G_i /V_{i-1}$.)

If $(P_i, T_i \cap \inn(P_i))$ is special, $T_i \cap \inn(P_i)$ contains exactly two vertices $w,r$ which have 
distance $\frac{|E(P_i)|}{2}$ in $P_i$. To construct $F_i$ we take $E(P_i)$, double all edges of the $w$-$r$-path in $P_i$
and remove both copies of one duplicated edge. 
If $P_i$ is closed, we take the $w$-$r$-path that does not contain the endpoint of $P_i$.
Then $|F_i| = \frac{3}{2}|\inn(P_i)| - \frac{1}{2}= \frac{3}{2}|\inn(P_i)| - \frac{1}{2} h_i$, implying $\text{gain}_i(F_i) = 0$.
See the brown ear in Figure \ref{fig:ear_induction_example_connectivity} (b) for an example.

If $(P_i, T_i \cap \inn(P_i))$ is good, we will apply Lemma \ref{lemma:enhanced_ear_induction}.
To this end, let $(P, T_P)$ be the instance $(G_i, T_i)/V_{i-1}$. 
(Then again, the circuit $P$ is $G_i /V_{i-1}$.)
We now define the set $U$:
Let $Q_1, \dots, Q_h$ be the clean ears entering internal vertices $w_1, \dots, w_h$ of $P_i$.

We then set $U$ to be the set of edges resulting from $\{\{r(w_1),w_1\}, \dots, \{r(w_h), w_h\}\}$
by contracting $V_{i-1}$.
Note that the vertex set of the circuit $P$ consists of $\inn(P_i)$ and 
the vertex arising from the contraction of $V_{i-1}$.
Since the vertices $r(w_1), \dots, r(w_h)$ are all contained in $V_{i}$ (by choice of the orientation of $G_{\gamma}$),
all endpoints of the edges in $U$ are vertices of $P = G_i /V_{i-1}$.
See Figure \ref{fig:ear_induction_example_connectivity} (c) for an example.

In order to apply Lemma \ref{lemma:enhanced_ear_induction}, we need that $(V(P), U)$ is a forest.
We orient every edge in $U$ resulting from $\{r(w_j),w_j\}$ from $r(w_j)$ to $w_j$.
Since all vertices $w_j$ for $j\in\{1,\dots, h\}$ are internal vertices of the ear $P_i$,
the vertex in $P$ arising from the contraction of $V_{i-1}$ has no incoming edge. 
Since $G_{\gamma}$ is a branching, no vertex in $P = G_i /V_{i-1}$ has more than one incoming edge in the oriented 
edge set $U$. 
Now suppose $(V(P), U)$ contains an undirected circuit.
Since every vertex in $(V(P), U)$ has at most one entering (directed) edge, the undirected circuit in $(V(P), U)$
must be also a directed circuit.

Now note that for $j,j'\in\{1,\dots,h\}$, the vertex $w_j \in \inn(P_i)$ can not be 
the root $r(w_{j'})$ of the connected component of $G_{\gamma}$ containing $w_{j'}$
since $w_j$ has an entering clean ear $Q_j$ in $G_{\gamma}$.
Thus, a vertex in $(V(P), U)$ that has an entering (directed) edge cannot have an outgoing edge.
This contradicts the fact that $(V(P), U)$ contains a directed circuit.
Hence, $(V(P), U)$ must be a forest and we can apply Lemma \ref{lemma:enhanced_ear_induction} to obtain 
a set $F\subseteq 2E(P) =2E(P_i)$ and a set $C\subseteq U$.

We set $F_i := F$ and set $C_i :=\{ Q_j : \{r(w_j), w_j\} \in C \}$.
Then by Lemma \ref{lemma:enhanced_ear_induction}, we have 
\begin{itemize}
 \item $\{ v\in \inn(P_i) : |\delta_{F_i}(v)|\text{ odd} \} = T_i \cap \inn(P_i)$,
 \item $\text{gain}_i(F_i) \ge \frac{1}{2} \max\{1, h_i-1\}$, and 
 \item $|C_i| = |C| \le 2 \left(\sfrac{3}{2}(|E(P)|-1) - \sfrac{1}{2} h_i - |F_i|\right) = 2\cdot \text{gain}_i(F_i)$.
\end{itemize}
It remains to prove that
$(V_i \cup \{v\in V(Q): E(Q) \subseteq E_{\gamma}(C_i) \}, E_{\gamma}(C_i) \cup F_i)/V_{i-1}$ is connected.
Recall that $E_{\gamma}(C_i)$ contains the edge set of all connected components of $G_{\gamma}$ that
contain the clean ears in $C_i$. 
Hence, for every edge $\{r(w_j),w_j\}\in C$, the set $E_{\gamma}(C_i)$ contains the edge set of the 
$r(w_j)$-$w_j$-path in $G_{\gamma}$. 
Since by Lemma \ref{lemma:enhanced_ear_induction}, $(V(P), F\cup C)$ is connected,
also $(V_i \cup \{v\in V(Q): E(Q) \subseteq E_{\gamma}(C_i) \}, E_{\gamma}(C_i) \cup F_i)/V_{i-1}$ is connected.
\endproof

\begin{lemma} \label{lemma:parity_1}
 $E_{\gamma} \cupp F_1 \cupp \dots \cupp F_l$ is a $T$-join.
\end{lemma}

\prove
The set $T_{i-1}$ is defined such that $F_i$ is a $(T_i \triangle T_{i-1})$-join. By induction on $l-i$, the set
$F_i \cupp \dots \cupp F_l$ is a $(T_{i-1} \triangle T_l)$-join.
For $i=1$, this implies that $F_1 \cupp \dots \cupp F_l$ is a $(T_0 \triangle T_l)$-join.
As $F_i$ is constructed such that $T_i \subseteq V_{i-1}$ for all $i\in \{1,\dots, l+1\}$, we have $T_0 \subseteq V_0$.
 
The set $T_0 \triangle T_l$ contains an even number of vertices (since a  $(T_0 \triangle T_l)$-join exists), 
and $|T_l|$ is even. Hence, $|T_0|$ must be even. 
Since $V_0$ has exactly one element, $T_0$ has at most one element. 
As $|T_0|$ is even, $T_0 = \emptyset$ and  $F_1 \cupp \dots \cupp F_l$ is a $T_l$-join.
By definition of $T_l$, adding the edges $E_{\gamma}$ to $F_1 \cupp \dots \cupp F_l$ results in a $T$-join.
\endproof

\begin{lemma} \label{lemma:connectivity_1}
  Let 
  \[\bar{E}_{\gamma} \ := \ \{ e\in E_{\gamma} : e \in E_{\gamma}(C_i) \text{ for }i=1,\dots,l\}.\]
  Then $\bigl(V_l \cup \{v\in \inn(Q) : E(Q) \subseteq \bar{E}_{\gamma}\},F_1 \cupp \dots \cupp F_l \cupp \bar{E}_{\gamma}\bigr)$ is connected.
\end{lemma}

\prove 
We prove by induction on $l-i$ that for every $i=1,\dots,l+1$, the graph resulting from
$(V_l \cup \{v\in \inn(Q) : E(Q) \subseteq \bar{E}_{\gamma}\},F_i \cupp \dots \cupp F_l \cupp \bar{E}_{\gamma})$ by contracting $V_{i-1}$ is connected.
For $i=1$ the set $V_{i-1}=V_0$ contains only one element, hence this completes the proof.

For $i=l+1$ the set $V_{i-1}=V_l$ contains the roots of all connected components of $G_{\gamma}$, 
hence $(V_l \cup \{v\in \inn(Q) : E(Q) \subseteq \bar{E}_{\gamma}\}, E_{\gamma})/ V_l$ is connected.
Now let $i \le l$. 
By induction hypothesis, for every vertex $v' \in V_l \setminus V_{i}$ the set $F_{i+1} \cup \dots \cup F_l \cup \bar{E}_{\gamma}$
contains the edge set of a $v'$-$w'$-path for some $w'\in V_i$.
Hence, it suffices to show that for every $v\in \inn(P_i)$ the set $F_i \cupp \dots \cupp F_l \cupp \bar{E}_{\gamma}$ contains a 
path from $v$ to a vertex in $V_{i-1}$.
This is the case since the graph $(V_i \cup \{v\in V(Q): E(Q) \subseteq E_{\gamma}(C_i)\}, E_{\gamma}(C_i) \cup F_i)/V_{i-1}$ is connected
by Lemma \ref{lemma:construction_F_i} (ii).
\endproof

\begin{lemma}\label{lemma:prelim_result_ear_ind_1}
Given a well-oriented ear-decomposition with long ears $P_1, \dots, P_l$
where all short ears are clean and the oriented ears are precisely the clean ears, 
we can compute a $T$-tour with at most
\begin{align*}
\sfrac{3}{2} (n-1) - \sfrac{1}{2} k_3  - \sfrac{1}{2} \sum_{i=1}^l \textnormal{gain}_i(F_i) 
  - \max\left\{0, k_{\textnormal{special}} - k_3 -   \sum_{i=1}^l  2 \cdot \textnormal{gain}_i(F_i) \right\}
 \end{align*}
edges, where $k_{\textnormal{special}}$ denotes the number of special pairs $(P_i, T_i \cap \inn(P_i))$.
\end{lemma}
\prove
  We call an ear $P_i$ good if $(P_i, T_i \cap \inn(P_i))$ is a good pair and call an ear $P_i$ special
  if $(P_i, T_i \cap \inn(P_i))$ is special.  
 By Lemma \ref{lemma:parity_1} and Lemma \ref{lemma:connectivity_1}, $E_{\gamma} \cupp F_1 \cupp \dots \cupp F_l$ is a $T$-tour
 and by  Lemma \ref{lemma:counting_ear_ind_1} we can bound the number of edges by
 \begin{align*}
   |E_{\gamma}| + \sum_{i=1}^l |F_i|  \ = \ \sfrac{3}{2} (n-1) - \sfrac{1}{2} k_3  - \sum_{i=1}^l \text{gain}_i(F_i).
 \end{align*}
 Note, that for any special ear $P$ and vertex $w\in \inn(P)$, the root $r(w)$ of the connected component of $G_{\gamma}$ 
 containing $w$ is contained in $\inn(P)$. (This follows from the definition of a special ear.)
 Thus, any connected component of $G_{\gamma}$ contains internal vertices of at most one special ear.
 
 \begin{figure}
  \begin{center}
   \begin{tikzpicture}
  \tikzstyle{vertex}=[circle,fill,minimum size=5.2,inner sep=0pt]
  \tikzstyle{pendant}=[circle,draw,minimum size= 5.2,inner sep=0pt] 
  \tikzstyle{main}=[line width=1]
  \tikzstyle{two}=[double,line width=1.0]
  \tikzstyle{short}=[line width=1.3]
  
  \begin{scope}[shift={(0,0)}]
  \node[vertex, white] (v0) at (0.4,0) {};
  \node[vertex, white] (v8) at (5.6,0) {};
  \foreach \x\y [count=\i] in { 0.3/1, 1/1.8, 1.9/2.3, 3/2.5, 4.1/2.3, 5/1.8, 5.7/1} {
     \node[vertex] (v\i) at (\x,\y) {};
  }
  \node[pendant] (p1) at (1.5,3.2) {};
  \node[vertex] (w1) at (2,4.1) {};
  \node[pendant] (p2) at (3,4.5) {};
  \node[vertex] (w2) at (4.1,4.4) {};
  \node[pendant] (p3) at (5.2,3.9) {};
  \node[vertex] (w3) at (6,2.9) {};
  \node[pendant] (p4) at (6.2,1.8) {};
  
  \draw[short, red,->] (w1) -- (p1);
  \draw[short, red,->] (p1) -- (v3);
  \draw[short, blue,->] (p2) -- (w1);
  \draw[short, blue,->] (w2) --(p2);
  \draw[short, orange,->] (w3) --(p3);
  \draw[short, orange,->] (p3) --(w2);
  \draw[short, darkgreen,->] (v7) -- (p4);
  \draw[short, darkgreen,->] (p4) -- (w3);
  
  \draw[two] (v3) --(v4) --(v5);
  \draw[two] (v6) --(v7);
  \draw[main] (v0) --(v1) --(v2) --(v3);
  \draw[main] (v7) --(v8);
  
  \node[below=1.5mm] (q) at (v3) {\small$w$};
   \node[left=0.5mm] (q) at (5.7,0.8) {\small$r(w)$};
  \end{scope}
  
  \begin{scope}[shift={(9,0)}]
  \node[vertex, white] (v0) at (0.4,0) {};
  \node[vertex, white] (v8) at (5.6,0) {};
  \foreach \x\y [count=\i] in {0.3/1, 1/1.8, 1.9/2.3, 3/2.5, 4.1/2.3, 5/1.8, 5.7/1} {
     \node[vertex] (v\i) at (\x,\y) {};
  }
  \draw[main] (v0) --(v1) --(v2) --(v3) --(v4) --(v5) --(v6) --(v7) --(v8);
  
  \node[pendant] (p1) at (1.5,3.2) {};
  \node[vertex] (w1) at (2,4.1) {};
  \node[pendant] (p2) at (3,4.5) {};
  \node[vertex] (w2) at (4.1,4.4) {};
  \node[pendant] (p3) at (5.2,3.9) {};
  \node[vertex] (w3) at (6,2.9) {};
  \node[pendant] (p4) at (6.2,1.8) {};
  
  \draw[two, red] (v3) -- (p1);
  \draw[two, blue] (w1) --(p2);
  \draw[two, orange] (w2) --(p3);
  \draw[two, darkgreen] (w3) -- (p4);
  
   \node[below=1.5mm] (q) at (v3) {\small$w$};
   \node[left=0.5mm] (q) at (5.7,0.8) {\small$r(w)$};
  \end{scope}
  
   \end{tikzpicture}
  \end{center}
  \caption{Modifying the $T$-tour for special ears as in the proof of Lemma
  \ref{lemma:result_ear_ind_1}. The edges shown in black are edges in $E(P_i)$, the colored edges
  are edges of (pendant) 2-ears that are part of the $r(w)$-$w$-path. The edges of different 2-ears are shown in different colors.
  The filled vertices are internal vertices of long ears, i.e. these vertices are contained in $V_l$.
  The non-filled vertices are internal vertices of (pendant) 2-ears.
  The left picture shows the edges of $E(P)$ and the $r(w)$-$w$-path used in the $T$-tour before modifying it,
  and the right picture shows the used edges after the modification.
  \label{figure:flipping_special_ears}}
 \end{figure}
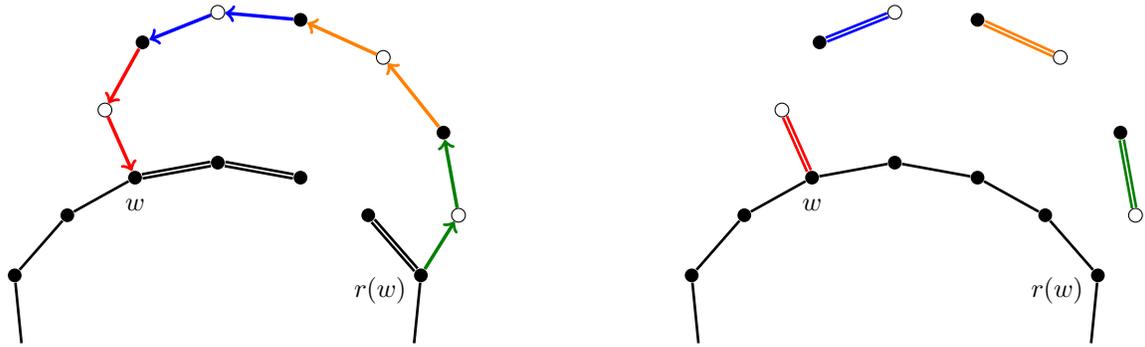

 We now modify the $T$-tour for every special ear $P_i$ as follows: The ear $P_i$ has exactly one entering clean ear $Q$.
 Let $w$ be the internal vertex of $P_i$ where $Q$ enters $P_i$. 
 By the definition of a special ear, we have $r(w)\in \inn(P_i)$ and  $w$ and $r(w)$ have distance $\frac{|E(P_i)|}{2}$ in $P_i$. 
 The connected component of $G_{\gamma}$ containing $w$ (and $r(w)$) contains no internal vertex of any special ear distinct from $P_i$
 since any connected component of $G_{\gamma}$ contains internal vertices of at most one special ear.
 If the connected component of $G_{\gamma}$ containing $w$ (and $r(w)$) contains neither an edge in $\bar{E}_{\gamma}$ nor the 
 edge set of a 3-ear, we modify our $T$-tour by replacing $F_i$ by $E(P_i)$ and replacing the $r(w)$-$w$-path in $E_{\gamma}$ by 
 two copies of every second edge on this path. 
 Note that the edge set of this path is the union of edge sets of 2-ears. (See Figure \ref{figure:flipping_special_ears}.)
 
 We now show that modifying the tour results in a $T$-tour with at least one edge less than before.
 Replacing the $r(w)$-$w$-path in $G_{\gamma}$ by two copies of every second edge on this path does not change 
 the parity of the vertex degrees at internal vertices of this path, but changes the parity of the vertex degrees of $r(w)$ and $w$.
 Recall that we constructed $F_i$ such that the edges incident to endpoints of $P_i$ are contained exactly once. 
 Thus, replacing $F_i$ by $E(P_i)$ does not change the parity of the degree of vertices not in $\inn(P_i)$.
 By definition of a special ear $T_i \cap \inn(P_i) = \{w, r(w)\}$, hence 
 $\{ v \in \inn(P_i) : |\delta_{F_i}(v)| \text{ odd}\} = \{w, r(w)\}$. 
 This shows that replacing $F_i$ by $E(P_i)$ changes the parity of the vertex degree exactly at the vertices $r(w)$ and $w$.
 After replacing both $F_i$ and the $r(w)$-$w$-path in $G_{\gamma}$ the parity of all vertex degrees is the same as before,
 hence the resulting (multi) edge set is a $T$-join. 

 Using Lemma \ref{lemma:connectivity_1}, we thus get that after replacing the $r(w)$-$w$-path in $G_{\gamma}$
 all vertices in $V_l \cup \{v\in \inn(Q) : E(Q) \subseteq \bar{E}_{\gamma}\}$ are still part of the same connected component.
 Moreover, we use at least one edge from every 2-ear and all edges from every 3-ear.
 Thus, we have indeed constructed a $T$-tour. 
 
 Note that replacing the $r(w)$-$w$-path in $G_{\gamma}$ as described above does not change 
 the total number of edges used from $2E_{\gamma}$.
 But we replaced $F_i$ by $E(P_i)$, and since $|F_i| = \frac{3}{2}|\inn(P_i)| - \frac{1}{2}$ (as we had $\text{gain}_i(F_i)=0$) and $|E(P_i)| \ge 6$
 (because $P_i$ is special), we decreased the number of edges by at least one.
 
 This shows that we can decrease the number of edges of the $T$-tour by one for the special ear $P_i$, 
 unless the connected component of $G_{\gamma}$ containing $w$ contains a 3-ear or an edge from $\bar{E}_{\gamma}$.
 Since every connected component of $G_{\gamma}$ contains internal vertices of at most one special ear,
 we can modify our $T$-tour (in the way described above) for at least $k_{\text{special}} - k_3 - \sum_{i=1}^l |C_i|$
 special ears, where $k_{\text{special}}$ denotes the number of special ears. 
 Thus, we obtain a $T$-tour with at most
 \begin{align*}
  & \sfrac{3}{2} (n-1) - \sfrac{1}{2} k_3  - \sum_{i=1}^l \text{gain}_i(F_i)
  - \max\left\{0,\ k_{\text{special}} - k_3 -   \sum_{i=1}^l |C_i|\right\}\\
  \le\ & \sfrac{3}{2} (n-1) - \sfrac{1}{2} k_3  - \sum_{i=1}^l \text{gain}_i(F_i)
  - \max\left\{0,\ k_{\text{special}} - k_3 -   \sum_{i=1}^l 2\cdot \text{gain}_i(F_i)\right\}
 \end{align*}
 edges; here we used Lemma \ref{lemma:construction_F_i} (iv). 
 \endproof
 
 \begin{lemma}\label{lemma:result_ear_ind_1}
  Given a well-oriented ear-decomposition with long ears $P_1,\dots, P_l$
  where all short ears are clean and the oriented ears are precisely the clean ears,
 % Given an ear-decomposition where all short ears are pendant and clean and the clean ears form a forest, 
 we can compute a $T$-tour with at most
 \begin{align*}
 \sfrac{3}{2} (n-1) - \sfrac{7}{20} k_3 - \sum_{i\in I} \max\left\{\sfrac{7}{20} (h_i -1),\ \sfrac{3}{20} \right\}
 \end{align*}
 edges, where $I := \left\{ i\in \{1,\dots, l\} : (P_i,T_i \cap \inn(P_i))\text{ is good or special} \right\}$.
 \end{lemma}
 \prove 
 \begin{align*}
  & \sfrac{1}{2} k_3  + \sum_{i=1}^l \text{gain}_i(F_i)
  +\max\left\{0,\ k_{\text{special}} - k_3 -   \sum_{i=1}^l 2\cdot \text{gain}_i(F_i)\right\} \\
  \ge\ & \sfrac{7}{10} \left( \sfrac{1}{2} k_3  + \sum_{i=1}^l \text{gain}_i(F_i) \right)
  + \sfrac{3}{20} \left( k_3  + \sum_{i=1}^l 2\cdot \text{gain}_i(F_i) \right) \\
  & + \sfrac{3}{20} \max\left\{0,\ k_{\text{special}} - k_3 -   \sum_{i=1}^l 2\cdot \text{gain}_i(F_i)\right\} \\
  \ge\ &  \sfrac{7}{10} \left( \sfrac{1}{2} k_3  + \sum_{i=1}^l \text{gain}_i(F_i) \right) + \sfrac{3}{20} k_{\text{special}} \\
  \ge\ & \sfrac{7}{20} k_3 + \sum_{i\in I} \max\left\{\sfrac{7}{20} (h_i -1),\ \sfrac{3}{20} \right\},
 \end{align*}
 where we used $\text{gain}_i(F_i) \ge \max\{1, h_i -1\}$ if $(P_i,T_i \cap \inn(P_i))$ is good and 
 $h_i = 1$ if $(P_i,T_i \cap \inn(P_i))$ is special.
 Together with Lemma \ref{lemma:prelim_result_ear_ind_1}, this implies that we can compute a $T$-tour with at most
 \begin{align*}
 \sfrac{3}{2} (n-1) - \sfrac{7}{20} k_3 - \sum_{i\in I} \max\left\{\sfrac{7}{20} (h_i -1),\ \sfrac{3}{20} \right\}
 \end{align*}
 edges.
 \endproof

\subsection{Using clean ears for parity correction}\label{section:clean_ears_parity}

In this section we propose a different kind of ear induction. 
For 4-ears, good and special ears we essentially use Lemma \ref{lemma:standard_ear_induction}.
For bad long ears, however, we will now gain something by using the only entering clean ear in a different way:
if this clean ear enters at $w$, we can ``flip'' all clean ears on the path from $r(w)$ to $w$, changing the 
parity at $w$ and at $r(w)$, similar to the treatment of special ears above. 
This flexibility allows to save something for each bad ear of length at least five.

We again consider the long ears in reverse order, starting from $P_l$.
Let $T'_l := T_l = T \triangle \{v \in V_l : |\delta_{E_{\gamma}}(v)|\text{ odd}\}$ and $T_l^{\gamma} := \emptyset$.
The sets $T_i^{\gamma}$ record the vertices whose parity we will change by flipping clean ears, 
and $T'_i$ contains the vertices requiring odd degree in the first $i$ (non-oriented) ears after these flips.

If $(P_i, T_i \cap \inn(P_i))$ is a bad pair and $P_i$ is not a 4-ear,
$P_i$ has exactly one entering clean ear $Q$ that enters $P_i$ at the middle vertex $w$ (and $r(w) \in V_{i-1}$); 
then let $W_i:= \{w\}$.
Otherwise let $W_i := \emptyset$.
We construct a multi-set $F'_i \subseteq 2E(P_i)$ such that 
\begin{equation}\label{eq:parity_ear_ind_2}
  \{ v\in \inn(P_i) \setminus W_i : |\delta_{F_i'}(v)| \text{ odd} \} \ = \ T_i'\cap (\inn(P_i)\setminus W_i) 
\end{equation}
and $(V_i, F'_i)/ V_{i-1}$ is connected.
(We will later describe in detail how we construct $F'_i$; see Lemma \ref{lemma:construct_F'_i}.)
If $W_i =\{w\}$ and $w\in \{ v\in \inn(P_i) : |\delta_{F_i'}(v)| \text{ odd} \} \triangle T'_i$, 
we define $T_{i-1}^{\gamma} := T^{\gamma}_i \triangle \{w,r(w)\}$ and
\[T'_{i-1} \ := \ T'_i \triangle \{ v\in V : |\delta_{F'_i}(v)|\text{ odd} \} \triangle \{w,r(w)\}.\]
Otherwise, let $T_{i-1}^{\gamma} := T^{\gamma}_i$ and
\[T'_{i-1} \ := \ T'_i \triangle \{ v\in V : |\delta_{F'_i}(v)|\text{ odd} \} .\]
We have $ T'_{i-1} \subseteq  V_{i-1}$ by \eqref{eq:parity_ear_ind_2} and since $r(w)\in V_{i-1}$ in the first case.
Note that $|T_{i-1}|$ is even because both $|T'_i|$ and $|\{ v\in V : |\delta_{F'_i}(v)|\text{ odd} \}|$ are even.

\begin{lemma}\label{lemma:connectivity_2}
 $(V_l, F'_1 \cupp \dots \cupp F'_l)$ is connected.
\end{lemma}
\prove 
 Since $(V_i, F'_i)/ V_{i-1}$ is connected for every $i\in \{1,\dots,l\}$, the multi-set $F_i'$ contains the edge set of a path
 from every internal vertex of $P_i$ to a vertex in $V_{i-1}$. By induction this implies that for every $i\in \{1,\dots,l\}$ the multi-set
 $F'_1 \cupp \dots \cupp F'_i$ contains the edge set of a path from every internal vertex of $P_i$ to the unique element of $V_0$.
 This proves that $(V_l, F'_1 \cupp \dots \cupp F'_l)$ is connected.
\endproof

\begin{lemma}\label{lemma:parity_2}
 For every $i\in\{1, \dots, l+1\}$ the multi-set $F'_i \cupp \dots \cupp F'_l$ 
 is a $(T_l \triangle T^{\gamma}_{i-1} \triangle T'_{i-1})$-join.
\end{lemma}
\prove We prove this by induction on $l-i$. For $i=l+1$ we have $T_l = T'_l$ and $T_l^{\gamma} = \emptyset$.
Hence, $T_l \triangle T^{\gamma}_{i-1} \triangle T'_{i-1} = \emptyset$.
Now let $i\le l$. 
By induction hypothesis $F'_{i+1} \cupp \dots \cupp F'_l$ is a $(T_l \triangle T^{\gamma}_i \triangle T'_i)$-join.
Thus, proving that  $F'_i \cupp \dots \cupp F'_l$ is a $(T_l \triangle T^{\gamma}_{i-1} \triangle T'_{i-1})$-join is equivalent 
to proving that $F'_i$ is a  
$((T_l \triangle T^{\gamma}_i \triangle T'_i) \triangle (T_l \triangle T^{\gamma}_{i-1} \triangle T'_{i-1}))$-join.
We have either 
\begin{itemize}
 \item $T'_{i-1}:= T'_i \triangle \{ v\in V : |\delta_{F'_i}(v)|\text{ odd} \} \triangle \{w,r(w)\}$ 
 and $T_{i-1}^{\gamma} := T^{\gamma}_i \triangle \{w,r(w)\}$, or
 \item $T'_{i-1}:= T'_i \triangle \{ v\in V : |\delta_{F'_i}(v)|\text{ odd} \}$ and $T_{i-1}^{\gamma} := T^{\gamma}_i$.
\end{itemize}
In any of the two cases we have
$$(T_l \triangle T^{\gamma}_i \triangle T'_i) \triangle (T_l \triangle T^{\gamma}_{i-1} \triangle T'_{i-1}) 
\ = \ (T^{\gamma}_i \triangle T^{\gamma}_{i-1}) \triangle (T'_i \triangle T'_{i-1}) \ = \ \{ v\in V : |\delta_{F'_i}(v)|\text{ odd} \}.$$
\endproof

\begin{lemma}\label{lemma:T_join_exists}
 For every $i\in\{0, \dots, l\}$ the edge set $E_{\gamma}$ contains a $T^{\gamma}_i$-join.
\end{lemma}
\prove
We again use induction on $l-i$.
For $i=l$, we have $T^{\gamma}_i =\emptyset$ and the statement clearly holds.
Let now $i < l$. We either have $T_i^{\gamma} = T^{\gamma}_{i+1}$ or $T_i^{\gamma} = T^{\gamma}_{i+1} \triangle \{w,r(w)\}$ 
for some vertex $w$.
If $T_i^{\gamma} = T^{\gamma}_{i+1}$, the set $E_{\gamma}$ contains a $T^{\gamma}_i$-join by induction hypotheis.
If $T_i^{\gamma} = T^{\gamma}_{i+1} \triangle \{w,r(w)\}$, the set $E_{\gamma}$ contains the edge set of a $w$-$r(w)$-path since 
$r(w)$ is the root of the connected component of $G_{\gamma} =(V, E_{\gamma})$. 
Moreover, by induction hypothesis, $E_{\gamma}$ contains a $T^{\gamma}_{i+1}$-join.
The symmetric difference of such a $T^{\gamma}_{i+1}$-join and a $w$-$r(w)$-path in $G_{\gamma}$ is a $T^{\gamma}_i$-join.
\endproof

For $i\in \{1,\dots,l\}$ and a multi-set $F\subseteq 2E(P_i)$ let
\[ \text{gain}'_i(F) \ := \ \sfrac{3}{2} |\inn(P_i)| - |F|. \]

\begin{lemma}\label{lemma:counting_ear_ind_2}
 We can construct a $T$-tour in $G$ with at most
 \[ \sfrac{3}{2} (n-1) + \sfrac{1}{2}k_2 + k_3 - \sum_{i=1}^l \textnormal{gain}'_i(F'_i). \]
 edges.
\end{lemma}

\prove 
By Lemma \ref{lemma:T_join_exists} the edge set $E_{\gamma}$ contains a $T^{\gamma}_0$-join $J$.
Note that we never add any internal vertex of a short ear to a set $T^{\gamma}_i$ for any $i$.
Hence, $T^{\gamma}_0$ contains no internal vertex of any short ear. 
This shows that for every short ear $Q$ either all edges of $Q$ are contained in $J$ or none of these edges.
We now define an edge set $H\subseteq 2E_{\gamma}$. If the edge set of a short ear $Q$ is contained in $J$, 
we add two copies of all elements of $E(Q)$ except one to $H$. 
Otherwise, we have $E(Q) \cap J = \emptyset$ and we add $E(Q)$ to $H$. 
See Figure \ref{fig:flipping_T_join}.

Since $H$ and $E_{\gamma}\triangle J$ are identical up to pairs of parallel edges, 
$|\delta_H(v)|$ is odd if and only if $|\delta_{E_{\gamma}\triangle J}(v)$ is odd for every vertex $v$, and this holds if and only if
$|\delta_{E_{\gamma}}(v)| + |\{v\} \cap T^{\gamma}_0|$ is odd.
Recall that $T_l = T \triangle \{ v\in V : |\delta_{E_{\gamma}}(v)|\text{ odd}\} \subseteq V_l$ and thus
$T^{\gamma}_0 \triangle \{ v\in V : |\delta_{E_{\gamma}}(v)|\text{ odd}\} = T_l \triangle T \triangle T^{\gamma}_0$.
Hence, $H$ is a $(T \triangle T_l \triangle T^{\gamma}_0)$-join and we have $|H| \le 2 k_2 + 4 k_3$.
Moreover, for every short ear $Q$ and every vertex $v\in \inn(Q)$, the edge set of some path from $v$ to a vertex in $V_l$
is contained in $H$. 
Together with Lemma \ref{lemma:connectivity_2} this shows that $(V, H \cupp F_1' \cupp \dots \cupp F_l')$ is connected.

Since $|T'_i|$ is even for every $i\in\{0,\dots, l\}$, $T_0' \subseteq V_0$ ,and $|V_0|=1$, we have $T'_0 = \emptyset$. 
Hence, by Lemma \ref{lemma:parity_2}, $F_1' \cupp \dots \cupp F_l'$ is a $(T_l \triangle T^{\gamma}_0)$-join.
As $H$ is a $(T \triangle T_l \triangle T^{\gamma}_0)$-join, $H \cupp F_1' \cupp \dots \cupp F_l'$ is a $T$-join and
thus a $T$-tour with 
\[ |H| +  \sum_{i=1}^l |F'_i| \ \le \ 2 k_2 + 4 k_3 + \sum_{i=1}^l |F'_i| \]
edges.
By definition of $\text{gain}'_i$ we have $|F'_i|=  \sfrac{3}{2} |\inn(P_i)| - \text{gain}'_i(F'_i)$.
Moreover, the number of internal vertices of short ears is $k_2+2k_3$.
\endproof

\begin{lemma}\label{lemma:difference_potential}
 Let $i\in\{1, \dots, l+1\}$. 
 Then for every $v\in T_{i-1} \triangle T'_{i-1}$ we have one of the following three properties:
 \begin{enumerate}[(a)]
  \item The vertex $v$ is contained in $(T_i \triangle T'_i)\cap V_{i-1}$.
  \item $v$ is an endpoint of $P_i$ and $|\delta_{F_i}(v)| + |\delta_{F'_i}(v)|$ is odd.
  \item We have $h_i =1$ and the unique clean ear entering $P_i$ enters $P_i$ at a vertex $w$ with
         $v=r(w)$. Moreover, $|\delta_{F_i}(w)| + |\delta_{F'_i}(w)|$ is odd.
 \end{enumerate}
\end{lemma}
\prove 
We have by construction of the sets $T_{i-1}$ and $T_{i-1}'$ that both of these sets are subsets of $V_{i-1}$.
Now let $v\in T_{i-1} \triangle T'_{i-1}$ such that  $v$ fulfills neither (a) nor (c). 
Then we have either $v\in  T_i \triangle T_{i-1}$ or $v \in T'_i \triangle T'_{i-1}$.
If  $v\in  T_i \triangle T_{i-1}$ but $v \notin T'_i \triangle T'_{i-1}$, we have
$|\delta_{F_i}(v)|$ odd and $|\delta_{F'_i}(v)|$ even.
Similarly, if  $v\in  T'_i \triangle T'_{i-1}$ but $v \notin T_i \triangle T_{i-1}$, we have
$|\delta_{F'_i}(v)|$ odd and $|\delta_{F_i}(v)|$ even.
In any of the two cases (b) holds.
\endproof

For $i\in \{1,\dots, l\}$ let 
\[ \Delta_i \ := \ |T_{i-1}\triangle T'_{i-1}| - |T_i\triangle T'_i|.\]
\begin{lemma}\label{lemma:diffT}
For every $i\in \{1,\dots, l\}$, we have $\Delta_i \le  2$ and $\Delta_i$ is even.
\end{lemma}
\prove
By Lemma \ref{lemma:difference_potential}, $|T_{i-1}\triangle T'_{i-1}| \le |T_i\triangle T'_i| + 3$. 
As the symmetric difference of two even-cardinality sets has always even cardinality, we have $\Delta_i$ even and
$|T_{i-1}\triangle T'_{i-1}| \le |T_i\triangle T'_i| + 2$.
\endproof

 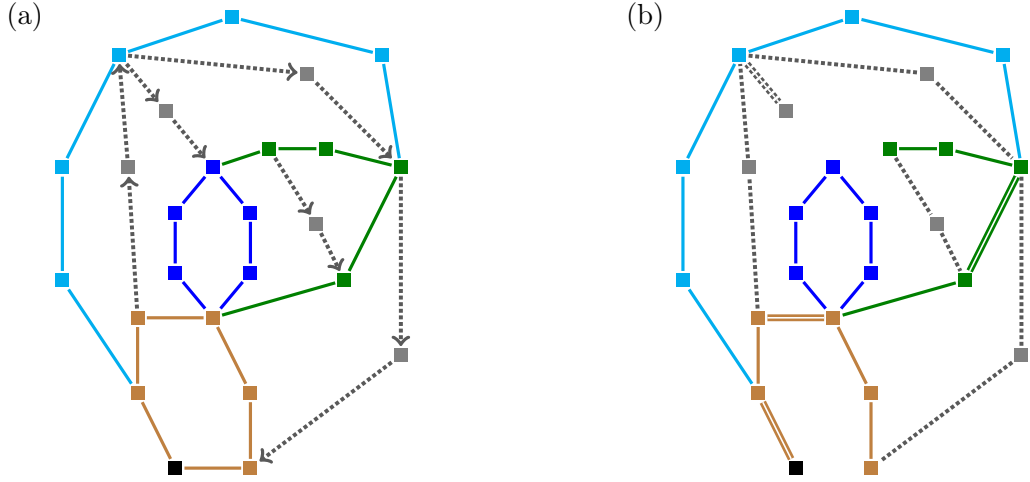
\begin{figure}
\begin{center}
 \begin{tikzpicture}[scale=0.5]

  \tikzstyle{tvertex}=[outer sep=1pt, darkgreen,circle,fill,minimum size=6,inner sep=0pt]
  \tikzstyle{vertex}=[outer sep =1pt, minimum size=5,fill,inner sep=0pt]
  \tikzstyle{forest}=[grey, line width=1.5, densely dotted]
  \tikzstyle{main}=[line width=1.2]
  \tikzstyle{two}=[double,line width=1.0]
  \tikzstyle{no}=[white]
  
\begin{scope}[shift={(0,0)}]
\node at (3,10) {(a)};
   \node[vertex] (a0) at (7, -2) {};
   \node[vertex, brown] (a1) at (9, -2) {};
   \node[vertex, brown] (a2) at (9, 0) {};
   \node[vertex, brown] (a3) at (8, 2) {};
   \node[vertex, brown] (a4) at (6, 2) {};
   \node[vertex, brown] (a5) at (6, 0) {};
   \draw[main, brown] (a0) -- (a1);
   \draw[main, brown] (a1) -- (a2);
   \draw[main, brown] (a2) -- (a3);
   \draw[main, brown] (a3) -- (a4);
   \draw[main, brown] (a4) -- (a5);
   \draw[main, brown] (a5) -- (a0);
   \node[vertex, blue] (b1) at (7, 3.2) {};
   \node[vertex, blue] (b2) at (7, 4.8) {};
   \node[vertex, blue] (b3) at (8, 6) {};
   \node[vertex, blue] (b4) at (9, 4.8) {};
   \node[vertex, blue] (b5) at (9, 3.2) {};
   \draw[main, blue] (a3) -- (b1);
   \draw[main, blue] (b1) -- (b2);
   \draw[main, blue] (b2) -- (b3);
   \draw[main, blue] (b3) -- (b4);
   \draw[main, blue] (b4) -- (b5);
   \draw[main, blue] (b5) -- (a3);
   \node[vertex, darkgreen] (c1) at (11.5, 3) {};
   \node[vertex, darkgreen] (c2) at (13, 6) {};
   \node[vertex, darkgreen] (c3) at (11, 6.5) {};
   \node[vertex, darkgreen] (c4) at (9.5, 6.5) {};
   \draw[main, darkgreen] (a3) -- (c1);
   \draw[main, darkgreen] (c1) -- (c2);
   \draw[main, darkgreen] (c2) -- (c3);
   \draw[main, darkgreen] (c3) -- (c4);
   \draw[main, darkgreen] (c4) -- (b3);
   \node[vertex, cyan] (d1) at (4, 3) {};
   \node[vertex, cyan] (d2) at (4, 6) {};
   \node[vertex, cyan] (d3) at (5.5, 9) {};
   \node[vertex, cyan] (d4) at (8.5, 10) {};
   \node[vertex, cyan] (d5) at (12.5, 9) {};
   \draw[main, cyan] (a5) -- (d1);
   \draw[main, cyan] (d1) -- (d2);
   \draw[main, cyan] (d2) -- (d3);
   \draw[main, cyan] (d3) -- (d4);
   \draw[main, cyan] (d4) -- (d5);
   \draw[main, cyan] (d5) -- (c2);
   \node[vertex, gray] (e) at (5.75, 6) {};
   \draw[forest,->] (a4) -- (e);
   \draw[forest,->] (e) -- (d3);
   \node[vertex, gray] (f) at (10.5, 8.5) {};
   \draw[forest,->] (d3) -- (f);
   \draw[forest,->] (f) -- (c2);
   \node[vertex, gray] (g) at (13, 1) {};
  \draw[forest,->] (c2) -- (g);
  \draw[forest,->] (g) -- (a1);
   \node[vertex, gray] (i) at (10.75, 4.5) {};
   \draw[forest,->] (c4) -- (i);
   \draw[forest,->] (i) -- (c1);
   \node[vertex, gray] (j) at (6.75, 7.5) {};
   \draw[forest,->] (d3) to (j);
   \draw[forest,->] (j) to (b3);
\end{scope}

\begin{scope}[shift={(16.5,0)}]
\node at (3,10) {(b)};
   \node[vertex] (a0) at (7, -2) {};
   \node[vertex, brown] (a1) at (9, -2) {};
   \node[vertex, brown] (a2) at (9, 0) {};
   \node[vertex, brown] (a3) at (8, 2) {};
   \node[vertex, brown] (a4) at (6, 2) {};
   \node[vertex, brown] (a5) at (6, 0) {};
   %\draw[main, brown] (a0) -- (a1);
   \draw[main, brown] (a1) -- (a2);
   \draw[main, brown] (a2) -- (a3);
   \draw[two, brown] (a3) -- (a4);
   \draw[main, brown] (a4) -- (a5);
   \draw[two, brown] (a5) -- (a0);
   \node[vertex, blue] (b1) at (7, 3.2) {};
   \node[vertex, blue] (b2) at (7, 4.8) {};
   \node[vertex, blue] (b3) at (8, 6) {};
   \node[vertex, blue] (b4) at (9, 4.8) {};
   \node[vertex, blue] (b5) at (9, 3.2) {};
   \draw[main, blue] (a3) -- (b1);
   \draw[main, blue] (b1) -- (b2);
   \draw[main, blue] (b2) -- (b3);
   \draw[main, blue] (b3) -- (b4);
   \draw[main, blue] (b4) -- (b5);
   \draw[main, blue] (b5) -- (a3);
   \node[vertex, darkgreen] (c1) at (11.5, 3) {};
   \node[vertex, darkgreen] (c2) at (13, 6) {};
   \node[vertex, darkgreen] (c3) at (11, 6.5) {};
   \node[vertex, darkgreen] (c4) at (9.5, 6.5) {};
   \draw[main, darkgreen] (a3) -- (c1);
   \draw[two, darkgreen] (c1) -- (c2);
   \draw[main, darkgreen] (c2) -- (c3);
   \draw[main, darkgreen] (c3) -- (c4);
   \draw[no] (c4) -- (b3);
   \node[vertex, cyan] (d1) at (4, 3) {};
   \node[vertex, cyan] (d2) at (4, 6) {};
   \node[vertex, cyan] (d3) at (5.5, 9) {};
   \node[vertex, cyan] (d4) at (8.5, 10) {};
   \node[vertex, cyan] (d5) at (12.5, 9) {};
   \draw[main, cyan] (a5) -- (d1);
   \draw[main, cyan] (d1) -- (d2);
   \draw[main, cyan] (d2) -- (d3);
   \draw[main, cyan] (d3) -- (d4);
   \draw[main, cyan] (d4) -- (d5);
   \draw[main, cyan] (d5) -- (c2);
   \node[vertex, gray] (e) at (5.75, 6) {};
   \draw[forest] (a4) -- (e);
   \draw[forest] (e) -- (d3);
   \node[vertex, gray] (f) at (10.5, 8.5) {};
   \draw[forest] (d3) -- (f);
   \draw[forest] (f) -- (c2);
   \node[vertex, gray] (g) at (13, 1) {};
   \draw[forest] (c2) -- (g);
   \draw[forest] (g) -- (a1);
   \node[vertex, gray] (i) at (10.75, 4.5) {};
   \draw[forest] (c4) -- (i);
   \draw[forest] (i) -- (c1);
   \node[vertex, gray] (j) at (6.75, 7.5) {};
   \draw[forest, two] (d3) to (j);
   \draw[no] (j) to (b3);
\end{scope}

 \end{tikzpicture}
 \end{center}
 \caption{The example from Figure \ref{fig:ear_induction_example_connectivity} with $T=\emptyset$. 
 (a): ears with oriented short ears (gray, dotted). (b): result of using short ears for parity.
 \label{fig:ear_induction_example_parity}}
 \end{figure}

We now show how to construct the $F_i'$.
For good or special ears we can simply apply Lemma \ref{lemma:standard_ear_induction},
but we will need a slightly refined version: in case the bound is tight, we want the parity at the endpoints
to match the previous construction.

\begin{lemma}\label{lemma:standard_ear_induction_with_parity}
 Let $P$ be a circuit with at least four edges and $T_P\subseteq V(P)$ with $|T_P|$ even. 
 Then there exists a $T_P$-join $F\subseteq 2 E(P)$ such that the graph $(V(P), F)$ is connected and
 \begin{equation}
 \label{eq:gainonehalf}
|F| \ \le \ \sfrac{3}{2} (|E(P)|-1) - \sfrac{1}{2}.
\end{equation} 
Moreover, if this bound is tight and $|E(P)|+|T_P|\ge 5$, then there is another $T_P$-join 
$F'\subseteq 2 E(P)$ such that the graph $(V(P), F')$ is connected and
$|F'|=|F|$ and the number of copies of any edge in $F\cupp F'$ is odd.
\end{lemma}

\prove
If $T_P=\emptyset$, then $F=E(P)$ does the job because $|E(P)| \le \frac{3}{2}|E(P)|-2$,
and the bound is tight only if $|E(P)|=4$.

Let now $T_P\ne\emptyset$. 
The vertices of $T_P$ subdivide $P$ into subpaths. Color these paths alternatingly red and blue.
Let $E_R$ and $E_B$ denote the set of edges of red and blue subpaths, respectively.
We define two $T_P$-joins, $F_R$ and $F_B$.
For the red solution $F_R$, take two copies of each edge in $E_R$ and 
one copy of each edge in $E_B$. Note that $E_R \ne \emptyset$, and remove one pair of parallel edges.
$F_B$ is formed by switching the roles of red and blue. 
This yields $F_R,F_B\subseteq 2 E(P)$ with 
\[ \sfrac{1}{2} \left( |F_R|+|F_B| \right) \ = \ \sfrac{1}{2} \left( 3|E_B|+3|E_R|-4 \right) 
\ \le \ \sfrac{1}{2} \left( 3|E(P)| - 4 \right) \ = \ \sfrac{3}{2} (|E(P)|-1) - \sfrac{1}{2}.\]
Note that the number of copies of any edge in $F_R\cupp F_B$ is odd.
Moreover, $|F_R|=|F_B|$ or the smaller one has fewer than $\sfrac{3}{2} (|E(P)|-1) - \sfrac{1}{2}$ edges.
\endproof

Recall that $W_i = \emptyset$ unless $(P_i, T_i \cap \inn(P_i))$ is a bad pair and $P_i$ is not a 4-ear. 
If $W_i\not=\emptyset$, the ear $P_i$ has exactly one entering clean ear, entering $P_i$ at $w$ and 
$r(w) \in V_{i-1}$; then $W_i= \{w\}$.

\begin{lemma}\label{lemma:construct_F'_i}
Let $i\in \{1,\dots, l\}$ and $T_i \subseteq V_i$ with $|T_i|$ even.
Then we can construct a multi-set $F'_i \subseteq 2 E(P_i)$ such that 
\begin{itemize}
 \item $ \{ v\in \inn(P_i) \setminus W_i : |\delta_{F_i'}(v)| \text{ odd} \} =  T_i' \cap (\inn(P_i)\setminus W_i)$,
 \item $(V_i, F'_i) / V_{i-1}$ is connected, and
\begin{empheq}[left={\textnormal{gain}'_i(F'_i) - \frac{1}{4} \Delta_i \ \ge \ \empheqlbrace}]{alignat=2}
& 1 &&\qquad  \textnormal{if  $|E(P_i)|\ge 5$ and $(P_i, T_i \cap \inn(P_i))$ is bad} \tag{a} \\
& \sfrac{1}{2} &&\qquad \textnormal{if either $|E(P_i)|\ge 5$ or $(P_i, T_i \cap \inn(P_i))$ is bad} \tag{b} \\
& 0 &&\qquad \textnormal{if $|E(P_i)|=4$ and $(P_i, T_i \cap \inn(P_i))$ is not bad} \tag{c} 
\end{empheq}
\end{itemize}
\end{lemma}

\prove
\begin{enumerate}
 \item[(a)] The vertices of $\inn(P_i)\cap T'_i$ subdivide $P_i$ into subpaths, alternatingly colored red and blue.
Let $E_R$ and $E_B$ denote the set of edges of red and blue subpaths, respectively.
Let $Q$ be the unique clean ear entering $P_i$ at $w\in \inn(P_i)$.
Let $E_1, E_2$ be the edge sets of the two paths in $P_i$ from $w$ to an endpoint of $P_i$. Then $E(P_i) = E_1 \cupp E_2$.
By definition of a bad pair, $|E_1|=|E_2|$ and $r(w)\notin \inn(P_i)$. Hence, by the choice of the orientation of $G_{\gamma}$, we have $r(w)\in V_{i-1}$.
Note that $W_i = \{w\}$.
We now distinugish two cases:
\\[2mm]
\textbf{Case 1:} 
$|E_R \cap E_1| \ne |E_B \cap E_1|$ or $|E_R \cap E_2| \ne |E_B \cap E_2|$  \\[2mm]
Witout loss of generality we may assume $|E_R \cap E_1| < |E_B \cap E_1|$. 
Then, we construct $F_i'$ from $E(P)$ by adding a second copy of $E_R \cap E_1$, adding a second copy of 
the edges in the smaller of the two sets $|E_R \cap E_2|$ and $|E_B \cap E_2|$, and removing one arbitrary duplicated edge,
if it exists. Then $|F_i'|\le|E(P)| \le \frac{3}{2}|\inn(P_i)| - \frac{3}{2}$ if there was no duplicated edge; otherwise
$|F_i'| \le |E(P_i)| + \lfloor\frac{1}{2} (|E_1| - 1)\rfloor + \lfloor \frac{1}{2} |E_2| \rfloor - 2 \le \frac{3}{2} |E(P_i)| - 3 \le \frac{3}{2} |\inn(P_i)| - \frac{3}{2}. $
Thus, in both cases, $\text{gain}'_i(F'_i)\ge\frac{3}{2}$ and $\Delta_i \le 2$.
\\[2mm]
\textbf{Case 2:} 
$|E_R \cap E_1| = |E_B \cap E_1|$ and $|E_R \cap E_2| = |E_B \cap E_2|$ \\[2mm]
By the definition of a bad pair, we have  $\inn(P_i) \cap T_i = \{w\}$, and $w$ is the middle internal vertex of $P_i$.
If in addition $|\inn(P_i) \cap (T'_i \triangle T_i)| \le 1$,
$E_R \cap E_1$, $E_B \cap E_1$, $E_R \cap E_2$, or $E_B \cap E_2$ must be empty, contradicting $|E_R \cap E_1| = |E_B \cap E_1|$ and $|E_R \cap E_2| = |E_B \cap E_2|$.
Hence, we may assume that $|\inn(P_i) \cap (T'_i \triangle T_i)|$ is at least two.

Similar to Case 1, we construct $F_i'$ from $E(P_i)$ by adding a second copy of either $E_R \cap E_1$ or $E_B \cap E_1$, 
adding a second copy of the edges of one the two sets $E_R \cap E_2$ and $E_B \cap E_2$, 
and removing both copies of an arbitrary duplicated edge.
We choose between  $E_R \cap E_1$ and $E_B \cap E_1$ and between $E_R \cap E_2$ and $E_B \cap E_2$ such that 
the endpoints of $P_i$ have the same parity of degree in $F'_i$ and in $F_i$. 
Then $|F_i'| = \frac{3}{2} |E(P)| - 2 = \frac{3}{2} |\inn(P_i)| - \frac{1}{2}$, so $\text{gain}'_i(F'_i)=\frac{1}{2}$.

Moreover, we get from Lemma \ref{lemma:difference_potential} that  
$|T_{i-1} \triangle T'_{i-1}| \le |T_i \triangle T'_i| + 1  - |(T_i \triangle T'_i)\cap \inn(P_i)|$.
Since $ | \inn(P_i) \cap (T'_{i+1} \triangle T_{i+1})|$ is at least two, 
we have $|T_{i-1} \triangle T'_{i-1}| \le |T_i \triangle T'_i| -1$.
As by Lemma \ref{lemma:diffT} the number $\Delta_i$ must be even, $\Delta_i \le -2$.
\item[(b)] 
First suppose that $|E(P_i)| \ge 5$ or $T'_i\not=\emptyset$. 
Then we apply Lemma \ref{lemma:standard_ear_induction_with_parity} to $(G_i, T'_i)/V_{i-1}$. 
We get a set $F_i'$ with $\text{gain}'_i(F_i') \ge \frac{1}{2}$. 
If $P_i$ is a circuit, we get from Lemma \ref{lemma:difference_potential}
that $\Delta_i \le 1$.
If $P_i$ is a path, we get from Lemma \ref{lemma:standard_ear_induction_with_parity} that one of the following holds:
\begin{itemize}
 \item We have $\text{gain}'_i(F_i') \ge 1$.
 \item For one endpoint of $P_i$ we can choose the parity of its degree in $F_i'$, implying $\Delta_i \le 1$
       by Lemma \ref{lemma:difference_potential}.
\end{itemize}
Using Lemma \ref{lemma:diffT} this implies that
($\text{gain}'_i(F_i') \ge 1$ and $\Delta_i \le 2$) or ($\text{gain}'_i(F_i') \ge \frac{1}{2}$ and $\Delta_i \le 0$).
In both cases we get $\text{gain}'_i(F'_i) - \frac{1}{4} \Delta_i \ge \sfrac{1}{2}$.

Now suppose that $|E(P_i)| =4$ and $T'_i=\emptyset$ and $(P_i, T_i \cap \inn(P_i))$ is bad; so $T_i=\emptyset$.
Then we set $F_i' := F_i$. Then we have  $T_{i-1} \triangle T'_{i-1} = T_i \triangle T'_i$ and thus $\Delta_i = 0$.
By Lemma \ref{lemma:construction_F_i}, we have 
$|F_i'| = |F_i| \le \sfrac{3}{2} |\inn(P_i)| - \sfrac{1}{2}$. Hence, $\text{gain}'_i(F'_i) \ge \frac{1}{2}$.
\item[(c)]
We apply Lemma \ref{lemma:standard_ear_induction} to $(G_i, T'_i)/V_{i-1}$ to obtain a multi-set 
$F_i'$ with $\text{gain}'_i(F'_i) \ge \frac{1}{2}$.
By Lemma \ref{lemma:diffT},  we always have $\Delta_i\le 2$.
Hence, $\text{gain}'_i(F'_i) - \frac{1}{4} \Delta_i \ge 0$. 
\endproof
\end{enumerate}

\begin{lemma}\label{lemma:result_ear_ind_2}
Given a well-oriented ear-decomposition with long ears $P_1, \dots, P_l$
where all short ears are clean and the oriented ears are precisely the clean ears,
% Given an ear-decomposition where all short ears are pendant and clean and the clean ears form a forest, 
we can compute a $T$-tour with at most
  \[  \sfrac{3}{2} (n-1) + \sfrac{1}{2} (k_2 + k_3 -  k_{\ge 5}) + \sfrac{1}{2} k_3   - \sfrac{1}{2} k_{\textnormal{bad}} \]
  edges, where $k_{\textnormal{bad}}$ is the number of bad pairs $(P_i, \inn(P_i)\cap T_i)$.
\end{lemma}

\prove 
Since $T_l = T_l'$ and $T_0=T_0'=\emptyset$, we have 
\begin{align}\label{eq:sum_delta_non_neg}
  \sum_{i=1}^l \Delta_i \ = \ \sum_{i=1}^l \left( |T_{i-1} \triangle T'_{i-1}| - |T_i \triangle T'_i| \right) 
  \ = \ |T_0 \triangle T'_0| - |T_l \triangle T'_l| \ = \ 0.
\end{align}
By construction of the sets $F_i'$ we have 
\begin{align*}
 \sum_{i=1}^l \left(\text{gain}'_i(F'_i) - \sfrac{1}{4} \Delta_i \right) \ \ge \ \sfrac{1}{2} k_{\ge 5} + \sfrac{1}{2} k_{\textnormal{bad}}.
\end{align*}
Using \eqref{eq:sum_delta_non_neg}, this implies
\begin{align*}
  \sum_{i=1}^l \text{gain}'_i(F'_i) \ \ge \ \sfrac{1}{2} k_{\ge 5} + \sfrac{1}{2} k_{\textnormal{bad}}.
\end{align*}
By Lemma \ref{lemma:counting_ear_ind_2} we can construct a $T$-tour in $G$ with at most
\begin{align*}
  & \sfrac{3}{2} (n-1) + \sfrac{1}{2}k_2 + k_3 - \sum_{i=1}^l \text{gain}'_i(F'_i) \\
  \le\ & \sfrac{3}{2} (n-1) + \sfrac{1}{2}k_2 + k_3 - \sfrac{1}{2} k_{\ge 5} - \sfrac{1}{2} k_{\textnormal{bad}} \\
  =\ & \sfrac{3}{2} (n-1) + \sfrac{1}{2} (k_2 + k_3 -  k_{\ge 5}) + \sfrac{1}{2} k_3   - \sfrac{1}{2} k_{\textnormal{bad}}.
\end{align*}
edges.
\endproof

We now combine Lemma \ref{lemma:result_ear_ind_1} and Lemma \ref{lemma:result_ear_ind_2}
to prove a bound on the number of edges of the better of the two $T$-tours 
resulting from the two different kinds of ear induction.

\begin{theorem}\label{thm:result_ear_induction}
 Let $G$ be a graph and $T\subseteq V(G)$ with $|T|$ even.
 Given a well-oriented ear-decomposition of $G$ where all short ears are clean
 %with $k$ ears, of which $k_4$ ears are 4-ears 
 and the oriented ears are precisely the clean ears, we can compute a $T$-tour in $G$ with at most
 \[ \sfrac{3}{2} (n-1)  - \sfrac{1}{26}  \pi + \sfrac{1}{26}(k_4 - 2 k_{\ge 5}) \]
 edges, where $\pi$ is the number of non-entered ears.
\end{theorem}
\prove
We apply Lemma \ref{lemma:result_ear_ind_1} and Lemma \ref{lemma:result_ear_ind_2} and take the shorter of the two $T$-tours.
This yields a $T$-tour with at most the following number of edges:
\begin{align*}  
\min &\biggl\{
 \sfrac{3}{2} (n-1) - \sfrac{7}{20} k_3 - \sum_{i\in I} \max\left\{\sfrac{7}{20} (h_i -1),\ \sfrac{3}{20} \right\},\  \\[1mm] 
 &\ \ \sfrac{3}{2} (n-1) + \sfrac{1}{2} (k_2 + k_3 -  k_{\ge 5}) + \sfrac{1}{2} k_3   - \sfrac{1}{2} k_{\textnormal{bad}}  \biggr\}. 
 \end{align*}
 edges, where $I= \{ i\in \{1,\dots, l\} : (P_i,T_i \cap \inn(P_i))\text{ is good or special}\}$. 
Taking $\frac{10}{13}$ of the first term and $\frac{3}{13}$ of the second term, we can bound it by
\begin{align*}
&\sfrac{3}{2} (n-1) - \sfrac{7}{26} k_3 - \sum_{i\in I} \max\left\{\sfrac{7}{26} (h_i -1),\ \sfrac{3}{26}\right\} 
 + \sfrac{3}{26} (k_2 + k_3 -  k_{\ge 5}) + \sfrac{3}{26} k_3   - \sfrac{3}{26} k_{\textnormal{bad}} \\
 =\ & \sfrac{3}{2} (n-1) - \sfrac{4}{26} k_3 - \sfrac{1}{26} \sum_{i=1}^l \max\{7(h_i -1), 3\} 
 + \sfrac{3}{26} (k_2 + k_3) - \sfrac{3}{26} k_{\ge 5},
\end{align*}
where we used $i\notin I$ if and only if $(P_i,T_i \cap \inn(P_i))$ is bad, and this implies $h_i =1$.
We get the following upper bound on the number of edges of our tour:
\begin{align*}
 & \sfrac{3}{2} (n-1) - \sfrac{1}{26} \sum_{i=1}^l \max\{7(h_i -1), 3\}  + \sfrac{3}{26} (k_2 + k_3) - \sfrac{3}{26} k_{\ge 5} \\
 \le\ & \sfrac{3}{2} (n-1) - \sfrac{1}{26} \sum_{i=1}^l \max\{4h_i - 1 , 0\}  + \sfrac{3}{26} (k_2 + k_3)- \sfrac{3}{26} k_{\ge 5} \\
 =\ & \sfrac{3}{2} (n-1) - \sfrac{1}{26} \sum_{i=1}^l \max\{4h_i , 1\}  + \sfrac{3}{26} (k_2 + k_3)+ \sfrac{1}{26}(k_4 - 2 k_{\ge 5})  \\
 =\ & \sfrac{3}{2} (n-1) - \sfrac{1}{26} \sum_{i=1}^l \max\{h_i , 1\} + \sfrac{1}{26}(k_4 - 2 k_{\ge 5})   \\
 =\ &\sfrac{3}{2} (n-1) - \sfrac{1}{26}\pi + \sfrac{1}{26}(k_4 - 2 k_{\ge 5}).
 \end{align*}
In the last inequality we used that every non-entered ear is a short ear or a long ear with $h_i =0$.
\endproof

\section{Computing the initial ear-decomposition }\label{section:ear_decomposition}

Previous papers that used ear-decompositions for approximation algorithms include \cite{CheSS01}, \cite{SebV12}, and \cite{HeeV17}.
They all exploit a theorem of \cite{Fra93}: one can compute an ear-decomposition with minimum number of even ears in polynomial time.
This minimum is denoted by $\varphi(G)$. Our ear-decompositions will also have only $\varphi(G)$ even ears, 
although (in contrast to the above-mentioned papers) we exploit this property only during the construction of the ear-decomposition.
The ear-decompositions in the above papers also have certain properties of 2-ears and 3-ears; we will additionally deal with 4-ears.
As in \cite{SebV12} we will compute a nice ear-decomposition. In particular, we make all short ears pendant.

We have seen in Theorem \ref{thm:result_ear_induction} that non-entered ears are cheap but 4-ears are expensive.
For pendant 4-ears we can apply Lemma \ref{lemma:standard_ear_induction} beforehand and apply Theorem \ref{thm:result_ear_induction}
to the rest.
Ideally, we would like to make all 4-ears pendant, but this is not always possible.

We distinguish four kinds of 4-ears: pendant, blocked, vertical, and horizontal 
(we will compute an ear decomposition in which every ear is of exactly one of these kinds);
see Figure \ref{fig4ears}:

\begin{figure}[ht]
\begin{center}
 \begin{tikzpicture}[scale=0.75]

  \tikzstyle{vertex}=[circle,fill,minimum size=5.2,inner sep=0pt]
  \tikzstyle{pendant}=[circle,draw,minimum size= 5.2,inner sep=0pt]
  \tikzstyle{degree2}=[draw,minimum size= 4.5,inner sep=0pt]
  \tikzstyle{edge}=[line width=1.5]
  \tikzstyle{path}=[line width=1.5]
  \tikzstyle{trivial}=[line width=1.5, densely dotted]
  \tikzstyle{deleted}=[gray, trivial]

%(a)
 \begin{scope}[shift={(0,0)}]
 \node at (-2.3,3) {(a)};
 
 \node[vertex] (a0)  at ( -0.6, 0) {};
 \node[red,vertex] (a1)  at ( -1, 1.1) {};
 \node[red,vertex] (a2)  at ( 0, 2) {};
 \node[red,vertex] (a3)  at ( 1, 1.1) {};
 \node[vertex] (a4)  at ( 0.6, 0) {};
 \draw[red,edge] (a0) to (a1);
 \draw[red,edge] (a1) to (a2);
 \draw[red,edge] (a2) to (a3);
 \draw[red,edge] (a3) to (a4);
 \draw[blue,path] (a1) to [out=120,in=90] (-2.4,1);
 \draw[blue,path] (-2.4,1) to[out=300,in=210] (a1);

 \begin{scope}[shift={(3,0)}]
 \node[vertex] (a0)  at ( -0.6, 0) {};
 \node[red,vertex] (a1)  at ( -1, 1.1) {};
 \node[red,vertex] (a2)  at ( 0, 2) {};
 \node[red,vertex] (a3)  at ( 1, 1.1) {};
 \node[vertex] (a4)  at ( 0.6, 0) {};
 \draw[red,edge] (a0) to (a1);
 \draw[red,edge] (a1) to (a2);
 \draw[red,edge] (a2) to (a3);
 \draw[red,edge] (a3) to (a4);
 \draw[blue,path] (a2) to [out=30,in=300] (1,3);
 \draw[blue,path] (1,3) to[out=120,in=120] (a2);
 \end{scope}
 \end{scope}

 %(b)
  \begin{scope}[shift={(7,0)}]
 \node at (-0.9,3) {(b)};
 
 \node[vertex] (a0)  at ( -0.6, 0) {};
 \node[red,pendant] (a1)  at ( -1, 1.1) {};
 \node[red,pendant] (a2)  at ( 0, 2) {};
 \node[red,pendant] (a3)  at ( 1, 1.1) {};
  \node[vertex] (a4)  at ( 0.6, 0) {};
 \draw[red,edge] (a0) to (a1);
 \draw[red,edge] (a1) to (a2);
 \draw[red,edge] (a2) to (a3);
 \draw[red,edge] (a3) to (a4);
 \draw[deleted] (a1) to (a3);
 
 \begin{scope}[shift={(3.5,0)}]
 \node at (-0.9,3) {(c)};
 \node[vertex] (a0)  at ( -0.6, 0) {};
 \node[red,pendant] (a1)  at ( -1, 1.1) {};
 \node[red,vertex] (a2)  at ( 0, 2) {};
 \node[red,pendant] (a3)  at ( 1, 1.1) {};
 \node[vertex] (a4)  at ( 0.6, 0) {};
 \draw[red,edge] (a0) to (a1);
 \draw[red,edge] (a1) to (a2);
 \draw[red,edge] (a2) to (a3);
 \draw[red,edge] (a3) to (a4);
\node[vertex] (b0)  at ( -0.2, 0) {};
\node[vertex] (c0)  at ( 0.2, 0) {};
\node[blue,pendant] (b1)  at ( -0.2, 1) {};
\node[darkgreen,pendant] (c1)  at ( 0.2, 1) {};
 \draw[blue,edge] (b0) to (b1);
 \draw[blue,edge] (b1) to (a2);
 \draw[darkgreen,edge] (c0) to (c1);
 \draw[darkgreen,edge] (c1) to (a2);
 \end{scope}
 \end{scope}

%(c)
  \begin{scope}[shift={(14,0)}]
 \node at (-0.9,3) {(d)};
 
 \node[vertex] (a0)  at ( -0.6, 0) {};
 \node[red,vertex] (a1)  at ( -1, 1.1) {};
 \node[red,degree2] (a2)  at ( 0, 2) {};
 \node[red,vertex] (a3)  at ( 1, 1.1) {};
  \node[vertex] (a4)  at ( 0.6, 0) {};
 \draw[red,edge] (a0) to (a1);
 \draw[red,edge] (a1) to (a2);
 \draw[red,edge] (a2) to (a3);
 \draw[red,edge] (a3) to (a4);
 \node[blue,degree2] (b1)  at ( 0, 0.7) {};
  \draw[blue,edge] (a1) to (b1);
 \draw[blue,edge] (b1) to (a3);
 \draw[deleted] (a1) to (a3);
 
 \begin{scope}[shift={(3,0)}]
 \node[vertex] (a0)  at ( -0.6, 0) {};
 \node[red,vertex] (a1)  at ( -1, 1.1) {};
 \node[red,degree2] (a2)  at ( 0, 2) {};
 \node[red,vertex] (a3)  at ( 1, 1.1) {};
 \node[vertex] (a4)  at ( 0.6, 0) {};
 \draw[red,edge] (a0) to (a1);
 \draw[red,edge] (a1) to (a2);
 \draw[red,edge] (a2) to (a3);
 \draw[red,edge] (a3) to (a4);
\node[blue,degree2] (b1)  at ( 0, 0.7) {};
\node[darkgreen,degree2] (c1)  at ( 0, 1.3) {};
 \draw[blue,edge] (a1) to (b1);
 \draw[blue,edge] (b1) to (a3);
 \draw[darkgreen,edge] (a1) to (c1);
 \draw[darkgreen,edge] (c1) to (a3);
 
 \end{scope}
 \end{scope}

 \end{tikzpicture}
 \caption{\label{fig4ears}{\small  (a) blocked 4-ears, (b) pendant 4-ear (c) vertical 4-ear, (d) horizontal 4-ears. 
 Filled circles denote arbitrary vertices, unfilled circles denote pendant vertices, unfilled squares denote degree-2 vertices.
 The endpoints of the 4-ears and endpoints of 2-ears that are not internal vertices of these 4-ears are shown at the bottom (black filled circles);
 some of these can be identical.
 Curves denote closed ears. Dotted edges are possible trivial ears connecting colored vertices.
 } }
\end{center}
\end{figure}
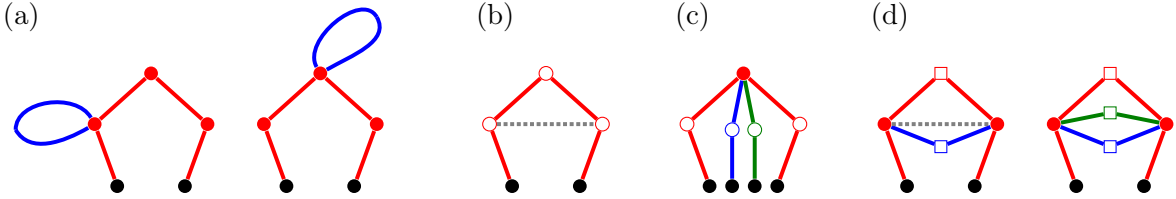

\begin{definition}
\label{def:4eartypes}
A 4-ear is called
\begin{itemize}
\item \emph{blocked} if a closed ear is attached to it.
\item \emph{vertical} if it is nonpendant,  its internal vertices are $v_1,v_2,v_3$ in this order, 
$v_1$ and $v_3$ are pendant and not adjacent, and the only nontrivial ears attached to $v_2$ are 2-ears whose
middle vertex is not adjacent to $v_1$ or $v_3$.
\item \emph{horizontal} if it is nonpendant, its internal vertices are $v_1,v_2,v_3$ in this order, $v_2$ has degree 2,
and all nontrivial ears attached to it are 2-ears with vertices $v_1,x,v_3$, where $x$ is a degree-2 vertex.
\end{itemize}
Call an ear \emph{outer} if it is a 2-ear, 3-ear, pendant 4-ear, vertical 4-ear, or horizontal 4-ear.
Call an ear \emph{inner} if it is a blocked 4-ear or has length at least 5.
\end{definition}

Later we will show that there are only few blocked 4-ears (Lemma \ref{lemma:numberofblockedears}).
Moreover, in Section \ref{section:improving_lower_bound} we make as many vertical 4-ears pendant as possible
and raise the lower bound for every 2-ear that is still attached to a vertical or horizontal 4-ear.

In this section, we prove the following:

\begin{theorem}
\label{earmain}
For any 2-vertex-connected graph $G$ we can construct an ear-decomposition in polynomial time that satisfies the following conditions:
\begin{enumerate}
\item[\rm (a)] 
All short ears are open and pendant.
\item[\rm (b)] 
Each 4-ear is blocked or pendant or vertical or horizontal; no closed 4-ear is attached to any closed 4-ear.

\item[\rm (c)] 
If there are internal vertices $v$ of an outer ear $P$ and $w$ of another outer ear $Q$ that are adjacent, 
then $P$ is attached to $Q$ at $w$, $Q$ is attached to $P$ at $v$, or both $v$ and $w$ are middle vertices of 4-ears. 
No 2-ear is attached to two outer 4-ears.
\end{enumerate}
\end{theorem}

For the proof, we will start with the following:
\begin{lemma}[\cite{CheSS01}] 
\label{openminphi}
For any given 2-vertex-connected graph $G$,
one can compute an open ear-decomposition with $\varphi(G)$ even ears in polynomial time.
\end{lemma}
We will maintain the following invariants at all times.
\begin{enumerate}
\item[\rm (d)] 
All short ears are open.
\item[\rm (e)] 
No closed ears are attached to short ears.
\item[\rm (f)] 
No closed 4-ears or nonpendant 3-ears are attached to any closed 4-ear. 
\item[\rm (g)] 
The number of even ears is $\varphi(G)$.
\end{enumerate}
Note that the result of Lemma \ref{openminphi} satisifes these invariants because all its ears are open,
including the first ear (which is always open by definition).
Now we apply a certain set of operations as long as possible.
Each of them maintains these invariants and decreases the following potential function: We lexicographically
\begin{enumerate}
 \item maximize the number of trivial ears,
 \item minimize the number of 4-ears, and
 \item minimize the number of trivial ears that are incident to middle vertices of outer 4-ears.
\end{enumerate}
Thus after fewer than $n^4$ steps none of the operations can be applied 
(where we use that the number of trivial ears is always at least $|E(G)|-2n$ and at most $|E(G)|-n$).
Then the properties (a), (b), and (c) will hold.
At any stage we put pendant ears at the end of the ear-decomposition in an order of nonincreasing length 
(in particular pendant 3-ears before pendant 2-ears), followed only by trivial ears.

\begin{lemma}
\label{eara}
Given an ear-decomposition with {\rm (d)--(g)} but not {\rm (a)}, 
we can compute an ear-decomposition with more trivial ears and {\rm (d)--(g)} in polynomial time.
\end{lemma}

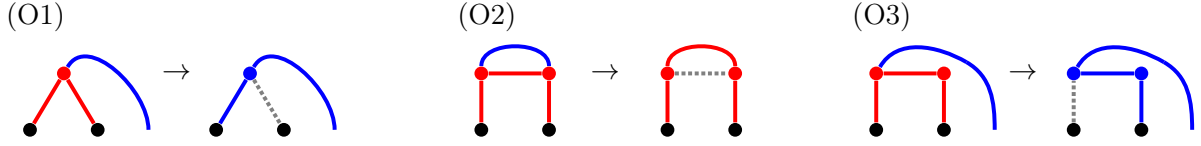
\begin{figure}[ht]
\begin{center}
 \begin{tikzpicture}[scale=0.75]

  \tikzstyle{vertex}=[circle,fill,minimum size=5.2,inner sep=0pt]
  \tikzstyle{pendant}=[circle,draw,minimum size= 5.2,inner sep=0pt]
  \tikzstyle{degree2}=[draw,minimum size= 4.5,inner sep=0pt]
  \tikzstyle{edge}=[line width=1.5]
  \tikzstyle{path}=[line width=1.5]
  \tikzstyle{trivial}=[line width=1.5, densely dotted]
  \tikzstyle{deleted}=[gray, trivial]

%(O1)
 \begin{scope}[shift={(-0.5,0)}]
 \node at (-0.5,2) {(O1)};
 
 \node[vertex] (a0)  at ( -0.6, 0) {};
 \node[red,vertex] (a1)  at ( 0, 1) {};
 \node[vertex] (a2)  at ( 0.6, 0) {};
 \draw[red,edge] (a0) to (a1);
 \draw[red,edge] (a1) to (a2);
 \draw[blue,path] (a1) to [out=60,in=90] (1.5,0);

\node at (2,1) {$\rightarrow$};

 \begin{scope}[shift={(3.3,0)}]
 \node[vertex] (a0)  at ( -0.6, 0) {};
 \node[blue,vertex] (a1)  at ( 0, 1) {};
 \node[vertex] (a2)  at ( 0.6, 0) {};
 \draw[blue,edge] (a0) to (a1);
 \draw[red,deleted] (a1) to (a2);
 \draw[blue,path] (a1) to [out=60,in=90] (1.5,0);
 \end{scope}
 \end{scope}

%(O2)
  \begin{scope}[shift={(7.5,0)}]
   \node at (-0.5,2) {(O2)};
 
 \node[vertex] (a0)  at ( -0.6, 0) {};
 \node[red,vertex] (a1)  at ( -0.6, 1) {};
 \node[red,vertex] (a2)  at ( 0.6, 1) {};
 \node[vertex] (a3)  at ( 0.6, 0) {};
 \draw[red,edge] (a0) to (a1);
 \draw[red,edge] (a1) to (a2);
 \draw[red,edge] (a2) to (a3);
 \draw[blue,path] (a1) to [out=90,in=90] (a2);

\node at (1.6,1) {$\rightarrow$};

 \begin{scope}[shift={(3.3,0)}]
 \node[vertex] (a0)  at ( -0.6, 0) {};
 \node[red,vertex] (a1)  at ( -0.6, 1) {};
 \node[red,vertex] (a2)  at ( 0.6, 1) {};
 \node[vertex] (a3)  at ( 0.6, 0) {};
 \draw[red,edge] (a0) to (a1);
 \draw[deleted] (a1) to (a2);
 \draw[red,edge] (a2) to (a3);
 \draw[red,path] (a1) to [out=90,in=90] (a2);
 \end{scope} 

 \end{scope}

%(O3)
 \begin{scope}[shift={(14.5,0)}]
 \node at (-0.5,2) {(O3)};
 
 \node[vertex] (a0)  at ( -0.6, 0) {};
 \node[red,vertex] (a1)  at ( -0.6, 1) {};
 \node[red,vertex] (a2)  at ( 0.6, 1) {};
 \node[vertex] (a3)  at ( 0.6, 0) {};
 \draw[red,edge] (a0) to (a1);
 \draw[red,edge] (a1) to (a2);
 \draw[red,edge] (a2) to (a3);
 \draw[blue,path] (a1) to [out=60,in=150] (1,1.2);
 \draw[blue,path] (1,1.2) to [out=330,in=90] (1.5,0);

\node at (2,1) {$\rightarrow$};

 \begin{scope}[shift={(3.5,0)}]
 \node[vertex] (a0)  at ( -0.6, 0) {};
 \node[blue,vertex] (a1)  at ( -0.6, 1) {};
 \node[blue,vertex] (a2)  at ( 0.6, 1) {};
 \node[vertex] (a3)  at ( 0.6, 0) {};
 \draw[deleted] (a0) to (a1);
 \draw[blue,edge] (a1) to (a2);
 \draw[blue,edge] (a2) to (a3);
 \draw[blue,path] (a1) to [out=60,in=150] (1,1.2);
 \draw[blue,path] (1,1.2) to [out=330,in=90] (1.5,0);
 \end{scope} 
 \end{scope}
 \end{tikzpicture}
 \caption{\label{fig:make2and3earspendant}{\small  Removing nonpendant short ears.
 (O1): removing a nonpendant 2-ear;
 (O2),(O3): removing a nonpendant 3-ear. 
 New trivial ears are dotted.
  } }
\end{center}
\end{figure}

\prove 
The following set of operations removes a nonpendant 2-ear or 3-ear and is illustrated in Figure \ref{fig:make2and3earspendant}.
Each operation increases the number of trivial ears and does not increase the number of even ears.

\begin{enumerate}
\item[(O1)]
If there is a nonpendant 2-ear $P$, let $Q$ be the first nontrivial ear attached to $P$. Note that $Q$ is open by (e).
We extend $Q$ by one of the edges of $P$ so that the resulting ear is open; $P$ vanishes; the other edge of $P$ becomes a trivial ear.
\end{enumerate}

Note that $Q$ must be odd because otherwise (O1) would reduce the number of even ears, contradicting (g).
Since $Q$ is not a 2-ear, the new ear is not short, and (f) is maintained.

If all 2-ears but not all 3-ears are pendant,
let $P$ be the first nonpendant 3-ear, and let $v_0,v_1,v_2,v_3$ be the vertices of $P$ in this order.
Note that $P$ is open by (d).
Let $Q$ be the first nontrivial ear attached to $P$, and without loss of generality $v_1$ is an endpoint of $Q$.
Note that $Q$ is open by (e). 
There are two cases.

\begin{enumerate}
\item[(O2)]
If $Q$ has endpoints $v_1$ and $v_2$, we replace the middle edge of $P$ by the edges of $Q$,
creating a new open ear with at least four edges and a trivial ear.
\end{enumerate}

\begin{enumerate}
\item[(O3)]
If $v_2$ is not an endpoint of $Q$, 
we extend $Q$ by the $v_1$-$v_3$-path in $P$; the remaining edge of $P$ becomes a trivial ear.
\end{enumerate}

The operation (O3) might create a closed ear of length at least 4, 
but then it is not attached to a short ear because $P$ was the first nonpendant short ear,
and it is not attached to a closed 4-ear because $P$ was not attached to a closed 4-ear by (f).
Moreover, if the new ear is a closed 4-ear, then $Q$ was a pendant 2-ear, and (by the choice of $Q$ and the order of the
ear-decomposition) only pendant 2-ears are attached to the new closed 4-ear. Therefore (f) is maintained.
\endproof

Before we get to condition (b), let us show a sufficient condition for 4-ears to be pendant or vertical or horizontal:

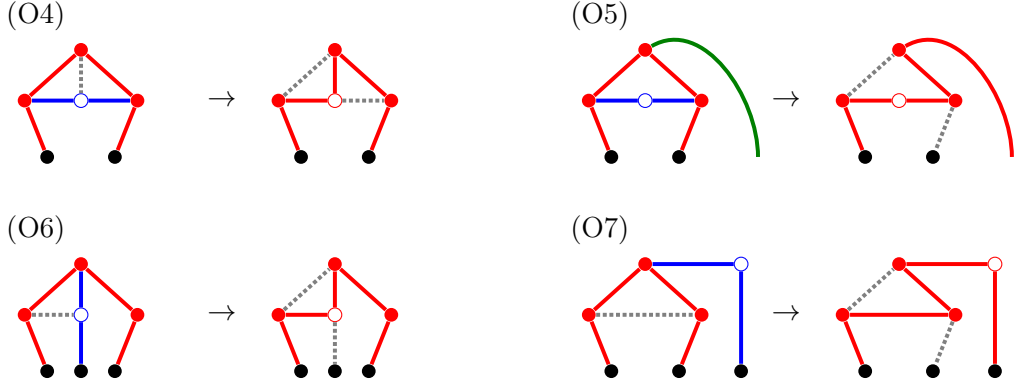
\begin{figure}[ht]
\begin{center}
 \begin{tikzpicture}[scale=0.75]

  \tikzstyle{vertex}=[circle,fill,minimum size=5.2,inner sep=0pt]
  \tikzstyle{pendant}=[circle,draw,minimum size= 5.2,inner sep=0pt]
  \tikzstyle{degree2}=[draw,minimum size= 4.5,inner sep=0pt]
  \tikzstyle{edge}=[line width=1.5]
  \tikzstyle{path}=[line width=1.5]
  \tikzstyle{trivial}=[line width=1.5, densely dotted]
  \tikzstyle{deleted}=[gray, trivial]

%(O4)
 \begin{scope}[shift={(0,-7.6)}]
 \node at (-0.8,2.5) {(O4)};
 
 \node[vertex] (a0)  at ( -0.6, 0) {};
 \node[red,vertex] (a1)  at ( -1, 1) {};
 \node[red,vertex] (a2)  at ( 0, 1.9) {};
 \node[red,vertex] (a3)  at ( 1, 1) {};
 \node[vertex] (a4)  at ( 0.6, 0) {};
 \draw[red,edge] (a0) to (a1);
 \draw[red,edge] (a1) to (a2);
 \draw[red,edge] (a2) to (a3);
 \draw[red,edge] (a3) to (a4);
 \node[blue,pendant] (b1)  at ( 0, 1) {};
 \draw[blue,edge] (a1) to (b1);
 \draw[blue,edge] (b1) to (a3);
 \draw[deleted] (a2) to (b1);

\node at (2.5,1) {$\rightarrow$};

 \begin{scope}[shift={(4.5,0)}]
 \node[vertex] (a0)  at ( -0.6, 0) {};
 \node[red,vertex] (a1)  at ( -1, 1) {};
 \node[red,vertex] (a2)  at ( 0, 1.9) {};
 \node[red,vertex] (a3)  at ( 1, 1) {};
 \node[vertex] (a4)  at ( 0.6, 0) {};
 \draw[red,edge] (a0) to (a1);
 \draw[deleted] (a1) to (a2);
 \draw[red,edge] (a2) to (a3);
 \draw[red,edge] (a3) to (a4);
 \node[red,pendant] (b1)  at ( 0, 1) {};
 \draw[red,edge] (a1) to (b1);
 \draw[deleted] (b1) to (a3);
 \draw[red,path] (a2) to (b1);
 \end{scope}
 \end{scope}

 %(O5)
 \begin{scope}[shift={(10,-7.6)}]
 \node at (-0.8,2.5) {(O5)};

 \node[vertex] (a0)  at ( -0.6, 0) {};
 \node[red,vertex] (a1)  at ( -1, 1) {};
 \node[red,vertex] (a2)  at ( 0, 1.9) {};
 \node[red,vertex] (a3)  at ( 1, 1) {};
 \node[vertex] (a4)  at ( 0.6, 0) {};
 \draw[red,edge] (a0) to (a1);
 \draw[red,edge] (a1) to (a2);
 \draw[red,edge] (a2) to (a3);
 \draw[red,edge] (a3) to (a4);
 \node[blue,pendant] (b1)  at ( 0, 1) {};
 \draw[blue,edge] (a1) to (b1);
 \draw[blue,edge] (b1) to (a3);
 \draw[darkgreen,path] (a2) to [out=30,in=90] (2,0);

\node at (2.5,1) {$\rightarrow$};

 \begin{scope}[shift={(4.5,0)}]
 \node[vertex] (a0)  at ( -0.6, 0) {};
 \node[red,vertex] (a1)  at ( -1, 1) {};
 \node[red,vertex] (a2)  at ( 0, 1.9) {};
 \node[red,vertex] (a3)  at ( 1, 1) {};
 \node[vertex] (a4)  at ( 0.6, 0) {};
 \draw[red,edge] (a0) to (a1);
 \draw[deleted] (a1) to (a2);
 \draw[red,edge] (a2) to (a3);
 \draw[deleted] (a3) to (a4);
 \node[red,pendant] (b1)  at ( 0, 1) {};
 \draw[red,edge] (a1) to (b1);
 \draw[red,edge] (b1) to (a3);
 \draw[red,path] (a2) to [out=30,in=90] (2,0);
 \end{scope}
 \end{scope}

%(O6)
 \begin{scope}[shift={(0,-11.4)}]
 \node at (-0.8,2.5) {(O6)};
 
 \node[vertex] (a0)  at ( -0.6, 0) {};
 \node[red,vertex] (a1)  at ( -1, 1) {};
 \node[red,vertex] (a2)  at ( 0, 1.9) {};
 \node[red,vertex] (a3)  at ( 1, 1) {};
 \node[vertex] (a4)  at ( 0.6, 0) {};
 \draw[red,edge] (a0) to (a1);
 \draw[red,edge] (a1) to (a2);
 \draw[red,edge] (a2) to (a3);
 \draw[red,edge] (a3) to (a4);
 \node[vertex] (b0)  at ( 0, 0) {};
 \node[blue,pendant] (b1)  at ( 0, 1) {};
 \draw[deleted] (a1) to (b1);
 \draw[blue,edge] (b1) to (a2);
 \draw[blue,edge] (b0) to (b1);

\node at (2.5,1) {$\rightarrow$};

 \begin{scope}[shift={(4.5,0)}]
 \node[vertex] (a0)  at ( -0.6, 0) {};
 \node[red,vertex] (a1)  at ( -1, 1) {};
 \node[red,vertex] (a2)  at ( 0, 1.9) {};
 \node[red,vertex] (a3)  at ( 1, 1) {};
 \node[vertex] (a4)  at ( 0.6, 0) {};
 \draw[red,edge] (a0) to (a1);
 \draw[deleted] (a1) to (a2);
 \draw[red,edge] (a2) to (a3);
 \draw[red,edge] (a3) to (a4);
  \node[vertex] (b0)  at ( 0, 0) {};
 \node[red,pendant] (b1)  at ( 0, 1) {};
 \draw[red,edge] (a1) to (b1);
 \draw[deleted] (b1) to (b0);
 \draw[red,path] (a2) to (b1);
 \end{scope}
 \end{scope}

 %(O7)
 \begin{scope}[shift={(10,-11.4)}]
 \node at (-0.8,2.5) {(O7)};

 \node[vertex] (a0)  at ( -0.6, 0) {};
 \node[red,vertex] (a1)  at ( -1, 1) {};
 \node[red,vertex] (a2)  at ( 0, 1.9) {};
 \node[red,vertex] (a3)  at ( 1, 1) {};
 \node[vertex] (a4)  at ( 0.6, 0) {};
 \draw[red,edge] (a0) to (a1);
 \draw[red,edge] (a1) to (a2);
 \draw[red,edge] (a2) to (a3);
 \draw[red,edge] (a3) to (a4);
 \draw[deleted] (a1) to (a3);
 \node[vertex] (b0)  at ( 1.7, 0) {};
 \node[blue,pendant] (b1)  at ( 1.7, 1.9) {};
 \draw[blue,edge] (a2) to (b1);
 \draw[blue,edge] (b0) to (b1);

\node at (2.5,1) {$\rightarrow$};

 \begin{scope}[shift={(4.5,0)}]
 \node[vertex] (a0)  at ( -0.6, 0) {};
 \node[red,vertex] (a1)  at ( -1, 1) {};
 \node[red,vertex] (a2)  at ( 0, 1.9) {};
 \node[red,vertex] (a3)  at ( 1, 1) {};
 \node[vertex] (a4)  at ( 0.6, 0) {};
 \draw[red,edge] (a0) to (a1);
 \draw[deleted] (a1) to (a2);
 \draw[red,edge] (a2) to (a3);
 \draw[deleted] (a3) to (a4);
 \draw[red,edge] (a1) to (a3);
  \node[vertex] (b0)  at ( 1.7, 0) {};
 \node[red,pendant] (b1)  at ( 1.7, 1.9) {};
 \draw[red,edge] (a2) to (b1);
 \draw[red,edge] (b0) to (b1);
 \end{scope}
 \end{scope}

 \end{tikzpicture}
 \caption{\label{fig4earswith2ears}{\small  
 (O4)--(O7): excluding 4-ears that are neither pendant nor vertical nor horizontal. 
  } }
\end{center}
\end{figure}

\begin{lemma}
\label{4earsverticalorhorizontal}
Let $P$ be a 4-ear in an ear-decomposition satisfying (a) and (g).
Let $v_0,v_1,v_2,v_3,v_4$ be the vertices of $P$ in this order (where $v_0$ and $v_4$ can be identical).
Then $P$ is pendant or vertical or horizontal if and only if
every nontrivial ear attached to $P$ is a 2-ear that is attached to $P$ at $v_1$ and $v_3$ or only at $v_2$.
\end{lemma}

\prove
Necessity follows directly from Definition \ref{def:4eartypes}, so we prove sufficiency.
So let $P$ be a 4-ear in an ear-decomposition satisfying (a) and (g) such that
every nontrivial ear attached to $P$ is a (pendant) 2-ear that is attached to $P$ at $v_1$ and $v_3$ or only at $v_2$.

First suppose that a 2-ear $Q$ with endpoints $v_1$ and $v_3$ exists.
Let $w$ be the middle vertex of $Q$.
If $P$ is not horizontal, at least one of the following two operations must apply:

\begin{enumerate}
\item[(O4)]
If there is an edge connecting $w$ and $v_2$,
we replace $P$ by a 5-ear, formed by the edges $\{v_0,v_1\}$, $\{v_2,v_3\}$ and $\{v_3,v_4\}$ of $P$, the edge $\{v_1,w\}$ of $Q$,
and the edge $\{w,v_2\}$.
\end{enumerate}

\begin{enumerate}
\item[(O5)]
If any (possibly trivial) ear $R$ has endpoints $x$ and $y$,
where $x\in\{v_2,w\}$ and $y\notin\{w,v_1,v_2,v_3\}$,
we replace $P$ by an ear of length at least 5, formed by $R$ and all but two of the edges of $P$ and $Q$; 
the remaining two edges become trivial ears.
\end{enumerate}

Each of (O4) and (O5) decreases the number of even ears, and therefore this cannot happen.
It remains to consider the case when all ears attached to $P$ are 2-ears attached to $P$ only at $v_2$.
If $P$ is not pendant, there is at least one such a 2-ear attached to $P$.

If $P$ also not vertical, this means that $v_1$ and $v_3$ are adjacent or 
the middle vertex $w$ of a 2-ear attached to $v_2$ is adjacent to $v_1$ or $v_3$.
Then we can apply one of the following operations:

\begin{enumerate}
\item[(O6)]
If $w$ is adjacent to $v_1$, we can replace $P$ by a 5-ear with edges 
$\{v_0, v_1\}, \{v_1, w\}, \{w, v_2\}, \{v_2, v_3\}$ and $\{v_3, v_4\}$. 
(By renumbering the vertices of $P$, the same operation applies if $w$ is adjacent to $v_3$.)
\end{enumerate}

\begin{enumerate}
\item[(O7)]
If $v_1$ and $v_3$ are adjacent, we can replace $P$ by a 5-ear formed by a 2-ear attached to $P$ at $v_2$ and
the edges $\{v_0, v_1\}, \{v_1, v_3\}, \{v_3, v_2\}$.
\end{enumerate}
Both (O6) and (O7) replace $P$ and an attached 2-ear by a 5-ear, again contradicting (g). 
\endproof

\begin{lemma}
\label{earb}
Given an ear-decomposition with {\rm (d)--(g)} but not {\rm (b)}, 
we can compute an ear-decomposition with {\rm (d)--(g)} in polynomial time
that either has more trivial ears or has the same number of trivial ears but fewer 4-ears.
\end{lemma}

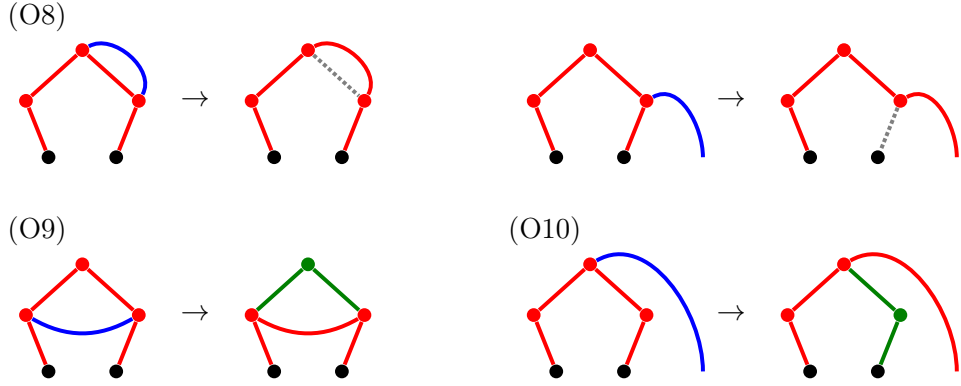
\begin{figure}[ht]
\begin{center}
 \begin{tikzpicture}[scale=0.75]

  \tikzstyle{vertex}=[circle,fill,minimum size=5.2,inner sep=0pt]
  \tikzstyle{pendant}=[circle,draw,minimum size= 5.2,inner sep=0pt]
  \tikzstyle{degree2}=[draw,minimum size= 4.5,inner sep=0pt]
  \tikzstyle{edge}=[line width=1.5]
  \tikzstyle{path}=[line width=1.5]
  \tikzstyle{trivial}=[line width=1.5, densely dotted]
  \tikzstyle{deleted}=[gray, trivial]

%(O8a)
 \begin{scope}[shift={(0,0)}]
 \node at (-0.8,2.5) {(O8)};
 
 \node[vertex] (a0)  at ( -0.6, 0) {};
 \node[red,vertex] (a1)  at ( -1, 1) {};
 \node[red,vertex] (a2)  at ( 0, 1.9) {};
 \node[red,vertex] (a3)  at ( 1, 1) {};
 \node[vertex] (a4)  at ( 0.6, 0) {};
 \draw[red,edge] (a0) to (a1);
 \draw[red,edge] (a1) to (a2);
 \draw[red,edge] (a2) to (a3);
 \draw[red,edge] (a3) to (a4);
 \draw[blue,path] (a2) to [out=30,in=60] (a3);

\node at (2,1) {$\rightarrow$};

 \begin{scope}[shift={(4,0)}]
 \node[vertex] (a0)  at ( -0.6, 0) {};
 \node[red,vertex] (a1)  at ( -1, 1) {};
 \node[red,vertex] (a2)  at ( 0, 1.9) {};
 \node[red,vertex] (a3)  at ( 1, 1) {};
 \node[vertex] (a4)  at ( 0.6, 0) {};
 \draw[red,edge] (a0) to (a1);
 \draw[red,edge] (a1) to (a2);
 \draw[deleted] (a2) to (a3);
 \draw[red,edge] (a3) to (a4);
 \draw[red,path] (a2) to [out=30,in=60] (a3);
 \end{scope}
 \end{scope}

 %(O8b)
  \begin{scope}[shift={(9,0)}]
 
  \node[vertex] (a0)  at ( -0.6, 0) {};
 \node[red,vertex] (a1)  at ( -1, 1) {};
 \node[red,vertex] (a2)  at ( 0, 1.9) {};
 \node[red,vertex] (a3)  at ( 1, 1) {};
 \node[vertex] (a4)  at ( 0.6, 0) {};
 \draw[red,edge] (a0) to (a1);
 \draw[red,edge] (a1) to (a2);
 \draw[red,edge] (a2) to (a3);
 \draw[red,edge] (a3) to (a4);
 \draw[blue,path] (a3) to [out=30,in=90] (2,0);

\node at (2.5,1) {$\rightarrow$};

 \begin{scope}[shift={(4.5,0)}]
 \node[vertex] (a0)  at ( -0.6, 0) {};
 \node[red,vertex] (a1)  at ( -1, 1) {};
 \node[red,vertex] (a2)  at ( 0, 1.9) {};
 \node[red,vertex] (a3)  at ( 1, 1) {};
 \node[vertex] (a4)  at ( 0.6, 0) {};
 \draw[red,edge] (a0) to (a1);
 \draw[red,edge] (a1) to (a2);
 \draw[red,edge] (a2) to (a3);
 \draw[deleted] (a3) to (a4);
 \draw[red,edge] (a3) to [out=30,in=90] (2,0);
 \end{scope}

 \end{scope}

%(O9)
 \begin{scope}[shift={(0,-3.8)}]
 \node at (-0.8,2.5) {(O9)};
 
 \node[vertex] (a0)  at ( -0.6, 0) {};
 \node[red,vertex] (a1)  at ( -1, 1) {};
 \node[red,vertex] (a2)  at ( 0, 1.9) {};
 \node[red,vertex] (a3)  at ( 1, 1) {};
 \node[vertex] (a4)  at ( 0.6, 0) {};
 \draw[red,edge] (a0) to (a1);
 \draw[red,edge] (a1) to (a2);
 \draw[red,edge] (a2) to (a3);
 \draw[red,edge] (a3) to (a4);
 \draw[blue,path] (a1) to [out=330,in=210] (a3);

\node at (2,1) {$\rightarrow$};

 \begin{scope}[shift={(4,0)}]
 \node[vertex] (a0)  at ( -0.6, 0) {};
 \node[red,vertex] (a1)  at ( -1, 1) {};
 \node[darkgreen,vertex] (a2)  at ( 0, 1.9) {};
 \node[red,vertex] (a3)  at ( 1, 1) {};
 \node[vertex] (a4)  at ( 0.6, 0) {};
 \draw[red,edge] (a0) to (a1);
 \draw[darkgreen,edge] (a1) to (a2);
 \draw[darkgreen,edge] (a2) to (a3);
 \draw[red,edge] (a3) to (a4);
 \draw[red,path] (a1) to [out=330,in=210] (a3);
 \end{scope}
 \end{scope}

 %(O10)
  \begin{scope}[shift={(9,-3.8)}]
 \node at (-0.8,2.5) {(O10)};
 
  \node[vertex] (a0)  at ( -0.6, 0) {};
 \node[red,vertex] (a1)  at ( -1, 1) {};
 \node[red,vertex] (a2)  at ( 0, 1.9) {};
 \node[red,vertex] (a3)  at ( 1, 1) {};
 \node[vertex] (a4)  at ( 0.6, 0) {};
 \draw[red,edge] (a0) to (a1);
 \draw[red,edge] (a1) to (a2);
 \draw[red,edge] (a2) to (a3);
 \draw[red,edge] (a3) to (a4);
 \draw[blue,path] (a2) to [out=30,in=90] (2,0);

\node at (2.5,1) {$\rightarrow$};

 \begin{scope}[shift={(4.5,0)}]
 \node[vertex] (a0)  at ( -0.6, 0) {};
 \node[red,vertex] (a1)  at ( -1, 1) {};
 \node[red,vertex] (a2)  at ( 0, 1.9) {};
 \node[darkgreen,vertex] (a3)  at ( 1, 1) {};
 \node[vertex] (a4)  at ( 0.6, 0) {};
 \draw[red,edge] (a0) to (a1);
 \draw[red,edge] (a1) to (a2);
 \draw[darkgreen,edge] (a2) to (a3);
 \draw[darkgreen,edge] (a3) to (a4);
 \draw[red,edge] (a2) to [out=30,in=90] (2,0);
 \end{scope}
 \end{scope}

 \end{tikzpicture}
 \caption{\label{figmake4earssatisfyb}{\small  Making sure that every 4-ear is blocked or pendant or vertical or horizontal.
 (O8): removing nontrivial ears attached to a 4-ear either at $v_1$ or at $v_3$; 
 (O9): removing ears of length at least 3 attached to a 4-ear at $v_1$ and $v_3$;
 (O10): removing ears of length at least 3 attached to a 4-ear only at $v_2$;
  } }
\end{center}
\end{figure}

\prove 
We may assume that (a) holds; otherwise we are done by Lemma \ref{eara}.
We have (f) but not (b), so let $P$ be a 4-ear that is neither blocked nor pendant nor vertical nor horizontal.
Let $v_0,v_1,v_2,v_3,v_4$ be the vertices of $P$ in this order (where $v_0$ and $v_4$ can be identical), and let $Q$ be the first ear attached to $P$. 
As $P$ is neither pendant nor blocked, $Q$ is nontrivial and open.

\begin{enumerate}
\item[(O8)]
If either $v_1$ or $v_3$ is an endpoint of $Q$,
we can replace one edge of $P$ by $Q$; this edge becomes a trivial ear. The new nontrivial ear has length at least 5.
\end{enumerate}

Note that if $Q$ has endpoints $v_0$ and $v_3$, or $v_1$ and $v_4$, this creates a closed ear.
Since all short ears were pendant, neither $v_0$ nor $v_4$ can be an internal vertex of a short ear, so (e) is maintained.

\begin{enumerate}
\item[(O9)]
If $Q$ has endpoints $v_1$ and $v_3$ and is not a 2-ear,
we replace $P$ by an ear of length at least 5, formed by the edges $\{v_0,v_1\}$ and $\{v_3,v_4\}$ of $P$ and the edges of $Q$;
in addition we replace $Q$ by a 2-ear that consists of the remaining edges of $P$.
\end{enumerate}

\begin{enumerate}
\item[(O10)]
If $Q$ is attached to $P$ only at $v_2$ and is not a 2-ear,
we replace $P$ by an ear of length at least 5, formed by the edges $\{v_0,v_1\}$ and $\{v_1,v_2\}$ of $P$ and the edges of $Q$;
in addition we replace $Q$ by a 2-ear with the remaining edges of $P$.
\end{enumerate}

Operations (O9) and (O10) do not change the number of trivial ears, but each of them decreases the number of 4-ears.

If none of the above operations apply, every nontrivial ear attached to $P$ is a pendant 2-ear, 
attached to $P$ at $v_1$ and $v_3$ or only at $v_2$. 
Then we are done by Lemma \ref{4earsverticalorhorizontal}.
\endproof

\begin{lemma}
\label{earc}
Given an ear-decomposition with {\rm (d)--(g)} but not {\rm (c)}, 
we can compute an ear-decomposition with {\rm (d)--(g)} in polynomial time
that either has more trivial ears, or has the same number of trivial ears but fewer 4-ears,
or has the same number of trivial ears and the same number of 4-ears but fewer trivial ears that are incident to middle vertices of outer 4-ears.
\end{lemma}

\begin{figure}[ht]
\begin{center}
 \begin{tikzpicture}[scale=0.75]

  \tikzstyle{vertex}=[circle,fill,minimum size=5.2,inner sep=0pt]
  \tikzstyle{pendant}=[circle,draw,minimum size= 5.2,inner sep=0pt]
  \tikzstyle{degree2}=[draw,minimum size= 4.5,inner sep=0pt]
  \tikzstyle{edge}=[line width=1.5]
  \tikzstyle{path}=[line width=1.5]
  \tikzstyle{trivial}=[line width=1.5, densely dotted]
  \tikzstyle{deleted}=[gray, trivial]

%(O11)
 \begin{scope}[shift={(-1,-3.5)}]
 \node at (-0.3,2) {(O11)};
 
 \node[vertex] (a0)  at ( -0.6, 0) {};
 \node[red,pendant] (a1)  at ( -0.6, 1.2) {};
 \node[red,pendant] (a2)  at ( 0, 0.6) {};
 \node[vertex] (a3)  at ( 0.6, 0) {};
 \draw[red,edge] (a0) to (a1);
 \draw[red,edge] (a1) to (a2);
 \draw[red,edge] (a2) to (a3);
  \node[vertex] (b0)  at ( 1.2, 0) {};
 \node[blue,pendant] (b1)  at ( 1.8, 0.6) {};
 \node[blue,pendant] (b2)  at ( 2.4, 1.2) {};
 \node[vertex] (b3)  at ( 2.4, 0) {};
 \draw[blue,edge] (b0) to (b1);
 \draw[blue,edge] (b1) to (b2);
 \draw[blue,edge] (b2) to (b3);
 \draw[deleted] (a1) to (b2);
 
\node at (3.2,1) {$\rightarrow$};

 \begin{scope}[shift={(4.5,0)}]
  \node[vertex] (a0)  at ( -0.6, 0) {};
 \node[red,pendant] (a1)  at ( -0.6, 1.2) {};
 \node[red,pendant] (a2)  at ( 0, 0.6) {};
 \node[vertex] (a3)  at ( 0.6, 0) {};
 \draw[deleted] (a0) to (a1);
 \draw[red,edge] (a1) to (a2);
 \draw[red,edge] (a2) to (a3);
  \node[vertex] (b0)  at ( 1.2, 0) {};
 \node[blue,pendant] (b1)  at ( 1.8, 0.6) {};
 \node[blue,pendant] (b2)  at ( 2.4, 1.2) {};
 \node[vertex] (b3)  at ( 2.4, 0) {};
 \draw[red,edge] (b0) to (b1);
 \draw[red,edge] (b1) to (b2);
 \draw[deleted] (b2) to (b3);
 \draw[red,edge] (a1) to (b2); 
 \end{scope}
 \end{scope}

%(O12)
 \begin{scope}[shift={(9,-3.5)}]
 \node at (-1.2,2) {(O12)};
    
 \node[vertex] (a0)  at ( -0.6, 0) {};
 \node[red,pendant] (a1)  at ( -1, 1.1) {};
 \node[red,vertex] (a2)  at ( 0, 2) {};
 \node[red,pendant] (a3)  at ( 1, 1.1) {};
 \node[vertex] (a4)  at ( 0.6, 0) {};
 \draw[red,edge] (a0) to (a1);
 \draw[red,edge] (a1) to (a2);
 \draw[red,edge] (a2) to (a3);
 \draw[red,edge] (a3) to (a4);
 
 \node[vertex] (d0)  at ( 2.4, 0) {};
 \node[red,pendant] (d1)  at ( 2, 1.1) {};
 \node[red,vertex] (d2)  at ( 3, 2) {};
 \node[red,pendant] (d3)  at ( 4, 1.1) {};
 \node[vertex] (d4)  at ( 3.6, 0) {};
 \draw[red,edge] (d0) to (d1);
 \draw[red,edge] (d1) to (d2);
 \draw[red,edge] (d2) to (d3);
 \draw[red,edge] (d3) to (d4);
 \node[vertex] (b0)  at ( 3, 0) {};
 \node[cyan,pendant] (b1)  at ( 3, 1) {};
 \draw[cyan,edge] (b0) to (b1);
 \draw[cyan,edge] (b1) to (d2);
 
 \node[blue,pendant] (b1)  at ( 1.5, 2) {};
 \node[vertex] (b0)  at ( 1.3, 0) {};
 \node[vertex] (b3)  at ( 1.7, 0) {};
 \draw[blue, edge] (b3) to (b1);
 \draw[blue, edge] (b0) to (b1);
 \draw[deleted] (b1) to (d2);
 \draw[deleted] (b1) to (a2);
 
 \node at (4.5,1.5) {$\rightarrow$};

 \begin{scope}[shift={(6,0)}]
  \node[vertex] (a0)  at ( -0.6, 0) {};
 \node[red,pendant] (a1)  at ( -1, 1.1) {};
 \node[red,vertex] (a2)  at ( 0, 2) {};
 \node[green,pendant] (a3)  at ( 1, 1.1) {};
 \node[vertex] (a4)  at ( 0.6, 0) {};
 \draw[red,edge] (a0) to (a1);
 \draw[red,edge] (a1) to (a2);
 \draw[green,edge] (a2) to (a3);
 \draw[green,edge] (a3) to (a4);
 
 \node[vertex] (d0)  at ( 2.4, 0) {};
 \node[darkgreen,pendant] (d1)  at ( 2, 1.1) {};
 \node[red,vertex] (d2)  at ( 3, 2) {};
 \node[red,pendant] (d3)  at ( 4, 1.1) {};
 \node[vertex] (d4)  at ( 3.6, 0) {};
 \draw[darkgreen,edge] (d0) to (d1);
 \draw[darkgreen,edge] (d1) to (d2);
 \draw[red,edge] (d2) to (d3);
 \draw[red,edge] (d3) to (d4);
 \node[vertex] (b0)  at ( 3, 0) {};
 \node[cyan,pendant] (b1)  at ( 3, 1) {};
 \draw[cyan,edge] (b0) to (b1);
 \draw[cyan,edge] (b1) to (d2);

 \node[red,pendant] (b1)  at ( 1.5, 2) {};
 \node[vertex] (b0)  at ( 1.3, 0) {};
 \node[vertex] (b3)  at ( 1.7, 0) {};
 \draw[deleted] (b0) to (b1);
 \draw[deleted] (b3) to (b1);
 \draw[red,edge] (a2) to (b1);
 \draw[red,edge] (b1) to (d2);   
 \end{scope}
 
 \end{scope}

%(O12)
 \begin{scope}[shift={(0,-7)}]
 \node at (-1.2,2) {(O12)};
  
 \node[vertex] (a0)  at ( -0.6, 0) {};
 \node[red,pendant] (a1)  at ( -1, 1.1) {};
 \node[red,vertex] (a2)  at ( 0, 2) {};
 \node[red,pendant] (a3)  at ( 1, 1.1) {};
 \node[vertex] (a4)  at ( 0.6, 0) {};
 \draw[red,edge] (a0) to (a1);
 \draw[red,edge] (a1) to (a2);
 \draw[red,edge] (a2) to (a3);
 \draw[red,edge] (a3) to (a4);
 
 \node[vertex] (d0)  at ( 2.4, 0) {};
 \node[red,pendant] (d1)  at ( 2, 1.1) {};
 \node[red,vertex] (d2)  at ( 3, 2) {};
 \node[red,pendant] (d3)  at ( 4, 1.1) {};
 \node[vertex] (d4)  at ( 3.6, 0) {};
 \draw[red,edge] (d0) to (d1);
 \draw[red,edge] (d1) to (d2);
 \draw[red,edge] (d2) to (d3);
 \draw[red,edge] (d3) to (d4);
 \node[vertex] (b0)  at ( 3, 0) {};
 \node[cyan,pendant] (b1)  at ( 3, 1) {};
 \draw[cyan,edge] (b0) to (b1);
 \draw[cyan,edge] (b1) to (d2);

 \node[blue,pendant] (b1)  at ( 1.5, 2) {};

 \draw[blue, edge] (a2) to (b1);
 \draw[blue, edge] (b1) to (d2);
 
 \node at (4.5,1.5) {$\rightarrow$};

 \begin{scope}[shift={(6,0)}]
  \node[vertex] (a0)  at ( -0.6, 0) {};
 \node[red,pendant] (a1)  at ( -1, 1.1) {};
 \node[red,vertex] (a2)  at ( 0, 2) {};
 \node[green,pendant] (a3)  at ( 1, 1.1) {};
 \node[vertex] (a4)  at ( 0.6, 0) {};
 \draw[red,edge] (a0) to (a1);
 \draw[red,edge] (a1) to (a2);
 \draw[green,edge] (a2) to (a3);
 \draw[green,edge] (a3) to (a4);
 
 \node[vertex] (d0)  at ( 2.4, 0) {};
 \node[darkgreen,pendant] (d1)  at ( 2, 1.1) {};
 \node[red,vertex] (d2)  at ( 3, 2) {};
 \node[red,pendant] (d3)  at ( 4, 1.1) {};
 \node[vertex] (d4)  at ( 3.6, 0) {};
 \draw[darkgreen,edge] (d0) to (d1);
 \draw[darkgreen,edge] (d1) to (d2);
 \draw[red,edge] (d2) to (d3);
 \draw[red,edge] (d3) to (d4);
 \node[vertex] (b0)  at ( 3, 0) {};
 \node[cyan,pendant] (b1)  at ( 3, 1) {};
 \draw[cyan,edge] (b0) to (b1);
 \draw[cyan,edge] (b1) to (d2);

 % \node[vertex] (b0)  at ( 1.3, 0) {};
 \node[red,pendant] (b1)  at ( 1.5, 2) {};
 % \node[vertex] (b2)  at ( 1.7, 0) {};
 % \draw[deleted] (b0) to (b1);
 % \draw[deleted] (b1) to (b2);
 \draw[red,edge] (a2) to (b1);
 \draw[red,edge] (b1) to (d2);
  
 \end{scope}
\end{scope}

%(O13)
 \begin{scope}[shift={(12.5,-7)}]
 \node at (-1,2) {(O13)};

 \node[vertex] (a0)  at ( -0.6, 0) {};
 \node[red,vertex] (a1)  at ( 0.6, 1.9) {};
 \node[vertex] (a2)  at ( 0.6, 0) {};
 \draw[red,path] (a1) to [out=180,in=90] (a0);
 \draw[red,edge] (a1) to (a2);
 \node[vertex] (b0)  at ( 1.4, 0) {};
 \node[blue,vertex] (b1)  at ( 2, 1.2) {};
 \node[vertex] (b2)  at ( 2.6, 0) {};
 \draw[blue,path] (b1) to [out=180,in=90] (b0);
 \draw[blue,path] (b1) to [out=0,in=90] (b2);
 \draw[deleted] (a1) to (b1);
  
 \node at (3,1.5) {$\rightarrow$};

 \begin{scope}[shift={(4,0)}]
  \node[vertex] (a0)  at ( -0.6, 0) {};
 \node[red,vertex] (a1)  at ( 0.6, 1.9) {};
 \node[vertex] (a2)  at ( 0.6, 0) {};
 \draw[red,path] (a1) to [out=180,in=90] (a0);
 \draw[deleted] (a1) to (a2);
 \node[vertex] (b0)  at ( 1.4, 0) {};
 \node[blue,vertex] (b1)  at ( 2, 1.2) {};
 \node[vertex] (b2)  at ( 2.6, 0) {};
 \draw[blue,path] (b1) to [out=180,in=90] (b0);
 \draw[blue,path] (b1) to [out=0,in=90] (b2);
 \draw[red,edge] (a1) to (b1);
\end{scope}
\end{scope}

 \end{tikzpicture}
 \caption{\label{figmakecfulfilled}{\small  Ensuring condition (c).
 (O11): removing 3-ears whose middle vertices are adjacent;
 (O12): removing 4-ears whose middle vertices are both adjacent to the middle vertex of a 2-ear;
 (O13): an internal vertex of an outer ear (red) that is adjacent to an endpoint of this ear and to an internal
 vertex of a different outer ear (blue, 2-ear or 4-ear).
  } }
\end{center}
\end{figure}
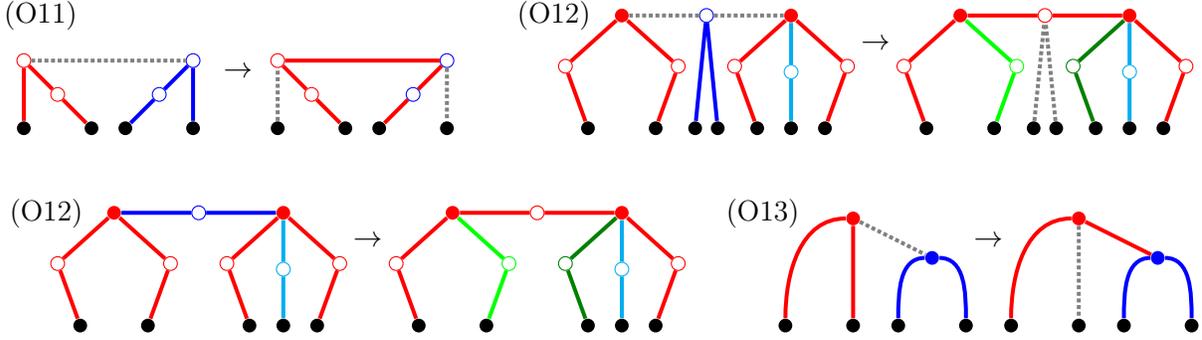

\prove 
We may assume that (a) and (b) hold, because otherwise Lemma \ref{eara} or Lemma \ref{earb} does the job.
First we consider adjacent 3-ears:

\begin{enumerate}
\item[(O11)]
If internal vertices of two (pendant) 3-ears are adjacent, we replace these ears
by a pendant 5-ear.
One trivial ear vanishes, but there are two new trivial ears.
\end{enumerate}

Next we consider 2-ears adjacent to 4-ears:

\begin{enumerate}
\item[(O12)]
If the middle vertices of two outer 4-ears are both adjacent to the middle vertex of a 2-ear, 
we replace these three ears by a 6-ear and two pendant 2-ears.
\end{enumerate}

(O12) does not change the number of trivial ears but removes two 4-ears.
Note that the new 5-ear or 6-ear in (O11) and (O12) can be closed, but then its endpoint was an endpoint of nontrivial ears before.
Due to (a) this means that (e) is preserved.

Now we may assume that (a) and (b) hold and neither (O11) nor (O12) applies.
Since (c) is violated, there are internal vertices $v$ of an outer ears $P$ and  $w$ of an outer ear $Q$ that are adjacent,
but $P$ is not attached to $Q$ at $w$, $Q$ is not attached to $P$ at $v$, 
and the vertices $v$ and $w$ are not both the middle vertex of a 4-ear.
Thus, after possibly exchanging the roles of $P$ and $Q$ and of $v$ and $w$,
the vertex $v$ is adjacent to an endpoint $x$ of $P$.
Since (O11) does not apply, $P$ and $Q$ cannot both be 3-ears. 
As all internal vertices of 3-ears are adjacent to an endpoint of the ear, we can assume that $Q$ is not a 3-ear.
(Otherwise, exchange the roles  of $P$ and $Q$ and of $v$ and $w$.)
Hence, the final case is covered by the following operation:

\begin{enumerate}
\item[(O13)]
If there are two outer ears $P$ and $Q$ none of which is attached to the other, and two internal vertices $v$ of $P$
and $w$ of $Q$ that are adjacent, and $v$ is adjacent to an endpoint $x$ of $P$, and $Q$ is not a 3-ear,
then we replace the edge $\{v,x\}$ by the edge $\{v,w\}$ in $P$ so that then $P$ is attached to $Q$.
If this violates (a) or (b), we then apply Lemma \ref{eara} or \ref{earb}. 
\end{enumerate}

Note that we avoid generating nonpendant 3-ears in order to maintain (f).
If (a) and (b) are not violated by (O13), this does not change the number of trivial ears or 4-ears.
In this case, $P$ is a 2-ear, $Q$ is an outer 4-ear, and $w$ is the middle vertex of $Q$.
Moreover, $v$ is not adjacent to the middle vertex of any other outer 4-ear since (O12) does not apply.
Therefore the new trivial ear is not incident to a middle internal vertex of an outer 4-ear 
and thus the operation (O13) reduces the number of trivial ears that are incident to middle vertices of outer 4-ears.
\endproof

Lemma \ref{openminphi}, \ref{eara}, \ref{earb}, and \ref{earc} imply Theorem \ref{earmain}.

\section{Optimizing outer ears and improving the lower bound}\label{section:improving_lower_bound}

The ear-decomposition from Theorem \ref{earmain} is the starting point for optimizing the outer ears.
While we do not touch pendant or horizontal 4-ears, we will change short ears and vertical 4-ears.
Like in \cite{SebV12}, our goal is that as many short ears as possible form a forest.
Because 2-ears entering 4-ears are not always useful, 
we will in addition try to make the vertical 4-ears pendant, by re-designing the 2-ears attached to them.
The two subpaths of a 4-ear from the middle vertex to an endpoint will be part of this optimization,
and might be replaced by attached 2-ears.

For every 2-ear that will not be part of the forest or remains attached to an outer 4-ear, we will raise the lower bound.
This includes in particular 2-ears attached to horizontal ears, which cannot be optimized.

\subsection{Matroid intersection}

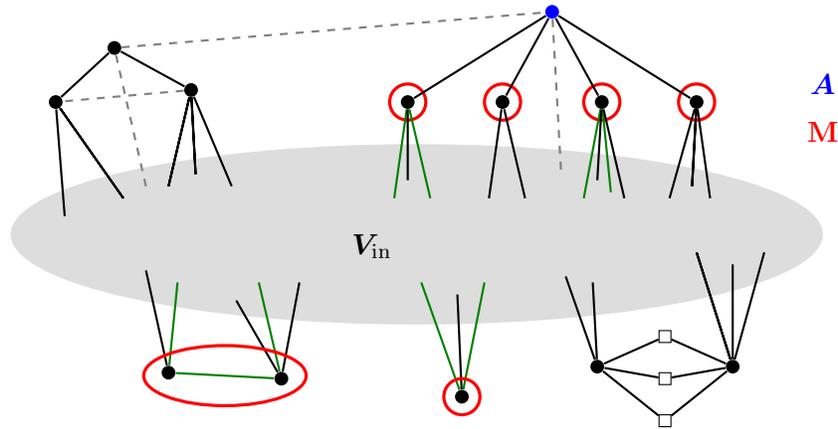
\begin{figure}
 \begin{center}
  \begin{tikzpicture}[xscale=0.6, yscale=0.8]
    \tikzstyle{vertex}=[circle,fill,minimum size=5.2,inner sep=0pt]
    \tikzstyle{pendant}=[circle,fill,minimum size= 5.2,inner sep=0pt]
    \tikzstyle{degree2}=[draw,minimum size= 4.5,inner sep=0pt]
    \tikzstyle{edge}=[thick, black]
    \tikzstyle{dual}=[fill=none, very thick]
    \tikzstyle{ear}=[thick, darkgreen]
    
    % inner ears
    \draw[fill=grey, draw=none, opacity=0.2] (0,-0.2) ellipse ({9} and {1.5});
   
   %vertical 4-ear
    \node[blue, vertex] (w) at (3,3.5) {};
    \node[pendant] (v1) at (-0.2,2) {};
    \node[pendant] (v2) at (1.9,2) {};
    \node[pendant] (v3) at (4.1,2) {};
    \node[pendant] (v4) at (6.2,2) {};

    \draw[dual, red] (v1) ellipse ({0.4} and {0.3});
    \draw[dual, red] (v2) ellipse ({0.4} and {0.3});
    \draw[dual, red] (v3) ellipse ({0.4} and {0.3});
    \draw[dual, red] (v4) ellipse ({0.4} and {0.3});
    
    \draw[edge] (w) --(v1);
    \draw[edge] (w) --(v2);
    \draw[edge] (w) --(v3);
    \draw[edge] (w) --(v4);
    \draw[ear] (-0.5, 0.4) --(v1) -- (0.3,0.4);
    \draw[edge] (v1) -- (-0.2, 0.7);
    \draw[edge] (1.6, 0.4) --(v2) -- (2.4,0.4); %(1.9, 0.7) --(v2) -- (2.4,0.4);
    \draw[ear] (3.7, 0.4) --(v3) -- (4.3, 0.5);
    \draw[edge]  (4.0, 0.7) -- (v3) -- (4.6,0.4);
    \draw[edge] (5.6, 0.4) --(v4) -- (6.1, 0.6) --(v4) -- (6.5,0.4);
    
    % horizontal 4-ear
    \node[degree2] (u1) at (5.5,-3.3) {};
    \node[degree2] (u2) at (5.5,-2.6) {};
    \node[degree2] (u3) at (5.5,-1.9) {};
    \node[vertex] (w1) at (7,-2.4) {};
    \node[vertex] (w2) at (4,-2.4) {};
    \draw[edge] (w1) -- (u1) -- (w2) -- (u2) -- (w1) -- (u3) -- (w2);
     \draw[edge] (7.7, -0.5) --(w1) -- (6.2, -0.5) --(w1) -- (7,-0.7);
     \draw[edge] (3.9, -1) --(w2) -- (3.3,-0.9);
     
     %3-ear
     \node[vertex] (z1) at (-5.5,-2.5) {};
     \node[vertex] (z2) at (-3,-2.6) {};
     \draw[edge] (-6, -0.8) -- (z1);
     \draw[ear] (-5.3, -1) -- (z1) -- (z2) -- (-3.5,-1); 
     \draw[edge] (-4, -1.3) -- (z2) -- (-2.6, -1);
     \draw[dual, red] (-4.25, -2.55) ellipse ({1.8} and {0.5});
     
     % 2-ear
     \node[vertex] (x) at (1, -2.9) {};
     \draw[ear] (0.1, -1) -- (x) -- (1.5,-1);
     \draw[edge] (x) -- (0.9, -1.2);
     \draw[dual, red] (x) ellipse ({0.4} and {0.3});  
     
     % pendant 4-ear
     \node[vertex] (y2) at (-8, 2) {};
     \node[vertex] (y1) at (-5, 2.2) {};
     \node[vertex] (y) at (-6.7, 2.9) {};
     \draw[edge] (-4.1, 0.6) -- (y1) -- (-4.9, 0.8) -- (y1) -- (-5.5,0.6) -- (y1) -- (y) 
        -- (y2) -- (-6.5, 0.4) -- (y2)  -- (-7.8,0.1);

     % trivial ears
     \draw[thick, gray, dashed] (w) -- (3.2,0.8);
     \draw[thick, gray, dashed] (y) -- (-6,0.6);
     \draw[thick, gray, dashed] (y1) --(y2);
     \draw[thick, gray, dashed] (y) --(w);
     
     \node (dummy) at (-1,-0.4) {{\boldmath $V_{\text{in}}$}};
    \node[red] (dummy) at (9,1.5) {\textbf{M}};
     \node[blue] (w') at (9,2.3) {{\boldmath $A$}};
  \end{tikzpicture}
 \end{center}
\caption{The outer ears in an ear-decomposition as in Theorem \ref{earmain}, and the sets $M$ and $A$. 
Green edges constitute a possible set of paths that is independent in both matroids.
Dashed edges show possible trivial ears.
\label{fig:overview_outer_ears}}
\end{figure}
Given an ear-decomposition as in Theorem \ref{earmain}, let
\begin{align*}
 M \ := \ &\big\{ \inn(P) : P \text{ short ear not attached to a horizontal 4-ear}\big\} \cupp  \\
&\big\{\{v\} : v\text{ a non-middle internal vertex of a vertical 4-ear}\big\}.
\end{align*}
Moreover, let $A$ denote the set of middle vertices of vertical 4-ears and for $a\in A$ let 
\[ M(a) \ := \ \bigl\{\{v\}\in M : \{v,a\}\in E(G) \bigr\}. \]
Note that $|M(a)| -2$ is the number of 2-ears attached to $a$.
Let 
\[ V_{\text{in}} \ := \ \bigl\{ v\in V(P) : P\text{ an inner ear} \bigr\}. \]
The sets $A$, $V_{\text{in}}$ and the elements of $M$ are pairwise disjoint. Moreover, the vertices 
of $G$ not contained in any of these sets are precisely the internal vertices of pendant or horizontal 4-ears and the internal
vertices of 2-ears attached to horizontal 4-ears.
See Figure \ref{fig:overview_outer_ears}.

Now let $\Pscr_f$ for $f\in M$ denote the set of paths in $G$ with the following two properties:
\begin{itemize}
 \item The set of internal vertices of the path is $f$. 
 \item The endpoints of the path are both contained in $V_{\text{in}}$.
\end{itemize}
We now define two matroids on the ground set $\bigcup_{f\in M} \Pscr_f$. 

The independent sets of the first matroid $\Mscr_1$ are given by all sets $\Iscr \subseteq \bigcup_{f\in M} \Pscr_f$
such that $\left( V(G) , \bigcup_{I\in \Iscr} E(I) \right)$ is a forest. 
It is clear that $\Mscr_1$ is a graphic matroid (every path can be represented by an edge connecting its endpoints).

In the second matroid $\Mscr_2$ a set $\Iscr \subseteq \bigcup_{f\in M} \Pscr_f$ is independent if and only if 
it fulfills the following two conditions:
\begin{enumerate}[(i)]
 \item $|\Iscr \cap \Pscr_f | \le 1$  for all $f\in M$ \\
 \item For every $a\in A$ we have 
 \[ \left| \bigcup_{f\in M(a)} \Pscr_f \cap \Iscr\ \right| \ \le \ |M(a)| - 2.\]
\end{enumerate}
Note that (i) and (ii) characterize the independent sets of a laminar matroid (see e.g.\ \cite{Fra11}).

We compute a maximum cardinality set $\Iscr \subseteq \bigcup_{f\in M} \Pscr_f$ such that $\Iscr$ is an 
independent set in both matroids defined above. 

The idea is that we want to choose as many short ears as possible to form a forest but leave two elements of each $M(a)$ unused;
they will form the 4-ear with middle vertex $a$. Thus the vertical 4-ears can also change.  
We will formally describe how the new ear-decomposition is formed in the proof of Theorem \ref{thm:raise_lb}.

\subsection{Improving the lower bound}

For $f\in M$ let $U_f \subseteq V_{\text{in}}$ be the set containing all neighbours of elements of $f$ in $V_{\text{in}}$. 
If $\Pscr_f \ne \emptyset$, this is the set of endpoints of paths in $\Pscr_f$. 
If $\Pscr_f =\emptyset$, then $U_f$ has a single element.
In particular, we have $U_f \ne \emptyset$ for all $f\in M$.

For a set $W \subseteq V_{\text{in}}$ let 
\[ \text{sur}(W) \ := \ |\{ f \in M : U_f \subseteq W \}| - (|W| -1) \]
be the surplus of $W$.
For a partition $\Wscr$ of $V_{\text{in}}$ and a vertex $a\in A$, let 
\[ \text{sur}(a, \Wscr) \ := \ 2 - \sum_{W\in \Wscr} \left|\{f \in M(a) :  U_{f} \subseteq W \}\right|. \]
Finally, for $A'\subseteq A$ let
\[ \mu(\Wscr, A') \ := \ \sum_{W\in \Wscr} \text{\rm sur}(W) + \sum_{a\in A'} \text{\rm sur}(a, \Wscr).\]

\begin{lemma}\label{lemma:matroid_intersection}
For a maximum cardinality set $\Iscr$ that is independent in both matroids, 
\[ |\Iscr| \ = \ |M| - \max\left\{ \mu(\Wscr, A') : A'\subseteq A,\ \Wscr\text{ partition of }V_{\textnormal{in}}\right\}. \]
\end{lemma}

\prove 
For any partition $\Wscr$ of $V_{\textnormal{in}}$ and any $A'\subseteq A$ we set
\begin{equation}
\label{eq:Qscr}
\Qscr_{\Wscr} \ := \ \left\{ P\in \bigcup_{f\in M} \Pscr_f : \exists W\in\Wscr \text{ such that both endpoints of $P$ belong to $W$} \right\}.
\end{equation}
and
\[M'_{\Wscr,A'} \ := \ \bigcup_{a\in A'} M(a) \cup \bigl\{ f\in M:  \Pscr_f \subseteq \Qscr_{\Wscr} \bigr\}. \]
Then, using the definitions of $\mu$ and sur:
% and \eqref{eq:Qscr}: 
\begin{equation}
\label{eq:mu}
\begin{aligned}
\mu(\Wscr, A') &\ = \
|\{ f\in M: \Pscr_f \subseteq \Qscr_{\Wscr} \}| - \sum_{W\in \Wscr} (|W|-1) + \sum_{a\in A'} \bigl( 2 - |\{ f\in M(a): \Pscr_f \subseteq \Qscr_{\Wscr}\}| \bigr) \\
&\ = \ |M'_{\Wscr,A'}| - \sum_{a\in A'} (|M(a)|-2) - \sum_{W\in \Wscr} (|W|-1).
\end{aligned}
\end{equation}
Moreover, every set $\Iscr$ that is independent in both matroids has at most $\sum_{W\in \Wscr}(|W|-1)$ elements of $\Qscr_{\Wscr}$ (due to $\Mscr_1$),
at most $|M(a)| -2$ elements of any $\bigcup_{f\in M(a)} \Pscr_f$ (due to $\Mscr_2$), and thus
$$|\Iscr| \ \le \  \sum_{W\in \Wscr} (|W|-1) + \sum_{a\in A'} (|M(a)| -2) + |M\setminus M'_{\Wscr,A'}|.$$ 
This shows ``$\le$''. 

Now let $r_1$ denote the rank function of the matroid $\Mscr_1$ and $r_2$ the rank function of the matroid $\Mscr_2$.
By the matroid intersection theorem (\cite{Edm70}),
\begin{equation}
\label{eq:matroidintersection}
\textstyle |\Iscr| \ = \ \min\left\{ r_1\bigl(\Qscr\bigr) + r_2\bigl(\bigcup_{f\in M} \Pscr_f \setminus \Qscr\bigr) 
   : \Qscr \subseteq \bigcup_{f\in M} \Pscr_f \right\}.
\end{equation}
Let $\Qscr$ be a set attaining the minimum, and among these a maximal one.
Let $\Wscr$ contain the intersections of $V_{\text{in}}$ with the vertex sets of the connected components of the graph
$G_{\Qscr} :=\left(V_{\text{in}}\cup \bigcup_{P\in \Qscr} V(P),\ \bigcup_{P\in \Qscr} E(P)\right)$.
Then $\Qscr=\Qscr_{\Wscr}$ by the maximality assumption.

The rank of $\Qscr$ in the graphic matroid $\Mscr_1$ is 
\begin{equation} \label{eq:rank_forest}
 r_1(\Qscr) \ = \ \sum_{W\in \Wscr} (|W|-1).
\end{equation}
Now let 
\[A'  \ := \ \bigl\{ a\in A: |\{f\in M(a): \Pscr_f \not\subseteq \Qscr\}| > |M(a)| -2 \bigr\}. \]

Then the rank of $\bigcup_{f\in M} \Pscr_f \setminus \Qscr$ in the laminar matroid $\Mscr_2$ is given by
\begin{equation} \label{eq:rank_laminar}
   r_2\bigl( {\textstyle \bigcup_{f\in M} \Pscr_f \setminus \Qscr } \bigr) 
  \ = \ |M\setminus M'_{\Wscr,A'}| +
\sum_{a\in A'} (|M(a)| -2).   
\end{equation}

\eqref{eq:matroidintersection}, \eqref{eq:rank_forest}, \eqref{eq:rank_laminar}, and \eqref{eq:mu} yield
$$|\Iscr| \ = \  \sum_{W\in \Wscr} (|W|-1) + |M\setminus M'_{\Wscr,A'}| + \sum_{a\in A'} (|M(a)| -2) \ = \ |M|-\mu(\Wscr,A').$$ 
\endproof

Denote by \lp\ the value of the following linear program:
 \begin{equation} \label{eq:subtour_lp}
 \begin{aligned}
 & \min x(E(G)) \\
 &s.t. & x(\delta(U)) &\geq 2 & & \text{for } \emptyset \subset U \subset V(G) \text{ with } |U\cap(\{s\}\triangle\{t\})| \text{ even},\\
 & & x(\delta(U)) &\geq 1  & & \text{for } U \subset V(G) \text{ with } |U\cap(\{s\}\triangle\{t\})| \text{ odd}, \\
 & & x_e &\geq 0 & & \text{for } e\in E(G).
\end{aligned}
\end{equation}
For every $s$-$t$-tour $F$, setting $x_e\in\{0,1,2\}$ to be the number of copies of $e$ in $F$ (for all $e\in E(G)$)
defines a feasible solution to \eqref{eq:subtour_lp}; hence $\lp$ is at most the number of edges in an optimum $s$-$t$-tour.

We now construct a dual solution to the LP \eqref{eq:subtour_lp}. The dual is given by 
 \begin{equation} \label{eq:dual_subtour_lp}
 \begin{aligned}
 & \max  \sum_{\emptyset \subset U \subset V(G)} 2y(U) - \sum_{ U : |U \cap(\{s\}\triangle\{t\})|\text{ odd}} y(U)  \hspace{-45mm}&\\[3mm]
 &s.t. & \sum_{U: e\in \delta(U)} y(U) &\le 1 & &\text{ for }e\in E(G) \\
 & & y(U) &\geq 0 & & \text{for } \emptyset \subset U \subset V(G).
\end{aligned}
\end{equation}

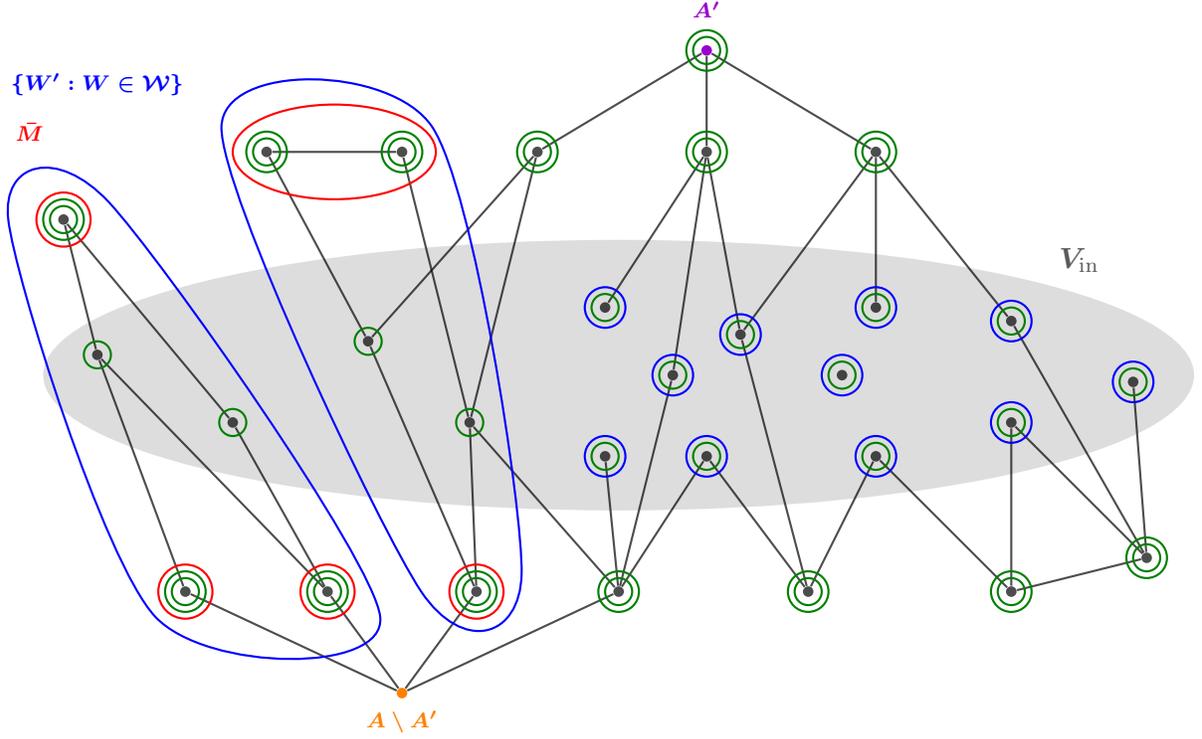
\begin{figure}
\begin{center}
 \begin{tikzpicture}[scale=0.9]
    \tikzstyle{vertex}=[circle,fill, minimum size=4,inner sep=0pt, opacity=0.7]
    \tikzstyle{pendant}=[circle,fill, minimum size=4,inner sep=0pt, opacity=0.7]
    \tikzstyle{edge}=[thick, black, opacity=0.7]
    \tikzstyle{dual}=[fill=none]
    
    \draw[fill=grey, draw=none, opacity=0.2] (0.2,0.7) ellipse ({8.5} and {2});
   
   %%%%% inner vertices %%%%%%%%%%%%%%%%%%%%%%%%%%%%%%%%%%%%%%%%%%%%%%%%%%%%%%%%%%%%%%%%%%%%%
   \node[vertex] (i1) at (-7.5, 1) {};
   \node[vertex] (i2) at (-5.5, 0) {};
   
   \node[vertex] (i3) at (-3.5, 1.2) {};
   \node[vertex] (i4) at (-2,0) {};
   
   \node[vertex] (i5) at (0,1.7) {};
   \node[vertex] (i6) at (2,1.3) {};
   \node[vertex] (i7) at (4,1.7) {};    
   \node[vertex] (i8) at (0,-0.5) {};
   \node[vertex] (i9) at (1.5,-0.5) {};  
   \node[vertex] (i10) at (6,1.5) {};
   
   \node[vertex] (i11) at (4,-0.5) {};
   \node[vertex] (i12) at (6,0) {};
   \node[vertex] (i13) at (7.8,0.6) {};   
   \node[vertex] (i14) at (3.5,0.7) {};
   \node[vertex] (i15) at (1,0.7) {};

    %%%%%%%%%% vertices and internal edges of outer ears %%%%%%%%%%%%%%%%%%%%%%%%%%%%%%%%%%%%%%
    % good 4-ear
    \node[vertex, fill=violet, opacity =1] (w1) at (1.5,5.5) {};
    \node[violet, above=3mm] (text) at (w1) {\scriptsize{\boldmath $ A'$}};
    \node[pendant] (u1) at (-1,4) {};
    \node[pendant] (u2) at (1.5,4) {};
    \node[pendant] (u3) at (4,4) {};  
    \draw[edge] (w1) --(u1);
    \draw[edge] (w1) --(u2);
    \draw[edge] (w1) --(u3);

    % bad 4-ear
    \node[vertex, fill=orange, opacity =1] (w2) at (-3,-4) {};
    \node[orange, below=1mm] (text) at (w2) {\scriptsize{\boldmath $A\setminus A'$}};
    \node[pendant] (v1) at (0.2,-2.5) {};
    \node[pendant] (v2) at (-1.9,-2.5) {};
    \node[pendant] (v3) at (-4.1,-2.5) {};
    \node[pendant] (v4) at (-6.2,-2.5) {};  
    \draw[edge] (w2) --(v1);
    \draw[edge] (w2) --(v2);
    \draw[edge] (w2) --(v3);
    \draw[edge] (w2) --(v4);
    
    % bad 3-ear
    \node[pendant] (b1) at (-3,4) {};
    \node[pendant] (b2) at (-5,4) {};
    \draw[edge] (b1) -- (b2);   
    % bad 2-ear 
    \node[pendant] (b) at (-8,3) {};  
    % good 3-ear
    \node[pendant] (g1) at (6,-2.5) {};
    \node[pendant] (g2) at (8,-2) {};
    \draw[edge] (g1) -- (g2);
    % good 2-ear 
    \node[pendant] (g) at (3,-2.5) {};
    
    %% edges to inner vertices %%%%%%%%%%%%%%%%%%%%%%%%%%%%%%%%%%%%%%%%%%%%%%%%%%%%%%%%%%
    \draw[edge] (i1) -- (b) --(i2);
    \draw[edge] (v4) -- (i1) -- (v3) -- (i2);
    
    \draw[edge] (i4) -- (u1) -- (i3) -- (b2);
    \draw[edge] (b1) -- (i4) -- (v2) -- (i3);
    \draw[edge] (i5) --(u2) -- (i6) -- (u3) -- (i7);
    \draw[edge] (u3) -- (i10) -- (g2);
    \draw[edge] (i8) -- (v1) -- (i9) -- (g);
    \draw[edge] (i13) -- (g2) -- (i12) -- (g1) -- (i11) -- (g);
    \draw[edge] (v1) -- (i15) -- (u2);
    \draw[edge] (g) -- (i6);
    \draw[edge] (v1) -- (i4);
    
    %%%%%%%%%%%%%%%%% dual solution %%%%%%%%%%%%%%%%%%%%%%%%%%%%%%%%%%%%%%%%%%%%%%%%%%%%%%%
    \node[grey] (dummy) at (7,2.4) {{\boldmath $V_{\text{in}}$}};
    \node[red] (text) at (-8.5,4.3) {\scriptsize{\boldmath $\bar{M}$}};
    \node[blue] (text) at (-7.5,5) {\scriptsize{\boldmath $\{W' : W \in \Wscr\}$}};
    
    % 2 circuits, outer
    \foreach \v in {v1, v2, v3, v4, u1, u2, u3, w1, b1, b2, b, g1, g2, g} {
       \draw[dual, darkgreen, thick] (\v) ellipse ({0.2} and {0.2});
       \draw[dual, darkgreen, thick] (\v) ellipse ({0.3} and {0.3});
    }   
    % 2 circuits, inner
    \foreach \v in {i5, i6, i7, i8, i9, i10, i11, i12, i13, i14, i15} {
       \draw[dual, darkgreen, thick] (\v) ellipse ({0.2} and {0.2});
       \draw[dual, blue, thick] (\v) ellipse ({0.3} and {0.3});
    }
    % 1 circuit
     \foreach \v in {i1, i2, i3, i4} {
       \draw[dual, darkgreen, thick] (\v) ellipse ({0.2} and {0.2});
    }
    
    % f\in \bar M
    \draw[red, thick] (-4,4) ellipse ({1.5} and {0.7});
     \foreach \v in {v2, v3, v4, b} {
       \draw[dual, red, thick] (\v) ellipse ({0.4} and {0.4});
    }
    
    % W' sets
    \node[right=4mm, above=2mm] (c1) at (b1) {};
    \node[left=6mm,above=2mm] (c2) at (b2) {};
    \node[left=6mm] (c3) at (v2) {};
    \node[right=6mm, above=1mm] (c4) at (v2) {};
    \draw [blue, thick] plot [smooth cycle] coordinates {(c1) (c2) (c3) (c4)};
    \node[right=7mm, below=2mm] (c5) at (v3) {};
    \node[left=4mm,below=2mm] (c6) at (v4) {};
    \node[left=6mm] (c7) at (b) {};
    \node[right=6mm, above=1mm] (c8) at (b) {};
    \draw [blue, thick] plot [smooth cycle] coordinates {(c5) (c6) (c7) (c8)};  
 \end{tikzpicture}
 \end{center}
 \caption{The dual solution $y'$ is shown in green, blue and red.
 Every red, blue or green line around a set $U$ indicates an (additional) value of $\sfrac{1}{4}$ of the corresponding dual
 variable $y'(U)$. Edges of the graph with both endpoints in $V_{\text{in}}$ are not shown. \label{fig:dual_solution}}
\end{figure}

Let $\Wscr$ be a partition of $V_{\text{in}}$ and $A' \subseteq A$ such that 
\begin{equation} \label{eq:lb_forest}
 |\Iscr| \ = \ |M| - \mu(\Wscr, A').
\end{equation}
(Such sets $\Wscr$ and $A'$ exist by Lemma \ref{lemma:matroid_intersection}.)
Let $\bar{M} \subseteq M$ be the set of all $f\in M$ that have the following two properties:
\begin{itemize}
 \item $U_f \subseteq W$ for some $W\in \Wscr$
 \item For every $a\in A'$ we have $f\not \in M(a)$.    
\end{itemize}

Let $V_{hor}$ be the union of the set of internal vertices of horizontal 4-ears and of 2-ears attached to horizontal 4-ears.
We first define a vector $y'$ that is a feasible solution to the dual LP \eqref{eq:dual_subtour_lp} if $V_{hor} = \emptyset$.
(This is a fact that we will prove later, in Lemma \ref{lemma:dual_solution_feasible}).
\begin{itemize}
 \item For every vertex $v\in V(G) \setminus V_{hor}$ we set 
 \begin{equation}
  y'(\{v\}) \ := \ \begin{cases}
               0, & \text{if } v\in A\setminus A' \\
               \sfrac{1}{4}, &\text{ if } v\in V_{\text{in}} \\
               \sfrac{1}{2}, &\text{ else}.
              \end{cases}
 \end{equation}
(green in Figure \ref{fig:dual_solution})
 \item  For every set $W\in \Wscr$ we define 
\[ W' \ := \ W \cup \{ v\in f :  f \in \bar{M}, U_f \subseteq W \} \]
and set  $y'(W') := \frac{1}{4}$. If $|W'| =1$, we instead increase $y'(W')$ by $\frac{1}{4}$ (from $\frac{1}{4}$ to $\frac{1}{2}$).
(blue in Figure \ref{fig:dual_solution})
\item We set $y'(f) := \frac{1}{4}$ for $f\in \bar M$. If $|f|=1$, we instead increase $y'(f)$ by $\frac{1}{4}$ (from $\frac{1}{2}$ to $\frac{3}{4}$). 
(red in Figure \ref{fig:dual_solution})
\end{itemize}
All other dual variables $y'(U)$ (for $\emptyset \subset U\subset V(G)$) are set to zero.

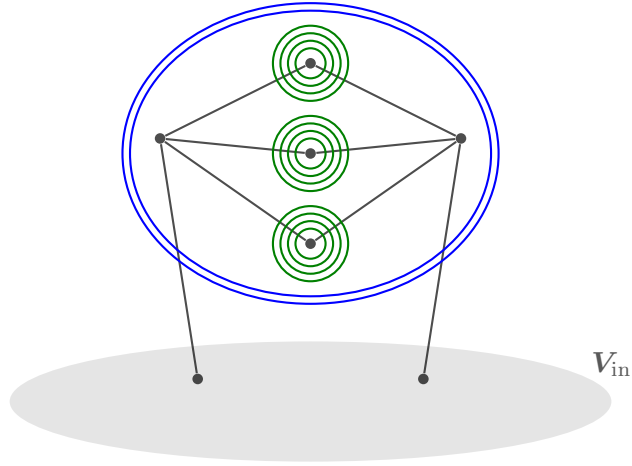
\begin{figure}
\begin{center}
 \begin{tikzpicture}
    \tikzstyle{vertex}=[circle,fill,minimum size=4,inner sep=0pt, opacity=0.7]
    \tikzstyle{pendant}=[circle,fill,minimum size=4,inner sep=0pt, opacity=0.7]
    \tikzstyle{edge}=[thick, black, opacity=0.7]
    \tikzstyle{dual}=[fill=none]
    
    \draw[fill=gray, draw=none, opacity=0.2] (0,0.7) ellipse ({4} and {0.8});
    \node[grey] (dummy) at (4,1.2) {{\boldmath $V_{\text{in}}$}};
      
     \node[vertex] (v) at (-2,4.2) {};
    \node[vertex] (w) at (2,4.2) {};
    \node[pendant] (u1) at (0,5.2) {};
    \node[pendant] (u2) at (0,4) {};
    \node[pendant] (u3) at (0,2.8) {};  
     \node[pendant] (iv) at (-1.5,1) {};
    \node[pendant] (iw) at (1.5,1) {}; 
    \draw[edge] (iv) -- (v) -- (u1) -- (w) -- (u2) -- (v) -- (u3) -- (w) -- (iw);
    
    % dual solution
    \draw[fill=none, draw=blue, thick] (0,4) ellipse ({2.4} and {1.9});
    \draw[fill=none, draw=blue, thick] (0,4) ellipse ({2.5} and {2});
    % 4 circuits
    \foreach \v in {u1, u2, u3} {
       \draw[dual, darkgreen, thick] (\v) ellipse ({0.2} and {0.2});
       \draw[dual, darkgreen, thick] (\v) ellipse ({0.3} and {0.3});
       \draw[dual, darkgreen, thick] (\v) ellipse ({0.4} and {0.4});
       \draw[dual, darkgreen, thick] (\v) ellipse ({0.5} and {0.5});
    }
 \end{tikzpicture}
 \end{center}
 \caption{The dual solution $y_{hor}$ for a horizontal 4-ear with one 2-ear attached to it.
 Every blue or green line around a set $U$ indicates an (additional) value of $\sfrac{1}{4}$ of the corresponding dual
 variable $y_{hor}(U)$.  \label{fig:dual_solution_horizontal}}
\end{figure}

To construct a feasible dual solution also if $V_{hor} \ne \emptyset$, we define a vector $y_{hor}$ and set 
$y := y' + y_{hor}$.
To define $y_{hor}$ we define for every horizontal 4-ear dual variables as follows:
Let $w$ be the middle vertex of $P$ and let $Q_1$, \dots, $Q_h$ be the 2-ears attached to $P$. 
Then we set 
\begin{align*}
  y_{hor}(\{w\}) & \ := \ 1 \\
  y_{hor}(\inn(Q_i)) & \ := \ 1 \hspace{5mm} (\text{for } i=1,\dots, h) \\
  y_{hor}(V(P) \cup V(Q_1) \cup \dots \cup V(Q_h)) & \ := \ \sfrac{1}{2}. 
\end{align*} 
See Figure \ref{fig:dual_solution_horizontal}.
All other dual variables $y_{hor}(U)$ (for $\emptyset \subset U\subset V(G)$) are set to zero.

\begin{lemma}\label{lemma:dual_solution_feasible}
 The vector $y = y' + y_{hor}$ is a feasible solution to the dual LP \eqref{eq:dual_subtour_lp} of 
 the LP \eqref{eq:subtour_lp}.
\end{lemma}
\prove
We clearly have $y(U) \ge 0$ for $ \emptyset \subset U \subset V(G)$. Moreover, we have for every vertex $v$
that is contained neither in $V_{hor}$ nor in $f\in \bar M$, that
\begin{equation}\label{eq:sum_single_vertex}
  \sum_{U: v\in U} y(U) \ \le \ \sfrac{1}{2}.
\end{equation}
Now let $e$ be an edge of $G$. 
\\[2mm]
\textbf{Case 1:} At least one of the endpoints of $e$ is contained in $V_{hor}$. \\[2mm]
If both endpoints of $e$ are contained in $V_{hor}$, one endpoint $v$ of $e$ is a non-middle internal 
vertex of a horizontal 4-ear $P$, and the other endpoint $w$ of $e$ is the middle vertex of either $P$ or a 2-ear attached to $P$
at $v$. (This follows from Theorem \ref{earmain} and the definitions of a horizontal 4-ear and $V_{hor}$.)
Then, $\sum_{U: e\in \delta(U)} y(U) = y(\{w\})= 1$.
It remains to consider the case $e\in \delta(V_{hor})$ and without loss of generality $w\in V_{hor}$.
By Theorem \ref{earmain}, this implies that $v\in V_{\text{in}}$. Moreover, by definition of horizontal 4-ears,
$w$ is a non-middle internal vertex of a horizontal 4-ear. (Otherwise, the edge $\{v,w\}$ could not exist.)
Using \eqref{eq:sum_single_vertex}, this implies
\[\sum_{U: e\in \delta(U)} y(U) \ \le \ \sum_{U: w \in U} y_{hor}(U) +  \sum_{U: v \in U} y(U) \ \le \ \sfrac{1}{2} + \sfrac{1}{2} \ = \ 1. \]
\\[2mm]
\textbf{Case 2:} None of the endpoints of $e$ is contained in $V_{hor}$. \\[2mm]
If none of the two endpoints of $e$ is contained in $\bigcup_{f\in \bar M} f$, we have by \eqref{eq:sum_single_vertex}, that 
$\sum_{U: e\in \delta(U)} y(U) \le 1$. So we may assume that $e=\{v, w\}$ with $w\in f \in \bar M$.
Then $\sum_{U: v \in U} y(U) = 1.$ 
By Theorem \ref{earmain}, either 
\begin{itemize}
 \item $v\in A$ and ${w} = f \in M(v)$, or
 \item $v\in V_{\text{in}}$, or
 \item $v\in f$ and $f=\{v,w\}$ is the set of internal vertices of a 3-ear.
\end{itemize}
If $v\in A$, by definition of $\bar M$, we have $v\in A\setminus A'$, since $f \in \bar M$.
Then $\sum_{U: v\in U} y(U) = 0$ and thus
\[ \sum_{U: e\in \delta(U)} y(U) \ \le \ \sum_{U: w \in U} y(U) \ = \ 1. \]
As $w\in f \in \bar M$, there exists a (unique) set $W\in \Wscr$ with $U_f \subseteq W$ and $w\in W'$. 
If $v\in V_{\text{in}}$, we have $v\in U_f \subseteq W \subseteq W'$. Hence,
\[ \sum_{U: e\in \delta(U)} y(U) \ \le \ \sum_{U: w \in U} y(U) +  \sum_{U: v \in U} y(U) - 2 y(W') \ \le \ 1 + \sfrac{1}{2} - \sfrac{1}{2} \ = \ 1. \]
Finally, if $v\in f$ and $f=\{v,w\}$ is the set of internal vertices of a 3-ear, we have 
\[  \sum_{U: e\in \delta(U)} y(U) \ = \ y(\{v\}) + y(\{w\}) \ = \ \sfrac{1}{2} + \sfrac{1}{2} \ = \ 1. \]
\endproof

\begin{theorem}\label{thm:lower_bound} 
If $\Iscr$ is a maximum cardinality set that is independent in both matroids, then
$$  \lp \ \ge \ n - 3  + \sfrac{1}{2} (k_2 + k_3 -|\Iscr|). $$
\end{theorem}

\prove
By Lemma \ref{lemma:dual_solution_feasible}, the vector $y =y' + y_{hor}$ is a feasible solution to the dual of \eqref{eq:subtour_lp}. 
Hence,
\[\lp \ \ge \ \sum_{\emptyset \subset U \subset V(G)} 2y(U) -  \sum_{ U : |U \cap(\{s\}\triangle\{t\})|\text{ odd}} y(U).\]
%We will first bound $ \sum_{\emptyset \ne U \subsetneq V} 2y'(U) $ and $  \sum_{\emptyset \ne U \subsetneq V} 2y_{hor}(U)$ separately.

Using the definitions of $\text{sur}$ and $\bar M$, we get
\begin{align*}
\sum_{a\in A'}\text{sur}(a, \Wscr) &\ = \ \sum_{a\in A'} \big( 2- |\{ f\in M(a) : U_f \subseteq W, W \in \Wscr \}| \big) \\
&\ = \ 2 |A'| - |\{ f\in M\setminus \bar M : U_f \subseteq W, W \in \Wscr \}|  
\end{align*}
and
\begin{align*}
\sum_{W\in \Wscr}\text{sur}(W) &\ = \ \sum_{W\in \Wscr} \big( |\{ f\in M : U_f \subseteq W\}| - (|W|-1) \big) \\
&\ = \ |\bar M| + |\{ f\in M\setminus \bar M : U_f \subseteq W, W \in \Wscr \}|  - | V_{\text{in}} | + |\Wscr|,
\end{align*}
which together with \eqref{eq:lb_forest} implies
\begin{equation}\label{eq:half_sur}
 \sfrac{1}{2} (|M|-\Iscr|) \ = \
 \sfrac{1}{2} \mu (\Wscr, A') \ = \ |A'| + \sfrac{1}{2} |\bar M| - \sfrac{1}{2} | V_{\text{in}} | + \sfrac{1}{2} |\Wscr|.
\end{equation}
By construction of $y'$, we have 
\begin{equation}\label{eq:total_value_y'}
\begin{aligned}
 \sum_{\emptyset \subset U \subset V(G)} 2y'(U) &\ = \ |V(G) \setminus V_{hor}| - |A\setminus A'| - \sfrac{1}{2} | V_{\text{in}}| 
  + \sfrac{1}{2} |\Wscr| + \sfrac{1}{2} |\bar M|\\
&\ = \ |V(G) \setminus V_{hor}| - |A| + \sfrac{1}{2} (|M|-|\Iscr|),
\end{aligned}
\end{equation}
where we used \eqref{eq:half_sur} in the second equality.

Furthermore, for a horizontal 4-ear $P$ with middle vertex $w$ and
2-ears $Q_1$, \dots, $Q_h$ attached to $P$, we have $|V(P) \cup V(Q_1) \cup \dots \cup V(Q_h)| = 3+h$ and thus
\begin{align*}
 & y_{hor}(\{w\}) + \sum_{i=1}^h y_{hor}(\inn(Q_i)) + y_{hor}(V(P) \cup V(Q_1) \cup \dots \cup V(Q_h)) \\
=\ & 1 + h + \sfrac{1}{2} \\
=\ & \sfrac{1}{2} |V(P) \cup V(Q_1) \cup \dots \cup V(Q_h)| + \sfrac{1}{2} h.
\end{align*}
This implies
\begin{equation} \label{eq:total_value_y_hor}
  \sum_{\emptyset \subset U \subset V(G)} 2 y_{hor}(U) \ = \ |V_{hor}| + |\{ Q : Q\text{ 2-ear attached to a horizontal 4-ear}\}|.
\end{equation}

Combining \eqref{eq:total_value_y'} and \eqref{eq:total_value_y_hor}, we get 
\[  \sum_{\emptyset \subset U \subset V(G)} 2y(U) \ = \
    n - |A| + \sfrac{1}{2} (|M|-|\Iscr|) + |\{ Q : Q\text{ 2-ear attached to a horizontal 4-ear}\}|. \]

Finally, we observe that we defined the dual solution $y$ such that for every vertex $v$ we have
$\sum_{U: v\in U} y(U) \le \frac{3}{2}$. In particular,
\[ \sum_{ U : |U \cap(\{s\}\triangle\{t\})|\text{ odd}} y(U) \ \le \ \sum_{U: s\in U} y(U) + \sum_{U: t\in U} y(U) \ \le \ 3. \]
This shows 
\begin{align*}
  \lp &\ \ge \ \sum_{\emptyset \subset U \subset V(G)} 2y(U) - \sum_{ U : |U \cap(\{s\}\triangle\{t\})|\text{ odd}} y(U)  \\[1mm]
  &\ \ge \ n - |A| + \sfrac{1}{2} (|M|-|\Iscr|) + |\{ Q : Q\text{ 2-ear attached to a horizontal 4-ear}\}|  - 3 \\[1mm]
  &\ = \ n-3 -|A|+\sfrac{1}{2} (2 |A| + k_2+k_3 - |\Iscr|)  
     + \sfrac{1}{2}|\{ Q : Q\text{ 2-ear attached to a horizontal 4-ear}\}| \\[1mm]
&\ \ge \ n-3 + \sfrac{1}{2} (k_2+k_3 - |\Iscr|)
\end{align*}
\endproof

\subsection{Optimizing outer ears}

After optimizing the outer ears via the matroid intersection approach described above, 
we will distinguish between primary ears (those that were inner ears, plus clean ears that form a forest)
and secondary ears. For secondary ears we will apply Lemma \ref{lemma:standard_ear_induction}, and for secondary short ears we raise the lower bound
using Theorem \ref{thm:lower_bound}. 
For primary ears we will apply Theorem \ref{thm:result_ear_induction}. Hence we need to bound the number of primary 4-ears,
which correspond exactly to the blocked 4-ears before optimization. Their number can be bounded easily as follows.

\begin{lemma}
\label{lemma:numberofblockedears}
Given an ear-decomposition as in Theorem \ref{earmain}, 
denote by $k_{4,\,\text{blocked}}$ and $k_{4,\,\text{non-blocked}}$ the number of blocked and non-blocked 4-ears, respectively. Then
$$k_{4,\,\text{blocked}} \ \le \ 2 k_{\ge 5} + k_{4,\,\text{non-blocked}}.$$
\end{lemma}

\prove
Recall that a 4-ear is blocked if a closed ear is attached to it. 
Since all short ears are open and no closed 4-ear is 
attached to any closed 4-ear, the only closed ears attached to a 4-ear can be ears of length at least five,
non-blocked 4-ears, and 4-ears to which a closed ear of length at least five is attached.
Using that every closed ear is attached to exactly one ear,
we get the result.
\endproof

We now describe in detail how we optimize the outer ears and summarize the results of this section in the following theorem.

\begin{theorem}\label{thm:raise_lb}
Given a 2-vertex-connected graph $G$ and $s,t\in V(G)$,
we can compute a well-oriented ear-decomposition $P_1,\ldots,P_l$ of $G$ and an index $p\le l$ with the following properties in polynomial time.
Call $P_1,\ldots,P_p$ the primary ears and $P_{p+1},\ldots,P_l$ the secondary ears. Then:
 \begin{itemize}
  \item The primary short ears are clean; an ear is oriented if and only if it is short and primary.
  \item   $k_{4,\,\textnormal{primary}} - 2 k_{\ge 5,\,\textnormal{primary}} \le k_{4,\,\textnormal{secondary}}$
  \item $\lp \ge n-3 + \frac{1}{2} k_{\textnormal{clean,\,secondary}}$
 \end{itemize}
where 
$k_{4,\,\textnormal{primary}}$ is the number of primary 4-ears,
$k_{4,\,\textnormal{secondary}}$ is the number of secondary 4-ears,
$k_{\ge 5,\,\textnormal{primary}}$ is the number of primary ears with at least five edges,
and $k_{\textnormal{clean,\,secondary}}$ is the number of secondary clean ears.
\end{theorem}

\prove
We first compute an ear-decomposition as in Theorem \ref{earmain}.
Then we compute a maximum independent set $\Iscr$ in both matroids $\Mscr_1$ and $\Mscr_2$ by a matroid intersection algorithm.
Then we modify the outer ears of our ear-decomposition as follows:
\begin{itemize}
 \item For every short ear $Q$ such that no internal vertex of $Q$ is adjacent to an internal vertex of an outer 4-ear and 
$\Iscr \cap \Pscr_{\inn(Q)} \ne \emptyset$, we replace $Q$ by the unique element of $\Iscr \cap \Pscr_{\inn(Q)}$.
\item Let $a\in A$ and  let $\{u\}, \{u'\}$ be two distinct elements of $M(a)$
such that both $\Iscr \cap \Pscr_{\{u\}}$ and  $\Iscr \cap \Pscr_{\{u'\}}$ are empty.
Then we replace the vertical 4-ear with middle vertex $a$ and the 2-ears attached to it as follows.
We choose a 4-ear with internal vertices $u,a,u'$ and with endpoints in $V_{\text{in}}$.
For every $\{v\} \in M(a) \setminus \{\{u\}, \{u'\}\}$ we choose a 2-ear with internal vertex $v$: 
if $\Iscr \cap \Pscr_{\{v\}} \ne \emptyset$, we choose the 2-ear to be the unique element 
of $\Iscr \cap \Pscr_{\{v\}}$;
otherwise, we choose the 2-ear to consist of the edge $\{a,v\}$ and an arbitrary
edge from $v$ to a vertex in $V_{\text{in}}$.
\end{itemize}
This modification of the ear-decomposition does not change any inner ear.
It does not change the total number of 4-ears.
%Note that each vertical 4-ear is replaced by a vertical or pendant 4-ear.
Moreover, all short ears are still pendant, and the (short) ears in $\Iscr$ form a forest.
We orient the clean ears in $\Iscr$ so that we have a well-oriented ear-decomposition.

We declare an ear as primary if all its vertices belong to $V_{\text{in}}$ or if it is a clean ear in $\Iscr$.
Other ears are secondary; their internal vertices do not belong to $V_{\text{in}}$. 
Since $\Iscr$ contains only paths with both endpoints in $V_{\text{in}}$, 
we can reorder the ears so that the first ears $P_1,\ldots,P_p$ are the primary ears (for some $p$).

Note that all ears of length at least five are primary, and the primary 4-ears
are exactly those that were blocked before the optimization.
Hence $k_{4,\text{primary}} - 2 k_{\ge 5,\text{primary}} \le k_{4,\text{secondary}}$ 
follows from Lemma \ref{lemma:numberofblockedears}.

%the only 4-ears to which an ear in $\Iscr$ is attached
%are 4-ears that were blocked in the initial ear-decomposition.
%To all other 4-ears only short ears not contained in $\Iscr$ are attached.
Finally, we have by Theorem \ref{thm:lower_bound} that 
\begin{align*}
 \lp &\ \ge \ n-3 + \sfrac{1}{2} (k_2+k_3 - |\Iscr|) \\[2mm]
&\ \ge \ n-3 + \sfrac{1}{2} k_{\textnormal{clean,\,secondary}},
\end{align*}
where the last inequality follows since clean secondary ears do not belong to $\Iscr$.
\endproof

\section{Ear-decompositions with many non-entered ears}\label{section:many_pendant}

Now we apply our ear induction to the optimized ear-decomposition, combining Theorems
\ref{thm:raise_lb} and \ref{thm:result_ear_induction}.

\begin{theorem}\label{thm:many_pendant}
 Given a graph $G$ and $s,t \in V(G)$, we can compute a well-oriented ear-decomposition of $G$ with $\pi$ non-entered ears,
 and an $s$-$t$-tour with at most 
 \[ \left(\sfrac{3}{2}  - \sfrac{1}{26}\cdot \sfrac{\pi}{n-1}\right)\lp + 3 \]
 edges in polynomial time.
\end{theorem}

\prove
We compute a well-oriented ear-decomposition of a graph $G$ as in Theorem \ref{thm:raise_lb}.
Let  $\pi$ be the number of non-entered ears and $V_{\text{primary}}$ the union of the vertex sets of all primary ears.
We apply Lemma \ref{lemma:standard_ear_induction} to all nontrivial secondary ears (in reverse order).
This yields a set $F'\subseteq \bigcup_{P \text{ secondary}}2E(P)$ which is a $T'$-join 
for some $T'$ with $T'\triangle\{s\}\triangle\{t\}\subseteq V_{\text{primary}}$
such that $(V(G),F')/V_{\text{primary}}$ is connected and
$$|F'| \ \le \ \sfrac{3}{2} \bigl( n-|V_{\text{primary}}| \bigr) - \sfrac{1}{2} k_{\text{nontrivial,\,secondary}} + k_{\text{clean,\,secondary}}.$$
%because compared to $\frac{3}{2} |\inn(P)|$ 
%we lose $\frac{1}{2}$ for every clean secondary ear and we gain $\frac{1}{2}$ for every other nonpendant secondary ear. 

Now let $T:=T'\triangle\{s\}\triangle\{t\}$, and we can apply 
Theorem \ref{thm:result_ear_induction} to the primary ears.
We get a $T$-join $F$ such that $(V_{\text{primary}}, F)$ is connected and
\begin{align*}
|F| &\ \le \ \sfrac{3}{2} \bigl( |V_{\text{primary}}| - 1 \bigr) 
- \sfrac{1}{26} k_{\text{non-entered,\,primary}} 
 + \sfrac{1}{26} (k_{4,\,\text{primary}} -2 k_{\ge 5,\,\text{primary}}) \\
& \ \le \  \sfrac{3}{2} \bigl( |V_{\text{primary}}| - 1 \bigr) 
- \sfrac{1}{26} k_{\text{non-entered,\,primary}} 
 + \sfrac{1}{26} k_{4,\,\text{secondary}}. 
\end{align*}
% Note that we gain $\frac{1}{26}$ for every primary ear to which only secondary ears are attached, even if it is not pendant,
% but we do not use this in the above bound.

Then $F'\cupp F$ is an $s$-$t$-tour in $G$ with at most 
\begin{align*}
 & \sfrac{3}{2} (n - 1) 
  - \sfrac{1}{2} k_{\text{nontrivial,\,secondary}} + k_{\text{clean,\,secondary}} 
  - \sfrac{1}{26} k_{\text{non-entered,\,primary}} + \sfrac{1}{26} k_{4,\,\text{secondary}} \\[2mm]
  \le\ & \sfrac{3}{2} (n-1)  - \sfrac{1}{26} \bigl( k_{\text{nontrivial,\,secondary}} + k_{\text{non-entered,\,primary}} \bigr)
  + \left(\sfrac{1}{2} + \sfrac{1}{26}\right) k_{\text{clean,\,secondary}}  \\[2mm]
  =\ & \sfrac{3}{2} (n-1)  -\sfrac{1}{26}\pi + \left(\sfrac{1}{2}+\sfrac{1}{26}\right)k_{\text{clean,\,secondary}} 
\end{align*}
edges.

By Theorem \ref{thm:raise_lb} we have $\lp \ge (n-1) + \sfrac{1}{2} k_{\text{clean,\,secondary}} - 2$.
Thus, we have (using $\pi\le n-1$)
\begin{align*}
 \left(\sfrac{3}{2}  - \sfrac{1}{26}\cdot \sfrac{\pi}{n-1}\right)\lp
 &\ \ge\ \sfrac{3}{2} (n-1)  -\sfrac{1}{26}\pi 
 + \left(\sfrac{3}{2} - \sfrac{1}{26} \sfrac{\pi}{n-1} \right) \cdot \left(\sfrac{1}{2} k_{\text{clean,\,secondary}} - 2 \right)\\
&\ \ge \ \sfrac{3}{2} (n-1)  -\sfrac{1}{26}\pi + (\sfrac{3}{4} - \sfrac{1}{52})k_{\text{clean,\,secondary}}  - 3.
\end{align*}
Hence we can bound the number of edges of our $s$-$t$-tour $F'\cupp F$ by
\[ \left(\sfrac{3}{2}  - \sfrac{1}{26}\cdot \sfrac{\pi}{n-1}\right)\lp + 3. \]
\endproof

\section{Ear-decompositions with few non-entered ears}\label{section:few_pendant}

In this section we show how to compute a cheap $s$-$t$-tour if our well-oriented ear-decomposition has only few non-entered ears and 
$s$ and $t$ have small distance.
The following lemma shows a new way to apply the removable-pairing technique of \cite{MomS16} to ear-decompositions.
In contrast to \cite{SebV12}, we do not require the graph after deleting the trivial ears to be 2-vertex-connected.
This requires a slight modification of their proof.

\begin{lemma}\label{lemma:MS_few_ears}
 Let $G$ be a graph with a well-oriented ear-decomposition, and $s,t\in V(G)$.
 Let $\pi$ be the number of non-entered ears.
 Then we can compute an $s$-$t$-tour with at most
 \[ \sfrac{4}{3}(n-1)+\sfrac{2}{3}\pi + \sfrac{1}{3}\textnormal{dist}(s,t) \]
 edges in polynomial time, where $\textnormal{dist}(s,t)$ denotes the distance of $s$ and $t$ in $G$.
\end{lemma}

\prove
Let $E'$ be the set of edges of nontrivial ears. Note that $(V(G),E')$ is 2-edge-connected, but in general not 2-vertex-connected.

For every non-entered ear choose one arbitrary edge of the ear and declare it removable.
For every entered ear $Q$ let $v\in\inn(Q)$ such that there is an oriented ear entering $Q$ at $v$;
declare the two edges of $Q$ that are incident to $v$ a removable pair.

Let $R$ be the set of all removable edges (including those in removable pairs),
and let $\Pscr$ be the set of removable pairs.
We have $|E'|-(n-1)$ nontrivial ears and 
\begin{equation}\label{eq:num_removable_edges}
 |R|=2(|E'|-(n-1))-\pi.
\end{equation}

Note that $(V(G),E'\setminus R')$ is connected for every subset $R'\subseteq R$ that contains at most one element of each removable pair.
(\cite{MomS16} called $(R,\Pscr)$ a removable pairing.)

We construct a 2-edge-connected auxiliary graph $G'$ with edge weights $c$ as follows.
We begin with $(V(G),E')$, and initially all removable edges have weight $-1$ and all other edges have weight 1.
Now consider the nontrivial ears in reverse order. Let $P_i$ be the current ear.
If $P_i$ is non-entered, we do nothing.
Otherwise there is a removable pair with edges $\{u,v\}$ and $\{v,w\}$ of $P_i$ (then $v\in\inn(P_i)$).
We insert a new vertex $v'$ and a new edge $\{v,v'\}$ with weight 0,
and replace the edges $\{u,v\}$ and $\{v,w\}$ by $\{u,v'\}$ and $\{v',w\}$, both with weight $-1$.
Note that the subgraph of oriented ears never changes.

To show that the graph remains 2-edge-connected we have to prove the new edge $\{v,v'\}$ is not a bridge
(because contracting this edge results in the previous graph, which was 2-edge-connected by induction).

To prove that the new edge $\{v,v'\}$ is not a bridge, we construct paths from $v$ to $r$ and from $v'$ to $r$, both not using this edge;
here $r$ denotes the initial vertex of the ear-decomposition (the vertex of $P_0$).
From $v'$ we follow the edge $\{v',w\}$ and then edges of $P_i$ to an endpoint $x$ of $P_i$. 
Since $x\in V_{i-1}$, there is a path from $x$ to $r$ in $G_{i-1}$ (which is a subgraph that we haven't changed yet).
From $v$ we follow the entering oriented ear (backwards) until we reach the root $r(v)$ of the connected component of oriented ears.
This root belongs to $V_i\setminus\{v\}$. If $r(v)\in\inn(P_i)\setminus\{v\}$, we follow edges of $P_i$ to an endpoint of $P_i$ without visiting $v$.
Then we reach a vertex in $V_{i-1}$ and again have path to $r$ within $G_{i-1}$ from there.

The result is a 2-edge-connected weighted graph $G'$, for which the contraction of the zero-weight edges
would result in $(V(G),E')$. The removable edges have weight $-1$, other edges have weight 1.
Finally, if $s\not=t$, we add a new edge $d$ with endpoints $s$ and $t$ and weight $\text{dist}(s,t)$ to $G'$ (possibly adding a parallel edge).
We call the result $G''$. Let $T'':=\{v\in V(G''): |\delta_{G''}(v)| \text{ odd}\}$.

Consider the vector $x\in\mathbb{R}^{E(G'')}$  whose entries are all $\frac{1}{3}$.
We claim that $x$ is in the $T''$-join polytope  
\begin{equation}
\label{eq:Tjoinpolytope}
\begin{aligned}
\Bigl\{ x\in [0,1]^{E(G'')} : |F|-x(F)+ x(\delta_{G''}(U)\setminus F) \ge 1 \ & \text{for all } U\subset V(G'') \text{ and } F\subseteq \delta_{G''}(U) \\
& \text{with } |U\cap T''|+|F| \text{ odd} \Bigr\},
\end{aligned}
\end{equation}
which is the convex hull of incidence vectors of simple $T''$-joins (by simple we mean that no edge is used twice); cf.\ (29.11) in \cite{Sch03}.
To see that $x$ is in \eqref{eq:Tjoinpolytope}, let $U\subset V(G'')$ and $F\subseteq \delta_{G''}(U)$ with $|U\cap T''|+|F|$ odd.
If $|\delta_{G''}(U)| \ge 3$, the inequality holds because every edge of $\delta_{G''}(U)$
contributes at least $\frac{1}{3}$ to the left-hand side. 
Otherwise $|\delta_{G''}(U)|=2$ because $G''$ is 2-edge-connected. Hence $|U\cap T''|$ is even by definition of $T''$.
Then $|F|$ is odd, so $|F|=1$.
Then $|F|-x(F)+ x(\delta_{G''}(U)\setminus F) = 1 - \frac{1}{3} + \frac{1}{3} = 1$.

We conclude that $x$ is in \eqref{eq:Tjoinpolytope}, and in fact in the face of this integral polytope
defined by $x(\delta(v'))=1$ for every new vertex $v'$ (they have degree three).
Faces of integral polytopes are integral, so $x$ is
in the convex hull of simple $T''$-joins that contain exactly one edge incident to each new vertex.

Hence there exists a simple $T''$-join $J''$ in $G''$ that contains exactly one edge incident to each new vertex, and with 
$$c(J'') \ \le \ c(x) \ = \ \sfrac{1}{3}|E'| -\sfrac{2}{3}|R| + \sfrac{1}{3} \text{dist}(s,t).$$
Such a $J''$ can be computed in polynomial time (using a standard reduction to weighted matching; see \cite{SebV12}).

Let $D$ be the edge set of a shortest $s$-$t$-path if $d\in J''$, and $D=\emptyset$ otherwise.
Let $T:=\{v\in V(G): |\delta_{E'}(v)| \text{ odd}\}\triangle\{s\}\triangle\{t\}$.
After contracting the zero-weight edges of $G''$,
$J''$ corresponds to a simple $T$-join $J$ in $(V(G),E'\cupp\{d\})$ with
$$|J\cap (E'\setminus R)| - |J\cap R| + |D| \ = \ c(J'') \ \le \ \sfrac{1}{3}|E'| -\sfrac{2}{3}|R| + \sfrac{1}{3} \text{dist}(s,t),$$
not containing both edges of any removable pair.
Then $(E'\setminus (J\cap R)) \cupp (J\cap (E'\setminus R)) \cupp D$ is an $s$-$t$-tour with at most
$|E'| + c(J'')$ edges. The result now follows from \eqref{eq:num_removable_edges}.
\endproof

\begin{theorem}\label{thm:approx_guarantee}
Given an instance of the $s$-$t$-path graph TSP with a 2-vertex-connected graph in which $s$ and $t$ have distance at most $0.3334\,\opt$,
where $\opt$ denotes the number of edges in the optimum solution,
we can compute an $s$-$t$-tour with at most $1.497\,\opt$ edges in polynomial time.
\end{theorem}

\prove
We apply Theorem \ref{thm:many_pendant} and obtain a well-oriented ear-decomposition and an $s$-$t$-tour.
If the number of non-entered ears is at least $\frac{13}{165}(n-1)$, then this $s$-$t$-tour has
at most $(\frac{3}{2}-\frac{1}{330})\lp + 3$ edges.
If $n> 99000$,  this is at most $1.497\,\lp \le 1.497\,\opt$ since $\lp\ge n-1$.
For $n \le 99000$ we can solve the instance by complete enumeration.

If the number of non-entered ears is at most $\frac{13}{165}(n-1)$, then  Lemma \ref{lemma:MS_few_ears}
yields an $s$-$t$-tour with at most $(\frac{4}{3}+\frac{26}{495})(n-1) + \frac{0.3334}{3}\, \opt<1.497\,\opt$ edges.
\endproof

\section{Solving instances with large distance of $s$ and $t$}\label{section:large_distance}

We show that for finding a polynomial time $\alpha$-approximation for the $s$-$t$-path graph TSP
for some constant $\alpha > 1$, it is sufficient to consider the special case 
where the distance of the vertices $s$ and $t$ is at most $\frac{1}{3}+ \delta$ times the length of an 
optimum solution for some arbitrary constant $\delta >0$. 

\begin{theorem}\label{thm:long_s_t_dist_easy}
  Let $\delta >0$ and $\alpha>1$ be constants. Assume there exists a polynomial time algorithm that computes 
   a solution with at most $\alpha\cdot\opt$ edges
  for instances of the 
  $s$-$t$-path graph TSP with a 2-vertex-connected graph $G$ and $d=\text{dist}(s,t) \le (\frac{1}{3}+\delta)\cdot\opt$. 
  Then there exists a polynomial time $\alpha$-approximation algorithm for the $s$-$t$-path graph TSP.
 \end{theorem}
 
Our proof of Theorem \ref{thm:long_s_t_dist_easy} also applies to weighted graphs.
We consider a graph $G$ with edge weights $c: E(G) \rightarrow \mathbb{R}_{\ge 0}$. 
Then the metric $s$-$t$-path TSP asks for a multi-subset $H$ of $E(G)$ minimizing $c(H) = \sum_{e\in H} c(e)$
such that all vertices in $V(G)\setminus (\{s\} \triangle \{t\})$ have even degree and vertices in $\{s\} \triangle \{t\}$ have 
odd degree. If $c(e) =1$ for all $e\in E(G)$ this is the $s$-$t$-path graph TSP.

\begin{theorem}\label{thm:weighted_long_s_t_dist_easy}
  Let $\delta >0$ and $\alpha>1$ be constants. Assume there exists a polynomial time algorithm that computes for instances of the 
  $s$-$t$-path TSP with $\text{dist}(s,t) \le (\frac{1}{3}+\delta)\cdot\opt$, a solution of length at most $\alpha\cdot\opt$. 
  Then there exists a polynomial time $\alpha$-approximation algorithm for the $s$-$t$-path TSP.
 \end{theorem}

We remark that Theorem \ref{thm:weighted_long_s_t_dist_easy} does not directly imply Theorem \ref{thm:long_s_t_dist_easy}.
The reason is that the given $\alpha$-approximation algorithm for instances with $d=\text{dist}(s,t) \le (\frac{1}{3}+\delta)\cdot\opt$
might be applied not only to the original given instance, but also to other instances.
However, our algorithm will apply the given $\alpha$-approximation algorithm only to subinstances of the original problem, i.e.
where we ask for an $s'$-$t'$-tour in a 2-vertex-connected subgraph $G[V']$ for $s',t' \in V' \subseteq V(G)$.
Since such subinstances of $s$-$t$-path graph TSP instances are again instances of the $s$-$t$-path graph TSP, 
our proof of Theorem \ref{thm:weighted_long_s_t_dist_easy} will also imply Theorem \ref{thm:long_s_t_dist_easy}.

Our algorithm uses an approximation algorithm for the $s$-$t$-path (graph) TSP 
as a subroutine to compute several tours, each of them visiting only a subset of the vertices. 
We then use dynamic programming to construct an $s$-$t$-tour by combining a selection of these tours. 
A similar dynamic program was used by \cite{Orienteering} to obtain the first constant-factor 
approximation algorithm for the orienteering problem, and also by \cite{TraV18} and \cite{Zen18} for the $s$-$t$-path TSP.

The approximation guarantee of our dynamic program will depend on the approximation guarantee of the algorithm used as a subroutine.
Assume our subroutine is a $\beta$-approximation algorithm for the $s$-$t$-path (graph) TSP and 
achieves an  approximation ratio of $\alpha$ in the special case where $s$ and $t$ are ``close'', where $1 < \alpha < \beta$.
Then our dynamic program will yield an approximation ratio better than that of the subroutine, i.e. better than $\beta$.
Using the dynamic program recursively as a subroutine for our algorithm, we obtain after a constant number of recursions
the approximation guarantee of $\alpha$ also for the general case of the $s$-$t$-path TSP.

We now describe the dynamic programming algorithm in detail. 
We first compute an embedding of the vertices that we will use later to derive lower bounds on the length of some edges.
Let for $v,w\in V(G)$, the distance ${\rm dist}(v, w)$ denote the length of a shortest $v$-$w$-path in $G$ with respect to the edge weights $c$.

\begin{lemma}\label{lemma:embedding}
 There exists a function $f : V(G) \rightarrow [0,{\rm dist}(s,t)]$ such that $f(s) = 0$, $f(t)={\rm dist}(s,t)$, and
 $c(\{v,w\}) \ge |f(v)-f(w)|$ for every edge $\{v,w\}$. Such an $f$ can be computed in polynomial time.
\end{lemma}
\prove 
We set $f(v):=\min \{{\rm dist}(s,v), {\rm dist}(s,t)\}$ for every vertex $v\in V(G)$. Then, $f(s) = 0$, $f(t)={\rm dist}(s,t)$, 
and $f(v)\in [0,{\rm dist}(s,t)]$ for every vertex $v$.
Let $\{v,w\}\in E(G)$. Then,
\begin{align*}
  f(w) + {\rm dist}(v,w) &\ = \ \min \{{\rm dist}(s,w)+ {\rm dist}(v,w), {\rm dist}(s,t) + {\rm dist}(v,w) \} \\
  &\ \ge \ \min \{{\rm dist}(s,v), {\rm dist}(s,t)\} \\ 
  &\ = \ f(v),
\end{align*} 
hence $|f(v)-f(w)| \le {\rm dist}(v,w) \le c(\{v,w\})$.
\endproof 

Let $f$ be a function as in Lemma \ref{lemma:embedding} and $V(G)=\{v_1,\dots,v_n\}$ such that
$s=v_1$, $t=v_n$ and $f(v_1)\le f(v_2) \le \dots \le f(v_{n-1}) \le f(v_n)$. 
We construct a directed auxiliary graph $D$, 
whose arcs will correspond to paths visiting a subset of the vertex set $V(G)$.
The vertex set of $D$ is given by
 \begin{align*}
  V(D) \ := \ &\bigl\{ (U,u,w) : U = \{v_1, \dots, v_k\}\text{ for some }k,\, \{u,w\}\in E(G),\,
                             u\in U,\, w\in V(G)\setminus U  \bigr\} \\
 &\cupp \bigl\{(\emptyset, \emptyset, s), (V(G), t, \emptyset) \bigr\}
 \end{align*}
and the arc set by
\begin{align*}
 A(D) \ := \ \bigl\{ ((U, u, w), (U', u', w')) :\ & U \subseteq U' , \ w,u' \in U' \setminus U \bigr\}.
\end{align*}

Note that $D$ has a polynomial number of vertices and arcs.
The next step of the algorithm is to compute weights $d$ for the arcs of $D$. 
The weight of an arc $((U, u, w), (U', u', w')) \in A(D)$ is given by the length 
of an approximately shortest $w$-$u'$-tour in $G[U' \setminus U]$
plus the cost $c(\{u', w'\})$, where we set $c(\{t, \emptyset\}):= 0$.

We obtain such an approximately shortest $w$-$u'$-tour in in $G[U' \setminus U]$ by applying a
$\beta$-approximation algorithm for the $s$-$t$-path (graph) TSP for some constant $\beta$.

Finally, we compute a shortest $(\emptyset, \emptyset, s)$-$(V(G), t, \emptyset)$-path in $D$.
We construct our $s$-$t$-tour to be the union of the tours defining the weights of the arcs of this path and the edges 
$\{u,w\}$ for every internal vertex $(U, u, w)$ of this path in $D$.
Let again $\opt$ denote the length of an optimum solution to a given instance of the $s$-$t$-path TSP. 
\begin{lemma}\label{lemma:improvement_dp}
 Let $\delta >0$ be a fixed constant.
 For an instance of the $s$-$t$-path TSP with a 2-vertex-connected graph, in which $s$ and $t$ have distance at least $(\frac{1}{3} + \delta)\opt$, 
 the algorithm described above yields an $s$-$t$-tour of length at most $\left(\beta -  \frac{3}{2}(\beta -1)\delta\right)\opt$.
\end{lemma}
\prove 
By our choice of the arc weights in $D$, the constructed tour has length $d(P)$,
where $P$ is a shortest $(\emptyset, \emptyset, s)$-$(V(G), t, \emptyset)$-path in $D$.
Thus, it suffices to prove that there exists a
$(\emptyset, \emptyset, s)$-$(V(G), t, \emptyset)$-path $P'$ in $D$ 
with $d(P')\le(\beta -  \frac{3}{2}(\beta -1)\delta)\opt$.

Consider the cuts $C_i := \delta(\{v_1, \dots, v_i\})$ for $i\in\{1,\dots,n-1\}$.
Note that for an edge $e=\{v_i,v_j\}\in E(G)$ with $i<j$, 
we have $\{ l  :  e\in C_l \} = \{ i, i+1,\dots, j-1 \}$.
By choice of the function $f$, this implies
\begin{equation}\label{eq:bound_edge_length}
  c(e) \ \ge \ f(v_j) - f(v_i) \ = \ \sum_{l=i}^{j-1} (f(v_{l+1}) -f(v_l)) \ = \ \sum_{l  :  e\in C_l} (f(v_{l+1}) -f(v_l)). 
\end{equation}
Now consider an optimum $s$-$t$-tour $H$. By (\ref{eq:bound_edge_length}), we have
\begin{equation}\label{eq:sum_cuts}
 c(H) \ \ge \ \sum_{e\in H} \sum_{l  :  e\in C_l} (f(v_{l+1}) -f(v_l)) \ = \ \sum_{l=1}^{n-1} |C_l\cap H|\cdot(f(v_{l+1}) -f(v_l)).
\end{equation}
Let $\Lscr := \bigl\{ C_l  :  |C_l \cap H| = 1,\, l\in \{1,\dots,n-1\}  \bigr\}$.
We now show that the summands corresponding to a cut $C_l \in \Lscr$ contribute at least 
$\frac{3}{2}\delta\cdot c(H)$ to \eqref{eq:sum_cuts}.
Each cut $C_l$ ($l\in \{1,\dots, n-1\}$) separates $s$ and $t$ and thus, $|C_l\cap H|$ is odd and at least one. 
This implies
\begin{align*}
  c(H) &\ \ge \ \sum_{l :  C_l\not \in \Lscr} 3 \cdot(f(v_{l+1}) -f(v_l))\ + \sum_{l  :  C_l \in \Lscr} (f(v_{l+1}) -f(v_l)) \\[2mm]
 &\ = \ 3 \sum_{l=1}^{n-1}(f(v_{l+1}) -f(v_l))\ \ -\ 2 \sum_{l  :  C_l \in \Lscr} (f(v_{l+1}) -f(v_l)) \\[2mm]
 &\ = \ 3 \cdot  {\rm dist}(s,t) \ \ -\ 2 \sum_{l :  C_l \in \Lscr} (f(v_{l+1}) -f(v_l)) \\[2mm]
 &\ \ge \ \left( 1 +3\,\delta\right) c(H)  \ -\ 2 \sum_{l  :  C_l \in \Lscr} (f(v_{l+1}) -f(v_l))
\end{align*}
showing that
\begin{equation} \label{eq:fraction_lonely_edges}
  \sum_{l  :  C_l \in \Lscr} (f(v_{l+1}) -f(v_l)) \ \ge\ \sfrac{3}{2}\delta \cdot  c(H).
\end{equation}
Assume for the sake of contradiction that some edge $e\in H$ is contained in at least two cuts $C_i, C_j \in \Lscr$ with $i<j$.
By definition of $\Lscr$ we have $|C_i \cap H| = |C_j \cap H| =1$ and hence, $C_i = \{e\} = C_j$. 
Then, $\delta(\{v_{i+1}, \dots, v_j\}) \cap H = (C_i \cup C_j)\setminus (C_i \cap C_j) = \emptyset$, 
but $\{v_{i+1}, \dots, v_j\} \not = \emptyset$,  a contradiction to $H$ visiting every vertex. 
Thus, every edge $e$ is contained in at most one cut $C\in \Lscr$.\\

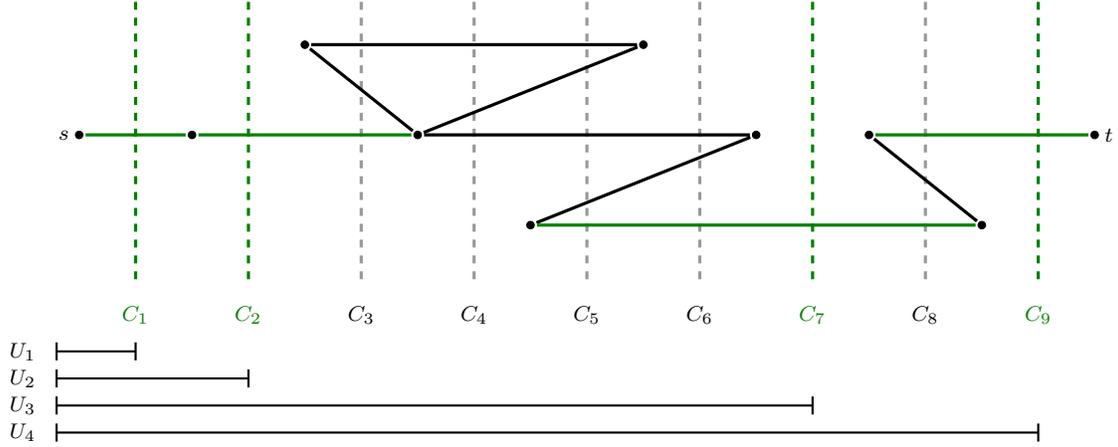
\begin{figure}
 \begin{center}
   \begin{tikzpicture}[xscale=1.5, yscale=1.2]
  %cuts
 \foreach \x in { 3, 4, 5, 6, 8} {
   \draw [dashed, very thick, gray, opacity =0.8] (\x +0.5,0.9) --(\x + 0.5,4);
   \node (c\x) at (\x +0.5, 0.5) {\scriptsize $C_{\x}$};
 }
 \foreach \x  in {1, 2, 7, 9} {
   \draw [dashed, very thick, darkgreen] (\x +0.5,0.9) --(\x + 0.5,4);
   \node[darkgreen] (c\x) at (\x +0.5, 0.5) {\scriptsize $C_{\x}$};
 }
 \begin{scope}[shift = {(0,0.5)}]
   %vertices
\foreach \x\y [count=\i] in {1/2, 2/2, 3/3, 4/2, 5/1, 6/3, 7/2, 8/2, 9/1, 10/2} {
   \node[circle, fill=black, draw=black, inner sep = 1, outer sep = 1] (v\i) at (\x,\y) {};
 }
 \node[left] (s) at (v1) {\scriptsize$s$};
 \node[right] (t) at (v10) {\scriptsize$t$};
 %path
 \draw[very thick, black] (v4) -- (v6) -- (v3);
 \draw[very thick, black] (v3)-- (v4) -- (v7) -- (v5);
 \draw[very thick, black] (v8) -- (v9);
 \begin{scope}[darkgreen, very thick] 
    \draw (v1) -- (v2) -- (v4);
    \draw  (v5)  -- (v9);
    \draw (v8) -- (v10);
 \end{scope}
 \end{scope}
 
%    \begin{scope}[shift = {(0,4.6)}]
%  \draw[black, thick] (1,-0.3) --(10,-0.3);
%  \draw[black, thick] (1,-0.4) --(1, -0.2);
%  \draw[black, thick] (10, -0.4) --(10,-0.2);
%  \node[above=0mm] (dummy) at (1, -0.2) {\scriptsize$0$};
%  \node[above=0mm] (dummy) at (10, -0.2) {\scriptsize${\rm dist}(s,t)$};
%    \end{scope}
   
  \begin{scope}[shift = {(0,-0.5)}]
 \draw[thick] (0.8,-0.3) --(9.5,-0.3);
 \draw[thick] (9.5, -0.4) --(9.5,-0.2);
 \draw[thick] (0.8, -0.4) --(0.8,-0.2);
 \node (dummy) at (0.5, -0.3) {\scriptsize$U_4$};
 \end{scope}
   \begin{scope}[shift = {(0,-0.2)}]
 \draw[thick] (0.8,-0.3) --(7.5,-0.3);
 \draw[thick] (7.5, -0.4) --(7.5,-0.2);
 \draw[thick] (0.8, -0.4) --(0.8,-0.2);
 \node (dummy) at (0.5, -0.3) {\scriptsize$U_3$};
 \end{scope}
  \begin{scope}[shift = {(0,0.1)}]
 \draw[thick] (0.8,-0.3) --(2.5,-0.3);
 \draw[thick] (2.5, -0.4) --(2.5,-0.2);
 \draw[thick] (0.8, -0.4) --(0.8,-0.2);
 \node (dummy) at (0.5, -0.3) {\scriptsize$U_2$};
 \end{scope}
   \begin{scope}[shift = {(0,0.4)}]
 \draw[thick] (0.8,-0.3) --(1.5,-0.3);
 \draw[thick] (1.5, -0.4) --(1.5,-0.2);
 \draw[thick] (0.8, -0.4) --(0.8,-0.2);
 \node (dummy) at (0.5, -0.3) {\scriptsize$U_1$};
 \end{scope}
 \end{tikzpicture}
 \end{center}
 \caption{ 
 The solid lines show the optimum tour $H$, where the green edges are the edges 
 $\{ \{u_{i}, w_{i}\}  :  i=1,\dots,p \}$.
 The dashed vertical lines indicate the cuts $C_1, \dots, C_{n-1}$, where the cuts in $\Lscr$ are drawn in green.
  \label{fig:opt_path}}
\end{figure}

Let $\{s\} \subseteq U_1 \subseteq \dots \subseteq U_p \subseteq V(G)\setminus \{t\}$ such that 
$ \{ \delta(U_1), \dots, \delta(U_p)\} = \Lscr$. Moreover, let 
$\{\{u_{i}, w_{i}\}\} = \delta(U_i) \cap H$ for all $i=1,\dots,p$.
See Figure \ref{fig:opt_path} for an illustration.
\\

\noindent \textbf{Claim:} The path $P^*$ in $D$ defined by the vertices $(\emptyset, \emptyset, s),
(U_{l_1}, u_1, w_1), \dots, (U_{l_p}, u_p, w_p),$ $(V(G), t, \emptyset)$ 
has length  $ d(P^*) \le (\beta -  \frac{3}{2}(\beta -1)\delta)\opt$.
\\

\noindent Since $|H\cap \delta(U_{i+1})| = |H\cap \delta(U_{i})| = 1$, the multi-graph
$H_i := H[U_{i+1}\setminus U_{i}]$ is connected for $i=1,\dots,p-1$.
Moreover, we have $H \cap \delta\left(U_{i+1}\setminus U_{i} \right) = 
\{ \{u_i, w_i \}, \{u_{i+1}, w_{i+1}\}\}$.
Hence $H_i = H[U_{i+1}\setminus U_{i}]$ is a $w_i$-$u_{i+1}$-tour in $G[U_{i+1}\setminus U_{i}]$.
By definition of the arc weights in $D$, it follows that
\[ d\left((U_i, u_i, w_i), (U_{i+1}, u_{i+1}, w_{i+1}) \right) \ \le \ c\left(u_{i+1}, w_{i+1} \right)
+ \beta \cdot c\left(H_i\right).  \]
Similarly,
\[ d\left((\emptyset, \emptyset, s), (U_{1}, u_{1}, w_{1}) \right) \ \le \ c\left(u_{1}, w_{1} \right)
+ \beta \cdot c\left(H[U_1]\right)  \]
and
\[ d\left((U_p, u_p, w_p), (V(G), t, \emptyset) \right) \ \le \ \beta \cdot c\left(H[V(G)\setminus U_p]\right).  \]
Summing up over all arcs of $P^*$, we get 
\begin{equation*}
 \begin{aligned}
   d(P^*) &\ \le \ \beta \cdot c(H) - (\beta -1) \cdot \sum_{i=1}^p c\left( u_i,w_i \right) \\
   &\ \le \ \beta \cdot c(H) - (\beta -1) \cdot \sum_{i=1}^p (f(w_i) - f(u_i))  \\
   &\ \le \ \beta \cdot c(H) - (\beta -1) \cdot \sum_{l  :  C_l\in\Lscr} (f(v_{l+1}) - f(v_{l})) \\
   &\ \le \ \beta \cdot c(H) - (\beta -1) \cdot \sfrac{3}{2}\delta \cdot  c(H),
 \end{aligned}
\end{equation*} 
where the last inequality follows from \eqref{eq:fraction_lonely_edges}.
 \endproof 

\par \noindent \hbox{\bf Proof of Theorem \ref{thm:long_s_t_dist_easy} and Theorem \ref{thm:weighted_long_s_t_dist_easy}:}\quad
 If $\alpha \ge 2$, there exists an $\alpha$-approximation for the $s$-$t$-path TSP; so we may assume $\alpha < 2$.  
 We define a sequence $\beta_i$ by $\beta_0:=2$ and for $i>0$ setting
\[ \beta_i \ := \ \max \left\{\alpha,\ \beta_{i-1} - \sfrac{3}{2}(\beta_{i-1} -1)\cdot \delta\right\}. \] 
Assume we have a polynomial time algorithm for the $s$-$t$-path (graph) TSP with approximation ratio 
at most $\beta_i$ for some $i\ge 0$.

Given an instance of the $s$-$t$-path (graph) TSP, we first decompose into 2-vertex-connected instances at the cut vertices 
(see Section \ref{section:preliminary_T_tours}). For each 2-vertex-connected sub-instance $(G', s', t')$
we compute a $\beta_{i+1}$-approximation as follows:

If the distance of $s'$ and $t'$ is at least $(\frac{1}{3}+\delta)\opt'$,
then applying the algorithm described above yields a $(\beta_{i} - \frac{3}{2}(\beta_{i} -1)\cdot \delta)$-approximation by 
Lemma \ref{lemma:improvement_dp}. Otherwise, by assumption we can compute an $\alpha$-approximation in polynomial time.
This shows that we can compute  a $\beta_{i+1}$-approximation in polynomial time. 

Combining the $\beta_{i+1}$-approximations for the 2-vertex-connected sub-instances yields a $\beta_{i+1}$-approximation for the 
original instance.

By induction, we get that for any constant $k$ there is a polynomial time $\beta_{k}$-approximation algorithm. 
Thus, proving the following claim completes the proof.
\\[2mm]
\textbf{Claim:} For $k= \left\lceil \frac{2-\alpha}{(\alpha - 1)\frac{3}{2}\delta}\right\rceil$ 
we have $\beta_k =\alpha$.\\

\noindent As $\alpha < 2$, we have $\beta_i \ge \alpha$ for all $i$. Moreover, we have 
$\frac{2-\alpha}{\alpha - 1} > 0$. Consider any $i\in \mathbb{N}$.
If $\beta_{i-1} = \alpha$, we have $ \beta_i = \max \left\{\alpha,\ \alpha - (\alpha -1)\cdot \frac{3}{2}\delta\right\} = \alpha.$
Otherwise, we have  
\[ \beta_i \ = \ \max \left\{\alpha,\ \beta_{i-1} - (\beta_{i-1} -1)\cdot \sfrac{3}{2}\delta \right\} \ \le \ \max \left\{\alpha,\ \beta_{i-1} - (\alpha -1)\cdot \frac{3}{2}\delta \right\}. \]
By induction we get $\beta_i \le \max \left\{ \alpha, 2 - i \cdot (\alpha -1)\cdot \frac{3}{2}\delta\right\}.$
Thus, for $k= \left\lceil\frac{2-\alpha}{(\alpha - 1)\frac{3}{2}\delta}\right\rceil$ we have
\begin{align*}
 \beta_k &\ \le \ \max \left\{ \alpha,\ 2 - k \cdot (\alpha -1)\cdot \sfrac{3}{2}\delta\right\} 
 \ \le \  \max \left\{\alpha,\ 2 - \sfrac{2-\alpha}{(\alpha - 1)\sfrac{3}{2}\delta} \cdot (\alpha -1)\cdot \sfrac{3}{2}\delta\right\} \
=\ \alpha.
\end{align*}
\endproof

\section{Conclusion}

Theorem \ref{thm:long_s_t_dist_easy} and Theorem \ref{thm:approx_guarantee} directly imply or main theorem:

\begin{theorem}
There is a polynomial-time $1.497$-approximation algorithm for the $s$-$t$-path graph TSP.
\endproof
\end{theorem}

Our main goal was to show that one can go below the integrality ratio lower bound, even while using the LP in the analysis.
We did not attempt to optimize the running time or the approximation ratio of our algorithm, and improvements are certainly possible
(but would probably make the proof more complicated). 
A more interesting question is whether our new techniques have further applications. 

Given that one can achieve a better approximation ratio for the $s$-$t$-path TSP than $\frac{3}{2}$, the integrality ratio,
it is natural to ask what is the best approximation ratio that can be obtained.
The only lower bound known today (unless P=NP) is $\frac{685}{684}$ (\cite{KarS15}).

Another interesting question is how the integrality ratio depends on the distance of $s$ and $t$.
For $d\in[0,\frac{1}{2}]$, let $\rho(d,n)$ 
denote the integrality ratio for 2-vertex-connected instances with at least $n$ vertices and $\text{dist}(s,t) \le d \cdot n$.
Let $\rho(d) = \lim_{n \rightarrow \infty} \rho(d,n)$.
The proof of Theorem \ref{thm:approx_guarantee} shows that $\rho(d) < \frac{3}{2}$ for all $d < \frac{1}{2}$, and 
the bound improves as $d$ decreases. More precisely, we get (cf.\ Figure \ref{fig:int_ratio_depending_on_d}):

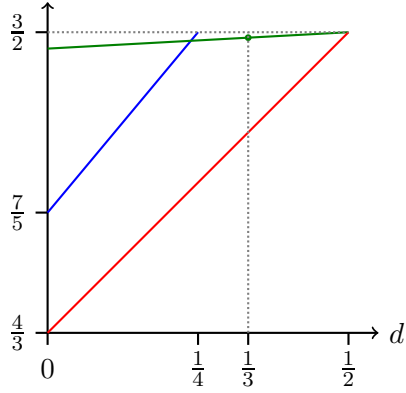
\begin{figure}
\begin{center}
  \begin{tikzpicture}[thick,xscale=0.8, yscale=0.8]
  \draw[->] (-0.2,0) to (5.5,0);
  \draw (-0.2,2) to (0,2);
  \draw (-0.2,5) to (0,5);
  \draw[->] (0,-0.2) to (0,5.5);
  \draw (2.5,-0.2) to (2.5,0);
  \draw (5,-0.2) to (5,0);
  \node at (-0.5,0) {$\frac{4}{3}$};
 \node at (-0.5,2) {$\frac{7}{5}$};
  \node at (-0.5,5) {$\frac{3}{2}$};
  \node at (0,-0.6) {0};
  \node at (2.5,-0.6) {$\frac{1}{4}$};
  \node at (5,-0.6) {$\frac{1}{2}$};
  \node at (5.8,0) {$d$};
  \draw[blue] (0,2) to (2.5,5);
  \draw[darkgreen] (0,4.727272) to (5,5);
  \node[fill,circle,darkgreen, inner sep=0.9] at (3.3333333,4.90909091) {};
  \draw[red] (0,0) to (5,5);
  \draw[gray, densely dotted] (0,5) to (5,5);
  \draw[gray, densely dotted] (3.3333333,0) to (3.3333333,5);
  \node at (3.3333333,-0.6) {$\frac{1}{3}$};
  \draw (3.3333333,-0.2) to (3.3333333,0);
\end{tikzpicture}
 \end{center}
 \caption{Lower bound (red) and upper bounds (green and blue) on the integrality ratio depending on $d=\frac{\text{dist}(s,t)}{n}$
 (for $n$ large enough).
 Section \ref{section:large_distance} shows that it suffices to consider instances with $d\le \frac{1}{3} + \epsilon$.
 \label{fig:int_ratio_depending_on_d}}
\end{figure}

\begin{theorem}
For all $d\in[0,\frac{1}{2}]$ we have $\frac{4+d}{3} \le \rho(d) \le \min\bigl\{\frac{82+d}{55},\,\frac{7+2d}{5}\bigr\}$.
\end{theorem}

\prove
For the first upper bound (green in Figure \ref{fig:int_ratio_depending_on_d}), 
we compute a well-oriented ear decomposition as in Theorem \ref{thm:raise_lb} and consider two different bounds.
Lemma \ref{lemma:MS_few_ears} yields an $s$-$t$-tour with at most $\frac{4}{3}(n-1)+\frac{2}{3}\pi + \frac{1}{3}dn$ edges.
Theorem \ref{thm:many_pendant} yields an $s$-$t$-tour with at most $(\frac{3}{2}-\frac{1}{26}\frac{\pi}{n-1})\lp + 3$ edges.
Taking $\frac{3}{55}$ times the first bound and $\frac{52}{55}$ times the second bound yields the upper bound $\frac{82+d}{55}\lp+3$.
This shows $\rho(d)\le\frac{82+d}{55}$.

If the distance from $s$ to $t$ is very small ($d<\frac{5}{21}$), a better upper bound $\rho(d)\le \sfrac{7+2d}{5}$
(blue in Figure \ref{fig:int_ratio_depending_on_d})
can be derived from \cite{SebV12}; see Theorem \ref{thm:bound_s_t_tour_sv} in the appendix.

\begin{figure}
\begin{center}
  \begin{tikzpicture}[thick,xscale=1.3, yscale=1.6]
  \begin{scope}[shift={(-4,0)}]
   \node[circle, minimum size=3, inner sep=1, draw] (v) at (1,0) {};
    \node[circle, minimum size=3, inner sep=1, draw] (w) at (3.5,0) {};
   \node[left=1pt] at (v) {$s\!=\!t\!=\!v$};
   \draw (v) to node[below] {$\frac{n}{3}+1$} (w); 
   \draw[bend left=80] (v) to node[above] {$\frac{n}{3}$} (w); 
   \draw[bend right=80] (v) to node[below] {$\frac{n}{3}$} (w); 
    \node[right=1pt] at (w) {$w$};
  \end{scope}

 \node[circle, minimum size=3, inner sep=1, draw] (v) at (1,0) {};
 \node[circle, minimum size=3, inner sep=1, draw] (w) at (4,0) {};
 \node[circle, minimum size=5, inner sep=1, fill] (s) at (1.6,1.2) {};
 \node[circle, minimum size=5, inner sep=1, fill] (t) at (1.6,-1.2) {};
 \node[above=1pt] at (s) { $s$};
 \node[below=1pt] at (t) { $t$};
 \node[left=1pt] at (v) {$v$};
 \node[right=1pt] at (w) {$w$};
 \draw (v) to node[left=2pt] {$\frac{dn}{2}$} (s);
 \draw (v) to node[left=2pt] {$\frac{dn}{2}$} (t);
 \draw (w) to node[above] {\qquad$\frac{n}{3}-\frac{dn}{6}$} (s);
 \draw  (w) to node[below] {\qquad$\frac{n}{3}-\frac{dn}{6}$} (t);
 \draw (v) to node[below=-1pt] {$1+\frac{n}{3}-\frac{2dn}{3}$\qquad\qquad} (w); 
 
 \begin{scope}[shift={(4,0)}]
  \node[circle, minimum size=5, inner sep=1, fill] (s) at (2,1.2) {};
  \node[circle, minimum size=5, inner sep=1, fill] (t) at (2,-1.2) {};
   \node[circle, minimum size=3, inner sep=1, draw] (v) at (1.9,0) {};
    \node[circle, minimum size=3, inner sep=1, draw] (w) at (2.1,0) {};
     \node[above=1pt] at (s) { $s$};
 \node[below=1pt] at (t) { $t$};
   \node[left=3pt] at (v) {$v$};
     \node[right=3pt] at (w) {$w$};
 \draw[bend left=70] (v) to node[left=2pt] {$\frac{n}{4}$} (s);
 \draw[bend right=70] (w) to node[right=2pt] {$\frac{n}{4}$} (s);
 \draw[bend left=70] (t) to node[left=2pt] {$\frac{n}{4}$} (v);
 \draw[bend right=70] (t) to node[right=2pt] {$\frac{n}{4}$} (w);
 \draw (v) to node[above] {\scriptsize$1$} (w); 
  \end{scope}
 \end{tikzpicture}

 \end{center}
 \caption{Middle picture: Instances with $\text{dist}(s,t)=dn$ whose integrality ratio tends to $\frac{4+d}{3}$ as $n\to\infty$. 
 Every line represents a path with the indicated number of edges.
 The left picture shows instances with $d=0$ whose integrality ratio tends to $\frac{4}{3}$ as $n\to\infty$;
 the right picture instances with $d=\frac{n}{2}$ whose integrality ratio tends to $\frac{3}{2}$ as $n\to\infty$.
 These two extreme cases are almost identical to the well-known examples that are the worst known for $s=t$ (left) and $s\not=t$ (right). 
 \label{fig:lower_bound_int_ratio}}
\end{figure}
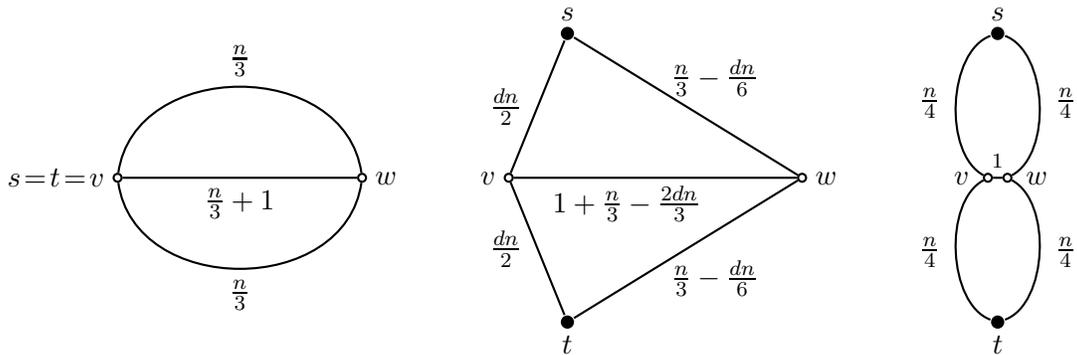

For the lower bound (red in Figure \ref{fig:int_ratio_depending_on_d}), 
note that there are infinitely many integers $n$ for which $\frac{dn}{6}$ and $\frac{n}{3}$ are integers.
We construct a graph $G$ with $n$ vertices (four of which are called $s,t,v,w$) and $n+1$ edges as follows:
join $s$ and $v$ by a path of length $\frac{dn}{2}$,  
join $t$ and $v$ by a path of length $\frac{dn}{2}$,  
join $s$ and $w$ by a path of length $\frac{n}{3}-\frac{dn}{6}$,  
join $t$ and $w$ by a path of length $\frac{n}{3}-\frac{dn}{6}$,  
and join $v$ and $w$ by a path of length $1+\frac{n}{3}-\frac{2dn}{3}$.
See Figure \ref{fig:lower_bound_int_ratio}, and observe that this graph indeed has $n$ vertices.

For this instance of the $s$-$t$-path TSP we have $\lp\le n+1$ because setting $x_e=1$ for all $e\in E(G)$ is a feasible solution.
However, any $s$-$t$-tour contains all but two edges, and some with two copies.
Since the minimum $(\{s\}\triangle\{t\}\triangle\{v\}\triangle\{w\})$-join has 
$\min\{\frac{dn}{2}+\frac{dn}{2}+1+\frac{n}{3}-\frac{2dn}{3},\, \frac{dn}{2}+\frac{n}{3}-\frac{dn}{6} \} = \frac{dn}{3}+\frac{n}{3}$ edges,
any $s$-$t$-tour has at least $\frac{(4+d)n}{3}-3$ edges.
For $n\to\infty$, the ratio converges to $\frac{4+d}{3}$.  
\endproof

One might conjecture that the lower bound (the red curve in Figure \ref{fig:int_ratio_depending_on_d}) is the true answer.
If one could always construct a solution with $\frac{4+d}{3} \lp$ edges in polynomial time, this would imply a $(\frac{13}{9}+\epsilon)$-approximation algorithm
for the graph $s$-$t$-path TSP via Theorem \ref{thm:long_s_t_dist_easy} for any $\epsilon > 0$.

   %%%%%%%%%%%%%%%%%%%%%%%%%%%%%%%%%%%%%%%%%%%%%%%%%%%%%%%%%%%%%%%%%%%%%%%%%%%%%%%%%%%%%%%%%%
\newcommand{\bib}[3]{\bibitem[\protect\citeauthoryear{#1}{#2}]{#3}} 

\newpage
\section*{Appendix}

Here we show how a short $s$-$t$-tour can be obtained from \cite{SebV12} if the distance of $s$ and $t$ is small.
The following lemma is a variation of the Claim in the proof of Theorem 10 of \cite{SebV12}.
All the numbers of lemmata and theorems in this appendix refer to that paper.
We use exactly the terminology of \cite{SebV12} without redefining it. The only difference is that 
the number of pendant ears was called $\pi$ in that paper, but we will call it $\pi^*$ to avoid confusion with the number of non-entered ears.

\begin{lemma}\label{lemma:tour_from_sv}
 Given a graph $G'$ with a nice ear-decomposition without 1-ears and  $s,t\in V(G')$.
 Let  $T=\{s\}\triangle\{t\}$ and $M= \{ \inn(P) : P \text{ clean ear}\}$.
 We can construct an $\emptyset$-tour $H_1$ of $G'$ and an $s$-$t$-tour $H_2$ of $G'$ such that
  \begin{equation}\label{eq:rhs_lemma_sv}
   \sfrac{1}{5} |H_1| + \sfrac{4}{5} |H_2|\ \le\ \sfrac{2}{5} L_{\mu}(G', M) + \sfrac{1}{5}L_{\varphi}(G') + \sfrac{4}{5} (|V(G')|-1).
   \end{equation}
\end{lemma}
\prove
We first prove the lemma for the case where $G'$ is 2-vertex connected.
Let $\pi^*$ be the number of pendant ears. 
First, by Theorem 7, we can compute an $s$-$t$-tour $H_2'$ with at most 
\begin{equation}\label{eq:bound_ear_induction_sv}
L_{\mu}(G',M)+ \sfrac{1}{2} L_{\varphi}(G') - \pi^* 
\end{equation}
 edges.
Now define a removable pairing $(R, \Pscr)$ as in the proof of Lemma 5.3. Add an edge $e' =\{s,t\}$ to $G'$ and declare it removable.
The proof of Theorem 9 yields a convex combination of $\emptyset$-tours in the extended graph such that 
$e'$ appears in exactly $\sfrac{2}{3}$ of these tours and removing $e'$ from these tours does not destroy connectivity and hence 
leaves $s$-$t$-tours.
The average number of edges of these $\emptyset$-tours and $s$-$t$-tours is $\sfrac{4}{3} |E(G')| - \sfrac{2}{3} |R|$.
Hence, we have an $\emptyset$-tour $H_1$ and an $s$-$t$-tour $H_2''$ with
\begin{equation}\label{eq:bound_ms_sv}
\sfrac{1}{3} |H_1| + \sfrac{2}{3} |H_2''|\ \le\ \sfrac{4}{3} |E(G')| - \sfrac{2}{3} |R|\ =\ \sfrac{4}{3}(|V(G')|-1) + \sfrac{2}{3} \pi^*.
\end{equation}
Letting $H_2$ be the smaller one of $H_2'$ and $H_2''$ and taking $\sfrac{2}{5}$ of \eqref{eq:bound_ear_induction_sv} and 
$\sfrac{3}{5}$ of \eqref{eq:bound_ms_sv} completes the proof if $G'$ is 2-vertex-connected.

For the general case, we decompose the graph at cut-vertices and apply induction on $|V(G')|$ as in the proof of Theorem 10.
Let $v\in V(G')$ be a cut-vertex, and $G_1$ and $G_2$ graphs with $G' = (V(G_1)\cup V(G_2), E(G_1) \cup E(G_2))$ and 
$V(G_1)\cap V(G_2)=\{v\}$.
Without loss of generality, $s\in V(G_1)$.
If $t \in V(G_1)$, then we set $s_1=s$, $t_1=t$ and $s_2=t_2=v$.
If $t\in V(G_2)$, we set $s_1=s$, $t_1=s_2=v$, and $t_2=t$. 
Let  $T_i=\{s_i\}\triangle\{t_i\}$ (for $i=1,2$).

Our nice ear-decomposition decomposes into nice ear-decompositions of $G_1$ and $G_2$.
Then we apply  the induction hypothesis to $(G_1,s_1,t_1)$ and $(G_2, s_2, t_2)$ and combine the resulting tours.
Note that $v$ is not an internal vertex of a short ear because all short ears are pendant. 
Let $M_i = \{ \inn(P) : P \text{ short ear with }\inn(P) \subseteq V(G_i)\setminus T_i \}$ for $i=1,2$.
Then $M$ is the disjoint union of $M_1$ and $M_2$.
Hence, as in the proof of Theorem 10, $L_{\mu}(G_1, M_1) + L_{\mu}(G_2, M_2) = L_{\mu}(G', M)$.
Moreover, $L_{\varphi}(G_1) + L_{\varphi}(G_2) =L_{\varphi}(G')$.
So the right-hand sides of \eqref{eq:rhs_lemma_sv} for $G_1$ and $G_2$ add up.
\endproof

\begin{theorem}\label{thm:bound_s_t_tour_sv}
 Given a graph $G$ and $s,t\in V(G)$, we can compute an $s$-$t$-tour with at most 
 $\sfrac{7}{5}\lp + \frac{2}{5}\text{dist}(s,t)$ edges in polynomial time. 
\end{theorem}
\prove
As in the proof of Theorem 10 we may assume that $G$ is 2-vertex connected and construct a nice ear-decomposition containing a maximum
earmuff for the eardrum $M$ associated with it and $T=\{s\}\triangle\{t\}$.
Let $G'$ result from $G$ by deleting the trivial ears. 
We have $L_{\mu}(G', M) = L_{\mu}(G,M) \le \lp$.
Moreover, 
\begin{equation}\label{eq:L_phi_bound}
L_{\varphi}(G')\ =\ L_{\varphi}(G)\ =\  n-1 +\varphi(G)\ \le\ 
\min \bigl\{ x(E(G)) : x(\delta(U)) \ge 2 \ (\emptyset \ne U \subsetneq V(G)), x\ge 0\bigr\}.\end{equation}
                                                                                             
Now let $P$ be a shortest $s$-$t$-path in $G$.
Then for any optimum solution $x^*$ to \eqref{eq:subtour_lp} and the incidence vector $\chi^P$, 
the vector $x^*+\chi^P$ is a feasible solution to 
the linear program on the right-hand side of \eqref{eq:L_phi_bound}, and thus $L_{\varphi}(G') \le \lp +\text{dist}(s,t)$.
We apply Lemma \ref{lemma:tour_from_sv} to $G'$ and get an $\emptyset$-tour $H_1$ and an $s$-$t$-tour $H_2$.
The cheaper one of the $s$-$t$-tours $H_1 \cupp E(P)$ and $H_2$ has at most 
\[ \sfrac{1}{5} (|H_1|+|E(P)|) + \sfrac{4}{5} |H_2|\ \le\ \sfrac{2}{5} L_{\mu}(G', M) + 
\sfrac{1}{5}L_{\varphi}(G') + \sfrac{4}{5} (n-1)+ \sfrac{1}{5}\text{dist}(s,t) \]
edges. Using $L_{\mu}(G', M)\ \le\ \lp$, $L_{\varphi}(G')\ \le\ \lp +\text{dist}(s,t)$ and $n-1 \le \lp$, the number of edges is at most 
\[ \sfrac{7}{5}\lp + \sfrac{2}{5}\text{dist}(s,t).\]
\endproof

\end{document}